\documentclass[onefignum,onetabnum]{siamart220329}

\usepackage{lipsum}
\usepackage{amsfonts}
\usepackage{graphicx}
\usepackage{epstopdf}
\usepackage{algorithmic}
\ifpdf
  \DeclareGraphicsExtensions{.eps,.pdf,.png,.jpg}
\else
  \DeclareGraphicsExtensions{.eps}
\fi

\newsiamremark{remark}{Remark}
\newsiamremark{hypothesis}{Hypothesis}
\crefname{hypothesis}{Hypothesis}{Hypotheses}
\newsiamthm{claim}{Claim}
\newsiamthm{assum}{Assumption}
\newsiamthm{scheme}{Scheme}
\newsiamthm{lem}{Lemma}
\newsiamthm{thm}{Theorem}
\headers{EnSF with inpainting for the SQG model}{S.~Liang, H.~Tran, F.~Bao, H.~G.~Chipilski, P.~J.~van Leeuwen, G.~Zhang}

\title{Ensemble score filter with image inpainting for data assimilation in tracking surface quasi-geostrophic dynamics with partial observations}

\author{Siming Liang\thanks{Department of Mathematics, Florida State University,
 Tallahassee, FL (\email{sliang@fsu.edu}, \email{fbao@fsu.edu}).}
 \and
Hoang Tran\thanks{Computer Science and Mathematics Division, Oak Ridge National Laboratory, Oak Ridge, TN (\email{tranha@ornl.gov}).}
\and 
Feng Bao\footnotemark[1]
 \and 
Hristo G.~Chipilski\thanks{Department of Scientific Computing, Florida State University, Tallahassee, FL (\email{hchipilski@fsu.edu}).}
\and 
\newline
Peter Jan van Leeuwen\thanks{Department of Atmospheric Science, Colorado State University, Fort Collins, CO (\email{Peter.vanLeeuwen@colostate.edu}).}
\and
Guannan Zhang\thanks{Corresponding author, Computer Science and Mathematics Division, Oak Ridge National Laboratory, Oak Ridge, TN (\email{zhangg@ornl.gov}).}
}

\usepackage{amsopn}

\usepackage{bm}
\usepackage{enumitem}
\usepackage{cite}
\usepackage{subcaption}
\usepackage{booktabs}
\makeatletter
\DeclareRobustCommand{\cev}[1]{%
  \mathpalette\do@cev{#1}%
}
\newcommand{\do@cev}[2]{%
  \fix@cev{#1}{+}%
  \reflectbox{$\m@th#1\vec{\reflectbox{$\fix@cev{#1}{-}\m@th#1#2\fix@cev{#1}{+}$}}$}%
  \fix@cev{#1}{-}%
}
\newcommand{\fix@cev}[2]{%
  \ifx#1\displaystyle
    \mkern#23mu
  \else
    \ifx#1\textstyle
      \mkern#23mu
    \else
      \ifx#1\scriptstyle
        \mkern#22mu
      \else
        \mkern#22mu
      \fi
    \fi
  \fi
}

\usepackage{color}

\allowdisplaybreaks
\begin{document}

\maketitle

\begin{abstract}
Data assimilation plays a pivotal role in understanding and predicting turbulent systems within geoscience and weather forecasting, where data assimilation is used to address three fundamental challenges, i.e., high-dimensionality, nonlinearity, and partial observations. Recent advances in machine learning (ML)-based data assimilation methods have demonstrated encouraging results. In this work, we develop an ensemble score filter (EnSF) that integrates image inpainting to solve the data assimilation problems with partial observations. 
The EnSF method, proposed in our previous work \cite{ensf_cmame}, exploits an exclusively designed training-free diffusion models to solve high-dimensional nonlinear data assimilation problems. Its performance has been successfully demonstrated in the context of having full observations, i.e., all the state variables are directly or indirectly observed. However, because the EnSF does not use a covariance matrix to capture the dependence between the observed and unobserved state variables, it is nontrivial to extend the original EnSF method to the partial observation scenario. In this work, we incorporate various image inpainting techniques into the EnSF to predict the unobserved states during data assimilation. At each filtering step, we first use the diffusion model to estimate the observed states by integrating the likelihood information into the score function. Then, we use image inpainting methods to predict the unobserved state variables. We demonstrate the performance of the EnSF with inpainting by tracking the Surface Quasi-Geostrophic (SQG) model dynamics under a variety of scenarios. The successful proof of concept paves the way to more in-depth investigations on exploiting modern image inpainting techniques to advance data assimilation methodology for practical geoscience and weather forecasting problems. 
\end{abstract}



\section{Introduction}\label{sec:intro}
Data assimilation plays a pivotal role in understanding and predicting turbulent systems within geoscience and weather forecasting \cite{doi:10.1137/1.9781611974546,Carrassi2017DataAI}, where chaotic dynamics and multiscale interactions create significant challenges for accurate state estimation. In atmospheric sciences, turbulent processes span vast spatial and temporal scales, from local wind patterns to global circulation systems, making it impossible to obtain complete observational coverage of the atmosphere's state at any given time. Data assimilation techniques bridge this gap by systematically combining sparse, noisy measurements from various sources with sophisticated numerical models that capture the underlying physics of atmospheric flow. 

The development of data assimilation methods faces three fundamental challenges. First, high-dimensionality of the state space poses a substantial computational burden because modern numerical models often involve millions or even billions of state variables, making direct manipulation of high-dimensional probability distributions computationally intractable. Second, nonlinear observations introduce difficulties in state estimation, as the relationship between the system state and measurements may be highly complex and non-Gaussian. These nonlinearities can arise from various sources, such as radar reflectivity data, where the observation operator involves complex physics. Third, partial observation is another obstacle because many real-world systems can only be observed sparsely in both space and time, with certain critical state variables remaining completely unobserved. This limitation is particularly evident in geophysical applications where vast regions may lack direct measurements, and crucial variables must be inferred indirectly from available observations, leading to substantial uncertainty in state estimation and forecast accuracy.

There are a variety of data assimilation approaches, each of which was designed to tackle specific aspects of the challenges, but none of them comprehensively addresses all three fundamental difficulties. Traditional ensemble Kalman filters (EnKF) \cite{https://doi.org/10.1029/94JC00572,DataAssimilationUsinganEnsembleKalmanFilterTechnique,10.5555/1206873} and their variants, e.g., the Local Ensemble Transform Kalman Filter (LETKF) \cite{hunt_et_al_2007,10.1145/3581784.3627047}, are capable of handling high-dimensional systems through efficient sample-based covariance approximation and localization techniques, but their underlying Gaussian assumptions severely limit their effectiveness with nonlinear observations. Particle filters (PF) \cite{pulido_vanLeeuwen_2019,crisan_doucet_2002,rojahn_et_al_2023 ,APF,particle-filter}, conversely, can theoretically handle arbitrary nonlinearity and non-Gaussian distributions, but suffer from the curse of dimensionality, requiring an exponentially growing number of particles in high-dimensional spaces to prevent degeneracy. Hybrid methods, such as the ensemble transform particle filter and nonlinear ensemble transform filter, attempt to bridge this gap by combining the dimensional scalability of EnKF with the nonlinear capabilities of PF, but still struggle with partial observations and often require careful tuning of localization parameters. Variational methods like 4D-Var provide a systematic framework for handling partial observations through the incorporation of model dynamics, but their optimization-based approach becomes computationally prohibitive for strongly nonlinear systems and relies heavily on the quality of the background error covariance specification. Even fully nonlinear particle flow methods rely on covariance information in the preconditioner used to accelerate convergence \cite{hu_vanLeeuwen_2021}.
Recent advances in machine learning (ML)-based data assimilation methods \cite{chen_et_al_2021,ensf_cmame,bao2023scorebased}, especially those employing deep neural networks, have demonstrated encouraging results in handling nonlinear dynamics and complex error structures. However, their practical implementation faces significant hurdles due to the complex training process. Moreover, due to the dynamical nature of data assimilation, ML-based method usually require re-training during the assimilation process, which adds another layer of complexity to their performances.

In this work, we develop an ensemble score filter (EnSF) with image inpainting to solve the data assimilation problem with partial observations. 
The EnSF method, proposed in our previous work \cite{ensf_cmame}, exploits exclusively designed training-free diffusion models to solve high-dimensional nonlinear data assimilation problems. Its performance has been successfully demonstrated in the context of having full observations, i.e., all the state variables are directly or indirectly observed. However, because EnSF does not use covariance information to capture the dependence between the observed and unobserved state variables, it is nontrivial to extend the original EnSF method to the partial observation scenario. In \cite{si2024latentensflatentensemblescore}, auto-encoders are used to encode the state space into a latent space then run EnSF in the latent space. It provides promising results in predicting unobserved state, but it requires an offline training process with sufficient high-quality training data, which may not be avaiable in real-world data assimilation problems. 
In this work, we incorporate various image inpainting techniques into the EnSF to estimate the unobserved state variables during data assimilation. At each filtering step, we first use the diffusion model to predict the observed states by integrating the likelihood information into the score function. Then, we use image inpainting methods to predict the unobserved states. We demonstrate the performance of EnSF with inpainting by tracking the Surface Quasi-Geostrophic (SQG) model dyanmics. The SQG  model serves as an invaluable testbed for developing and evaluating data assimilation methods due to its unique characteristics that bridge theoretical analysis and practical applications. 

The rest of the paper is organized as follows. In Section \ref{sec:prob}, we introduce the data assimilation with partial observations using the SQG model. In Section \ref{sec:methodology}, we provide details of the proposed EnSF with inpainting. A variety of numerical experiments are provided in Section \ref{sec:exp}, and concluding remarks are given in Section \ref{sec:con}.

\section{Data assimilation for the SQG model}\label{sec:prob}
We set up the nonlinear data assimilation with partial observations for tracking the dynamics of the SQG model. The process of combining numerical predictions with measured observations allows us to find the best possible estimate of how a stochastic dynamic system evolves over time. Data assimilation has found widespread use across disciplines, from predicting atmospheric conditions to monitoring moving objects. While the underlying stochastic dynamical system operates continuously in time, practical implementation requires working with discrete time steps since measurements are typically taken at specific intervals rather than continuously. Specifically, we are interested in the following discrete stochastic dynamical system:
\begin{equation}\label{eq:filtering-state} 
    \text{\bf State:}\quad X_{n+1} =\; \mathcal{F}(X_{n})+\omega_{n},
\end{equation}
where $F: \mathbb{R}^d \mapsto \mathbb{R}^d$ is a nonlinear physical model, e.g., the SQG model described in Section \ref{sec:SQG}, that maps the state from the time step $n$ to the time step $n+1$, and the random variable $\omega_n \in \mathbb{R}^d$ represents the model error of $\mathcal{F}$ (i.e., we assume the model is not perfect). We divide the state vector $X_n$ into two subsets, denoted by
\begin{equation}\label{eq:state_obs}
    X_n = (X_n^{\rm obs}, X_n^{\rm unobs}), 
\end{equation}
where $X_n^{\rm obs}, X_n^{\rm unobs}$ represent the observed and unobserved state variables, respectively. 
To correct the model error $\omega_n$ and keep tracking the dynamics of $X_n$, we use a sequence of observations given by
\begin{equation}\label{observations}
\text{\bf Observation:}\;\;  Y_{n} = \mathcal{H}(X_n^{\rm obs}) + \varepsilon_n,
\end{equation}
where $\mathcal{H}$ is the observation operator mapping the observed states $X_n^{\rm obs}$ to observation data $Y_n$ and $\varepsilon_n$ is the observation error. 

The study approaches data assimilation through sequential Bayesian inference, aiming to compute probabilistic state estimates. The central task involves calculating the posterior filtering distribution of $X_{n+1}$, conditioned on the complete observation history $Y_{1:n} := {Y_1, \ldots, Y_n}$. The process operates iteratively, namely, when new measurements become available, the forecast-based prior distribution undergoes Bayesian updating to generate a posterior distribution that optimally combines model predictions with observational evidence to characterize the system state.
Mathematically, at each iteration of the sequential Bayesian inference, we need to perform the following two steps:
\vspace{0.2cm}
\begin{itemize}[leftmargin=15pt]\itemsep0.2cm
    \item {\it The prediction step.} The transition from time step $n$ to $n+1$ involves evolving the posterior filtering distribution $p_{X_{n} | Y_{1:n}}$ through the Chapman-Kolmogorov equation, producing a forecast that precedes the assimilation of the newest measurement $Y_{n+1}$, i.e.,
\begin{equation}\label{eq:filtering-prior-ck}
p_{X_{n+1} | Y_{1:n}}(x_{n+1}) = \int p_{X_{n+1} | X_n}(x_{n+1} | x_n) p_{X_n | Y_{1:n}}(x_n) d x_{n},
\end{equation}
where $p_{X_{n+1} | X_n}(x_{n+1} | x_n)$ is the transition probability derived from the state dynamics in Eq.~\eqref{eq:filtering-state}. The distribution $p_{X_{n+1} | Y_{1:n}}$ in Eq.~\eqref{eq:filtering-prior-ck} is referred to as the {\it prior filtering distribution} at the $(n+1)$-th time step. The outcome of this step is an ensemble of samples of $X_{n+1} | Y_{1:n}$, denoted by
\begin{equation}\label{eq:prior_sample}
    \mathcal{D}_{n+1}^{\rm prior} := \left\{x_{n+1|n}^1, \ldots, x_{n+1|n}^K\right\}.
\end{equation}
\item {\it The update step.} 
We apply Bayes' theorem to update the prior filtering distribution in Eq.\eqref{eq:filtering-prior-ck} by assimilating the latest measurement $Y_{n+1}$ defined in Eq.\eqref{observations}, yielding the posterior filtering distribution through:
\begin{equation}\label{eq:filtering-bayesian-update}
\underbrace{p_{X_{n+1} | Y_{1:n+1}} (x_{n+1})}_{\rm Posterior} \,\propto\, \underbrace{p_{X_{n+1} | Y_{1:n}}(x_{n+1})}_{\rm Prior} \;\underbrace{p_{Y_{n+1} | X_{n+1}}(x_{n+1})}_{\rm Likelihood},
\end{equation}
where the likelihood $p_{Y_{n+1} | X_{n+1}}(x_{n+1})$ is defined based on the observation model Eq.~\eqref{observations}, i.e., 
\begin{equation}\label{eq:likelihood}
    p_{Y_{n+1} | X_{n+1}}(x_{n+1}) \propto \exp\Big[ - 1/2(y_{n+1}- \mathcal{H}(x_{n+1}^{\rm obs}))^\top R^{-1} (y_{n+1}- \mathcal{H}(x_{n+1}^{\rm obs})) \Big],
\end{equation}
under the assumption that the observation noise $\varepsilon_n$ follows the Gaussian distribution $\mathcal{N}(0,R)$. The outcome of this step is an ensemble of samples of $X_{n+1} | Y_{1:n+1}$, denoted by
\begin{equation}\label{eq:post_sample}
    \mathcal{D}_{n+1}^{\rm posterior} := \left\{x_{n+1|n+1}^1, \ldots, x_{n+1|n+1}^K\right\},
\end{equation}
which then will serve as the starting point for the next Bayesian iteration.
\end{itemize}

\subsection{The surface quasi-geostrophic (SQG) model}\label{sec:SQG}
The physical problem under consideration is based on the benchmark model for the surface quasi-geostrophic (SQG) dynamics \cite{tulloch_smith_2009a}. The SQG model belongs to a special class of quasi-geostrophic models in which a fluid in which potential vorticity (PV) on fluid parcels is conserved, and which is bounded between two flat, rigid surfaces \cite{tulloch_smith_2009a}. Despite its idealized nature, the system is capable of simulating turbulent motions similar to those occurring in real geophysical flows \cite{smith_et_al_2023}. It is also suitable for data assimilation studies because the SQG flow is inherently chaotic and sensitive to initial condition errors \cite{rotunno_snyder_2008,durran_gingrich_2014}. We adopt the SQG formulation proposed by \cite{tulloch_smith_2009b} where the dynamics reduce to the nonlinear Eady model that uses an f-plane approximation (the Coriolis parameter set to a constant value) with uniform stratification and shear. In this setting, the governing equations simplify to the advection of scaled potential temperature $\theta(x,y,z,t)$ on the bounding surfaces $z=0$ and $z=H$,
\begin{equation} \label{eq:SQG_dynamics}
\frac{\partial \theta}{\partial t} = - J(\psi, \theta) - U \frac{\partial \theta}{\partial x} - v \Theta_y.
\end{equation}
where $J$ is the two-dimensional Jacobian operator defined by
\[
J:f,g \mapsto \frac{\partial f}{\partial x}\frac{\partial g}{\partial y} - \frac{\partial f}{\partial y}\frac{\partial g}{\partial x}
\]
for two differentiable functions $f$ and $g$. The geostrophic streamfunction $\psi(x,y,z,t)$ is coupled to the two velocity components ($u$ and $v$) and the scaled potential temperature via 
\[
(u,v,\theta) = \left(-\frac{\partial \psi}{\partial y},\frac{\partial \psi}{\partial x},f \frac{\partial \psi}{\partial z}\right).
\]
Notice that the simplified formulation of \cite{tulloch_smith_2009b} uses a mean baroclinic zonal wind $U(z)$ which only depends on height and follows the gradient-wind balance. In view of this approximation, the meridional component of the wind $v$ acts on the mean gradient of potential temperature $ \Theta_y = -{dU}/{dz}$. 
Those equations are solved numerically by first applying a fast Fourier transform (FFT) to map model variables to spectral space. They are integrated forward with a 4$^{\text{th}}$-order Runge Kutta scheme that uses a $2/3$ dealiasing rule and implicit treatment of hyperdiffusion. For more details, readers are directed to the open-source GitHub repository of the model, which can be accessed via \url{https://github.com/jswhit/sqgturb}.

The SQG model presents an ideal framework for studying data assimilation with partial observations in geophysical flows, combining physical realism with computational tractability. The model's ability to generate turbulent behavior representative of real atmospheric dynamics is evidenced by its kinetic energy density spectrum with a -5/3 slope, which matches field measurements from atmospheric campaigns \cite{nastrom_gage_1985}. This turbulent character, while making the system challenging to predict due to rapid error growth as identified by Lorenz \cite{lorenz_1969}, makes it particularly valuable for studying data assimilation under partial observations. 

The fundamental challenge in tracking turbulent systems lies in the practical impossibility of measuring all relevant state variables, as sensors are often limited to discrete spatial locations or specific physical quantities. This partial observability becomes especially critical in turbulent flows where complex multi-scale interactions and sensitive dependence on initial conditions govern the dynamics. The SQG model, despite being simpler than operational Numerical Weather Prediction (NWP) systems, captures these essential characteristics of atmospheric turbulence that limit weather predictability to approximately two weeks \cite{durran_gingrich_2014}. Its ability to generate realistic turbulence while remaining computationally manageable makes it an excellent testbed for investigating how unobserved components influence system evolution through nonlinear coupling. Furthermore, the high-dimensional nature of the SQG model reflects the real-world challenge where the number of degrees of freedom typically exceeds available measurements, creating an underdetermined problem where multiple state configurations could explain the same partial observations.

\section{Ensemble score filter (EnSF) with inpainting}\label{sec:methodology}
We discuss details of the proposed method in this section.
In Section \ref{sec:ensf_overview}, we briefly overview the EnSF method studied in \cite{ensf_cmame,bao2023scorebased,bao2024nonlinearensemblefilteringdiffusion} with the emphasis on the limitation of the current EnSF in handling the scenario of having partial observation. The challenge is addressed by two types of inpainting methods, i.e., the PDE-based inpainting approach discussed in Section \ref{sec:PDE_inpaint}, and the dictionary-learning-based inpainting method discussed in Section \ref{sec:DL_inpaint}.

\subsection{Overview of EnSF}\label{sec:ensf_overview}
In EnSF, we employ a generative AI framework based on the score-based diffusion to model the the prior filtering distribution in Eq.~\eqref{eq:filtering-bayesian-update}. Our approach involves constructing a probabilistic mapping that connects the prior filtering distribution in Eq.\eqref{eq:filtering-bayesian-update} to a normalized Gaussian distribution. This transformation is implemented through a bidirectional stochastic differential equation system - comprising both forward and backward components - which operates over a synthetic temporal interval defined as $t \in \mathcal{T} = [0,1)$:
\begin{equation}\label{eq:sdes}
\begin{aligned}
   & \text{Forward SDE:}\;\; dZ_{n+1,t} = b_t Z_{n+1,t}\, dt + \sigma_t dW_t,\\[2pt]
   & \text{Reverse SDE:}\;\;\; d{Z}_{n+1,t} = \left[ b_t{Z}_{n+1,t} - \sigma_t^2 S_{n+1|n}(Z_{n+1,t}, t)\right] dt + \sigma_t d\cev{W}_t,
\end{aligned}
\end{equation}
where $S_{n+1|n}(Z_{n+1,t}, t)$ is the score function of the prior distribution in Eq.~\eqref{eq:filtering-bayesian-update}, $W_t$ and $\cev{W}_t$ are the forward and backward Brownian motions, respectively, $b_t$ and $\sigma_t$ are the drift and diffusion coefficients, respectively, and the subscript $(\cdot)_{n+1}$ indicates that the SDE is defined for the $(n+1)$-th filtering step. Following our previous work \cite{ensf_cmame}, 
$b_t$ and $\sigma_t$ in Eq.~\eqref{eq:sdes} are defined by 
\begin{equation}\label{eq:cof}
\begin{aligned}
b_t = \frac{{\rm d} \log \alpha_t}{{\rm d} t},\;\;\; \sigma_t^2 = \frac{{\rm d} \beta_t^2}{{\rm d}t} - 2 \frac{{\rm d}\log \alpha_t}{{\rm d}t} \beta_t^2, 
\end{aligned}
\end{equation}
where $\alpha_t$ and $\beta_t$ are defined by
\begin{equation}\label{eq:alpha-beta}
   \alpha_t = 1-t, \;\; \beta_t^2 = t.
\end{equation}
By initializing the forward SDE with the condition \( Z_{n+1,0} = X_{n+1} \mid Y_{1:n} \) from Eq.~\eqref{eq:filtering-bayesian-update}, the prior filtering distribution \( p_{X_{n+1} \mid Y_{1:n}} \) can be mapped to the standard Gaussian distribution \( \mathcal{N}(0, \mathbf{I}_d) \). Consequently, if the score function \( S_{n+1\mid n}(Z_{n+1,t}, t) \) is available, samples from the prior filtering distribution can be obtained by first drawing from the standard Gaussian and then solving the reverse SDE.

Since the filtering distribution evolves dynamically over time, using learning-based methods to estimate the score function (e.g., employing neural networks for score learning) is computationally impractical due to the need for frequent re-training \cite{bao2023scorebased}. In the EnSF framework, the score function is explicitly derived:
\begin{equation}\label{eq:score11}
\begin{aligned}
& S_{n+1|n}(z_{n+1,t}, t) \\[4pt]
  =\,&   \nabla_z \log \left(\int_{\mathbb{R}^d} q_{{Z}_{n+1,t} | Z_{n+1,0}}({z}_{n+1,t} | z_{n+1,0}) q_{Z_{n+1,0}}(
z_{n+1,0}) dz_{n+1,0}\right)\\
%
=\, &   \int_{\mathbb{R}^d}  - \frac{z_{n+1,t}- \alpha_t z_{n+1,0}}{\beta^2_t} w({z}_{n+1,t},  z_{n+1,0})  q_{Z_{n+1,0}}(z_{n+1,0})dz_{n+1,0},\\
\end{aligned}
\end{equation}
where the weight function $w({z}_{n+1,t},  z_{n+1,0})$ is defined by
\begin{equation}\label{eq:weight}
w({z}_{n+1,t},  z_{n+1,0}) := \frac{ q_{Z_{n+1,t}|Z_{n+1,0}}({z}_{n+1,t} | z_{n+1,0}) }{\int_{\mathbb{R}^d} q_{Z_{n+1,t}|Z_{n+1,0}}({z}_{n+1,t} | z'_{n+1,0})  q_{Z_{n+1,0}}(
z'_{n+1,0}) dz'_{n+1,0}},
\end{equation}
ensuring that \(\int_{\mathbb{R}^d} w({z}_{n+1,t}, z_{n+1,0}) q_{Z_{n+1,0}}(z_{n+1,0}) dz_{n+1,0} = 1\). Eq.~\eqref{eq:score11} leverages the fact that the conditional distribution \(q_{{Z}_{n+1,t} | Z_{n+1,0}}({z}_{n+1,t}| z_{n+1,0})\) follows a Gaussian distribution \(\mathcal{N}(\alpha_t Z_{n+1,0}, \beta_t^2 \mathbf{I}_d)\) (refer to \cite{ensf_cmame} for further details). The integrals in Eq.~\eqref{eq:score11} are taken over the initial distribution \(q_{Z_{n+1,0}}\), which corresponds to the prior filtering distribution \(X_{n+1} | Y_{1:n}\). Thus, the Monte Carlo method is employed to approximate the score function using prior samples. We refer to \cite{ensf_cmame} for details about the MC estimator of the score function. 

The key step in EnSF is how to update the score function and generate the posterior ensemble by incorporating the new observational data $Y_{n+1}$. The current version of EnSF updates the prior score $S_{n+1|1:n}$ with the new observation $Y_{n+1}$ according to the Bayes's rule from Eq.~\eqref{eq:filtering-bayesian-update} to approximate the posterior score $S_{n+1|1:n+1}$, i.e.,
\begin{equation}\label{eq:EnSF-posterior-score}
    \hat{S}_{n+1|1:n+1}(z, t) = \hat{S}_{n+1|1:n}(z, t) +  h(t) \nabla_{z} \log p_{Y_{n+1} | X_{n+1}}(z),
\end{equation}
where 
\(\hat{S}_{n+1|1:n}\) represents the Monte Carlo (MC) estimator for the prior score as defined in Eq.~\eqref{eq:score11}, \(p_{Y_{n+1} \mid X_{n+1}}(\cdot)\) denotes the likelihood function from Eq.~\eqref{eq:filtering-bayesian-update}, and \(h: \mathbb{R} \to \mathbb{R}\) is a continuous time-damping function that regulates the diffusion of information from \(Y_{n+1}\) within the diffusion framework. In the current EnSF approach, the function \(h(t)\) is monotonically decreasing on \([0,1]\) (e.g., \(h(t) = 1-t\)) and satisfies \(h(1) = 0\) and \(h(0) = 1\). This ensures that the likelihood information is progressively incorporated into the score function as the reverse SDE is solved. 

Substituting the definition of the likelihood function in Eq.~\eqref{eq:likelihood} into Eq.~\eqref{eq:EnSF-posterior-score}, we have the following expression for the likelihood score function
\begin{equation}\label{eq:EnSF-posterior-score1}
\begin{aligned}
    \nabla_{z} \log p_{Y_{n+1} | X_{n+1}}(z) & =
    \begin{pmatrix}
      \nabla_{z^{\rm obs}} \log p_{Y_{n+1} | X_{n+1}}(z)\\[5pt]
      \nabla_{z^{\rm unobs}} \log p_{Y_{n+1} | X_{n+1}}(z)
    \end{pmatrix}\\[4pt]
   &  =
    \begin{pmatrix}
            2 \Big(\frac{\partial \mathcal{H}}{\partial z^{\rm obs}} \Big)^{\top} R^{-1} (y_{n+1} - \mathcal{H}(z^{\rm obs}))\\[8pt]
            0
    \end{pmatrix},
\end{aligned}
\end{equation}
where $y_{n+1}$ is the measurement of $Y_{n+1}$, $z^{\rm obs}$ and $z^{\rm unobs}$ correspond to the observed states $X^{\rm obs}_{n+1}$ and the unobserved states $X^{\rm unobs}_{n+1}$, respectively. It is evident that the gradient of the log likelihood with respect to the unobserved state is always zero, which means that the unobserved states $X_{n+1}^{\rm unobs}$ are not updated by the observation data $y_{n+1}$ during the update of the score function in Eq.~\eqref{eq:EnSF-posterior-score}. Even though the value of $X_{\rm unobs}$ will change during the prediction step in Eq.~\eqref{eq:filtering-prior-ck} by pushing the ensemble through the physical model, numerical experiments show that the accuracy of EnSF is lower in tracking the unobserved states than that in tracking the observed states. 
\begin{figure}[h!]
    \centering
    \includegraphics[width=0.9\linewidth]{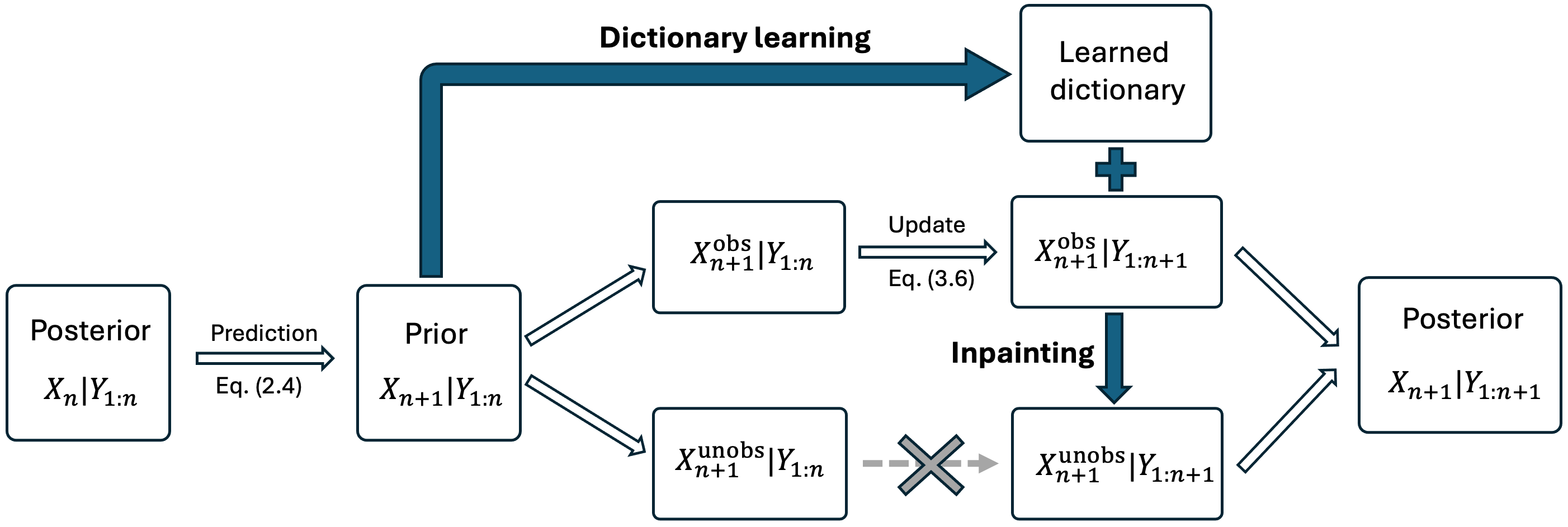}
    \caption{The proposed workflow.}
    \label{fig:workflow}
\end{figure}
We propose using image inpainting techniques during each filtering step to update unobserved states based on updated observed states, as illustrated in Figure \ref{fig:workflow}. During each filtering iteration's update step, we first apply the scheme in Eq.~\eqref{eq:EnSF-posterior-score} to solve the reverse SDE in Eq.~\eqref{eq:sdes}, updating the observed state variables from $X_{n+1}^{\rm obs}|Y_{1:n}$ to $X_{n+1}^{\rm obs}|Y_{1:n+1}$. For updating the unobserved states $X_{n+1}^{\rm unobs}|Y_{1:n}$, we explore two different inpainting approaches. The first technique uses a PDE model to describe the image and fills missing pixels (unobserved states) through the PDE solution. The second technique utilizes the full prior state $X_{n+1}|Y_{1:n}$ as a reference image to construct a sparse prior of the posterior state $X_{n+1}|Y_{1:n+1}$, assuming that they share the same sparse representation in a given dictionary, and then optimizes to find the best approximation of unobserved states on the dictionary-defined manifold. These two inpainting techniques are detailed in Sections \ref{sec:PDE_inpaint} and Section \ref{sec:DL_inpaint}, respectively.

\subsection{PDE-based inpainting}\label{sec:PDE_inpaint}
PDE-based (also known as diffusion-based) inpainting methods operate solely on a masked image itself, i.e., the observed states $X_n^{\rm obs}$ in Eq.~\eqref{eq:state_obs}, and use mathematical models inspired by physical processes to diffuse the data from observed regions into missing regions, i.e., unobserved states $X_n^{\rm unobs}$ in Eq.~\eqref{eq:state_obs}. To ensure that the inpainted state vector $X_n$ is consistent with $X_n^{\rm obs}$ at known pixels, $X_n^{\rm obs}$ is used to form a boundary/initial condition of the PDE. Guided by knowledge of image priors (smoothness, edge, sharpness, etc.), different PDE models can be considered (linear, nonlinear, isotropic, or anisotropic) to favor the propagation in particular directions or to take into account the curvature of the structure present in a local neighborhood, with the goal of creating an observed state as physically plausible as possible. In this work, we incorporate two PDE-based image inpainting techniques proposed in \cite{https://doi.org/10.1155/2018/3950312,Bertalmo2001NavierstokesFD} into the EnSF to recover the unobserved states $X_{n}^{\rm unobs}$ from the observed states $X_n^{\rm obs}$.

The Navier-Stokes inpainting method \cite{Bertalmo2001NavierstokesFD} uses ideas from classical fluid dynamics to propagate isophote lines continuously from the observed region into the unobserved one. The main idea is to treat the image intensity as a stream function for a two-dimensional incompressible flow. Then the Laplacian of the image intensity plays the role of the vorticity of the fluid. In particular, denote $w = \Delta X_n^{\rm unobs}$, we solve a vorticity transport equation, which is based on the vorticity-streamfunction form of the Navier-Stokes equation, for $w$ and $ X_n^{\rm unobs}$
\begin{equation}\label{eq:NS-inpainting}
\begin{aligned}
&\frac{\partial w}{\partial t} + v\cdot \nabla w  =  \nu \nabla  \cdot (\nabla w), 
\\
& v = \nabla^\bot X_n^{\rm unobs},\, \Delta X_n^{\rm unobs} = w,
\end{aligned}
\end{equation}
where $\nabla^\bot$ denotes the perpendicular gradient, $v = \nabla^\bot X_n^{\rm unobs}$ defines the velocity field that is recovered by solving the Poisson equation $\Delta X_n^{\rm unobs} = w$, 
and $\nu$, traditionally being the fluid viscosity in Navier-Stokes equations. Inheriting the well-developed theoretical and numerical framework for Navier-Stokes, it is fairly easy to implement the solver for \eqref{eq:NS-inpainting} efficiently and analyze transport of information from the observed into the inpainting region. 
This method excels at reconstructing geometric structures and extending edges smoothly, making it particularly effective for image and video inpainting applications. The approach balances local diffusion with global transport, ensuring coherence in the reconstructed regions.

The biharmonic inpainting method \cite{https://doi.org/10.1155/2018/3950312} is grounded in the mathematical properties of the fourth-order biharmonic PDE to ensure smooth extension of $X_n^{\rm obs}$ to the unobserved region. Specifically, this method solves the following equation for $X_n^{\rm unobs}$
\begin{equation}\label{eq:biharmonic}
\Delta^2 X_n^{\rm unobs} = 0,\ \text{ on missing region}, 
\end{equation}
equipped with boundary condition that $X_n^{\rm unobs}$ and $\Delta X_n^{\rm unobs}$ must match with the value and Laplacian of observed state ($X_n^{\rm obs}$ and $\Delta X_n^{\rm obs}$) on the boundary.  
The approach has been proved to be cubic inpainting, and is particularly well-suited for reconstructing smooth data, where preserving the smoothness and continuity in the gradient or curvature of the image is essential. By utilizing the inherent smoothness of biharmonic functions, this method stands out as a robust and mathematically elegant solution to the inpainting problem, offering a balance between computational simplicity and high-quality results.

\begin{figure}[h!]
    \centering
    \includegraphics[width=0.8\textwidth]{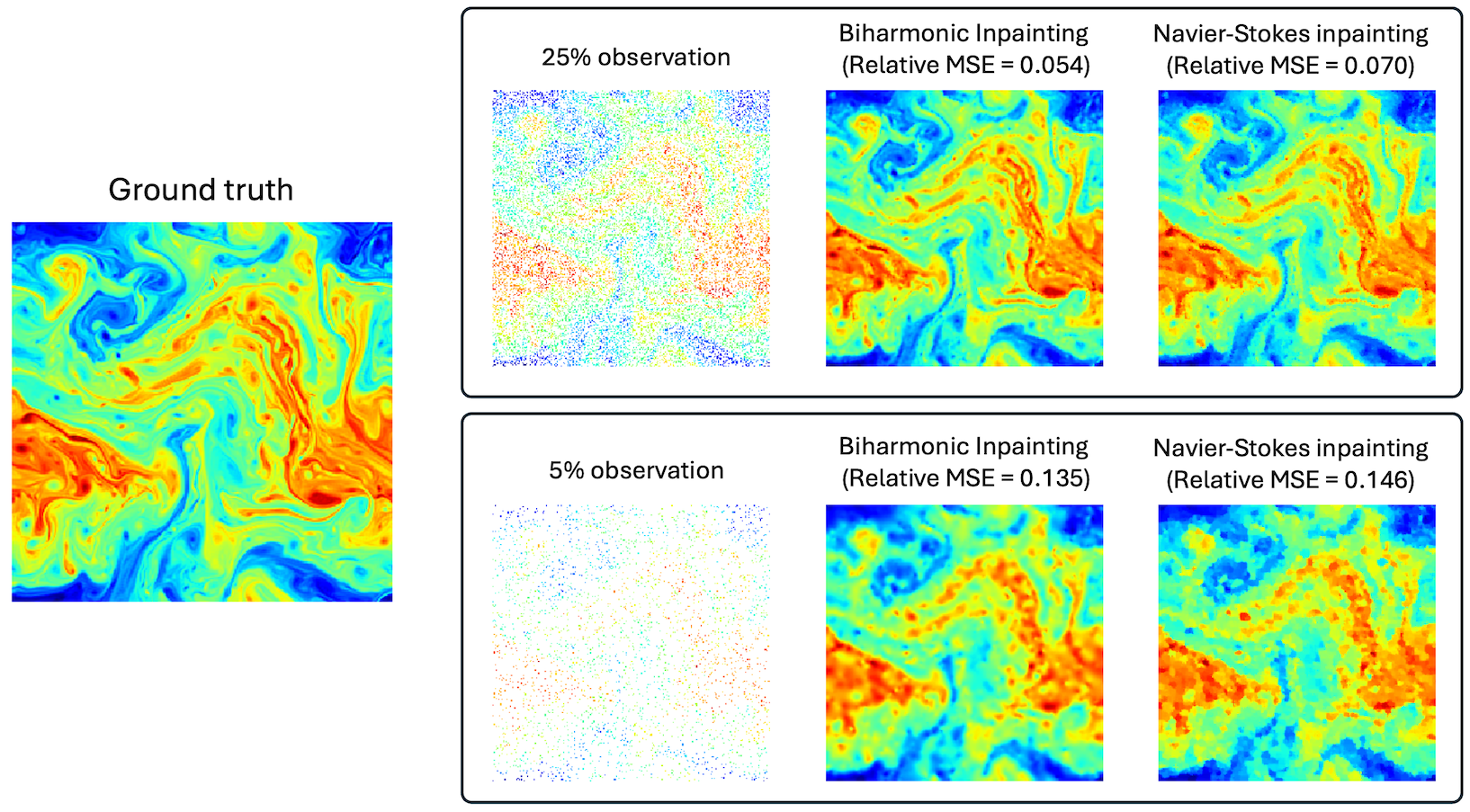}
    \vspace{-0.2cm}
    \caption{Illustration of PDE-based inpainting methods based on the Navier-Stokes equation and biharmonic equation for one snapshot of the SQG model. Both methods provide satisfactory reconstruction results when having 25\% of the state observed. The reconstruction accuracy deteriorates when having only 5\% of the state observed, which motivated us to purse dictionary-learning-based inpainting in Section \ref{sec:DL_inpaint}. }
    \label{fig:pde_inpaint}
\end{figure}

We illustrate the performance of the Navier-Stokes and biharmonic inpainting methods in recovering one snapshot of the SQG model in Figure \ref{fig:pde_inpaint}. We observe both methods provide satisfactory reconstruction results when having 25\% of the state observed. The reconstruction accuracy deteriorates when having only 5\% of the state observed, which motivated us to explore dictionary-learning-based inpainting in Section \ref{sec:DL_inpaint}. The Navier-Stokes method demonstrates particular strength in preserving the flow characteristics and vorticity patterns, while the biharmonic approach excels in maintaining smoothness across the inpainting region. However, both methods exhibit limitations in capturing fine-scale features and sharp gradients when the observation density becomes sparse. This degradation in performance is especially pronounced in regions where complex fluid dynamics occur, such as areas of high turbulence or strong vortex interactions, highlighting the need for more sophisticated reconstruction techniques that can better handle limited observational data.

\subsection{Dictionary-learning-based inpainting}\label{sec:DL_inpaint}
The dictionary learning based methods in image inpainting seek to identify a dictionary (possibly redundant), which the images can be sparsely represented with \cite{Mairal200853}. In the context of data assimilation, we employ sparse representations to address the fundamental challenge of partial state observability. First, we form a dictionary of functions, i.e., a basis,  that captures the underlying state space structure. We can utilize predefined dictionaries, which are capable to model various analytical and geometrical natures of the images, such as smooth, piece-wise smooth, or edge-dominated. Common examples of such dictionaries include the Discrete Cosine Transform (DCT) \cite{watson1994image}, wavelets of various sorts \cite{mallat1999wavelet}, curvelets \cite{candes2002recovering}, contourlets \cite{https://doi.org/10.1002/cpa.10116}, etc. These are well-studied and optimized, and thus can be easily implemented with existing highly effective algorithms. The predefined dictionary can be further adapted and reduced using the full posterior state estimates from one or multiple previous Bayesian iterations, for example, only the top elements that express these states are retained. 
\vspace{0.2cm}
\begin{itemize}[leftmargin=20pt]\itemsep0.2cm
    \item Step 1: Perform DCT on the each prior sample $x_{n+1|n}^k \in \mathcal{D}^{\rm prior}_{n+1}$ in Eq.~\eqref{eq:prior_sample}, i.e.,
    \[
    {C}_k = \texttt{DCT} \left(x_{n+1|n}^k\right) \;\; \text{ for }\; k = 1, \ldots, K,
    \]
    where ${C}_k$ is the coefficient matrix of the discrete cosine expansion. 
    \item Step 2: Build a dictionary, i.e., a cosine basis, by thresholding the coefficient matrix $C_k$, i.e., 
    \[
    C_k^\nu = C_k \odot M_k^{\nu}, 
    \]
    where $\odot$ is element-wise product of two matrices, and $M_k^{\nu} =[M_k^{ij}]$ is a mask matrix defined by $M_k^{ij} = 1$ if $|C_k^{ij}| > \nu$ and $M_k^{ij} = 0$ if $|C_k^{ij}| \le \nu$ with $\nu>0$ being the threshold. 
    \item Step 3: Perform gradient descent to update the coefficient matrix $C_k^\nu$, i.e., 
    \[
    \hat{C}_k^{\nu} = \operatorname*{argmin}\limits_{C_k^{\nu}}\left\{ \left\| \mathcal{H}^{\rm obs}(\texttt{iDCT}(C_k^{\nu})) - x_{n+1|n+1}^{{\rm obs},k}\right\|_2^2 + \gamma\|\texttt{iDCT}(C_k^{\nu})\|_{\rm TV}\right\},
    \]
    where $\texttt{iDCT}$ is the inverse DCT transform\footnote{We use the pytorch implementation of DCT at \url{https://github.com/zh217/torch-dct}, which supports back propagation to compute the gradient.}, $\mathcal{H}^{\rm obs}$ is the observation operator defined in Eq.~\eqref{observations}, $x_{n+1|n+1}^{{\rm obs},k}$ is observed states updated by solving the reverse SDE in Eq.~\eqref{eq:sdes} using the score function in Eq.~\eqref{eq:EnSF-posterior-score}, and $\|\cdot\|_{\rm TV}$ represent the total variation norm for regularization. 
    \item Step 4: Perform inverse DCT, i.e., $\texttt{iDCT}(\hat{C}_k^{\nu})$, using the updated coefficient matrix $\hat{C}_k^{\nu}$ to obtain the inpainted full state $x_{n+1|n+1}^k$ and put it in the posterior ensemble set $\mathcal{D}_{n+1}^{\rm posterior}$ in Eq.~\eqref{eq:post_sample}. 
\end{itemize}
\vspace{0.2cm}
\begin{figure}[h!]
    \centering
    \includegraphics[width=0.8\textwidth]{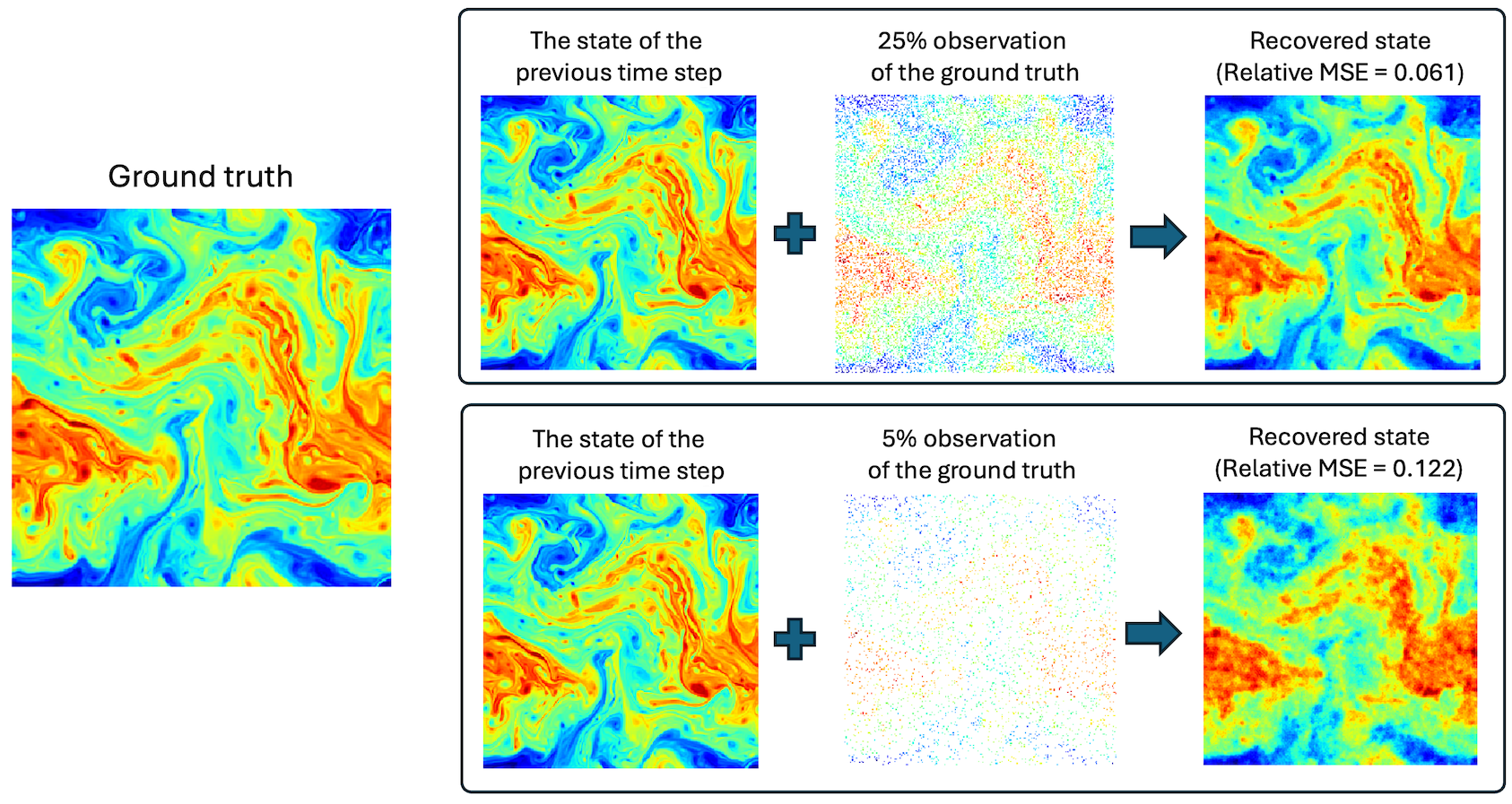}
    \vspace{-0.2cm}
    \caption{Illustration of dictionary-learning-based inpainting for one snapshot of the SQG model. When having 25\% obserable states, there is not much improvement compared to the PDE-based methods presented in Figure \ref{fig:pde_inpaint}. On the other hand, when having only 5\% of observed states, the dictionary-learning-based method provides better results due to the use of the dictionary. This is consistent with the numerical experiment in Section \ref{sec:exp}.}
    \label{fig:dct_inpaint}
\end{figure}

Dictionary-learning-based inpainting method effectively transforms the partial observability problem into a coefficient estimation problem, where the learned dictionary provides a robust framework for state inpainting. By using the dictionary as a bridge between fully-known previous states and partially-known current states, we can reconstruct the unobserved components of the state vector while maintaining consistency with both the learned state space structure and the available observations. The sparse representation approach offers several key advantages over direct image inpainting methods in the context of data assimilation. While direct inpainting typically relies on local spatial correlations or predefined interpolation schemes, sparse-based method offer greater flexibility via a dictionary that can exploit the full posterior states of previous assimilation cycles to capture complex, non-local patterns and structures. This temporal knowledge transfer is particularly crucial when observations of the current state are extremely sparse, as the dictionary learned from the previous posterior state provides a rich prior that compensates for the limited current observations. This learned representation is especially valuable when the underlying state exhibits recurring patterns or multiscale features that simple interpolation methods might miss. Moreover, the dictionary-based approach provides a more robust framework for handling observation noise and uncertainty, as the learned elements inherently incorporate the statistical properties of the state space. The method also adapts dynamically to the specific characteristics of the system being studied, rather than applying generic inpainting rules. This adaptive nature, combined with the ability to encode both local and global state dependencies in the dictionary elements, enables more accurate reconstruction of unobserved states compared to traditional inpainting techniques, particularly in scenarios where current observations are highly limited.

{
\subsection{Discussion on the connection between inpainitng and traditional approaches in data assimilation}
In traditional data assimilation methods such as (ensemble) Kalman filters and variational methods the information from observations is spread out over the model state via a prior covariance matrix. This covariance matrix describes how the state variables covary, such that a change of an observed variable enforces an update of other variables proportional to the covariance between the observed variable and the other variables. The prior covariance matrix encodes the linearized physical relations between all state variables, while the relations can be more general. In more advanced data assimilation methods, such as MCMC methods, particle filters, and the recently developed particle flow filters, the full joint probability distribution between an observed variable and unobserved variables is used in the update. 

The bi-harmonic inpainting can be seen as a first-order approximation of a covariance operator as a recursive filter, in which the operation of a covariance matrix on a vector is written as a sum of derivatives of that vector with appropriate coefficients, see e.g., \cite{James2003}. This identification shows a direct way to make the bi-harmonic inpainting more accurate by including a diffusion matrix $D$, resulting in an inpainting of the form $
\nabla \left(D \nabla X_n^{\rm unobs} \right)
$
in which $D$ can be constant over space, or represent physical flow-dependent structures, e.g. alignment along large gradients. 

The Navier-Stokes inpainting can be seen as applying an approximation of the evolution equations of the state, with boundary and internal fixed points defined by the observed variables. In a sense, it mimics how point-wise observation information is propagated through the system in real time. Since the evolution equation used is close to the actual evolution equation of the state, it largely explores the physical relations in the system, which can be nonlinear.

The dictionary-based method can be viewed as imposing climatological relations between observed and unobserved variables. However, the method is more advanced than a standard climatological covariance matrix in a method such as 3DVar because the relations can be nonlinear, and they will be state dependent. There is a connection with analog-based Ensemble Kalman filters \cite{TheAnalogDataAssimilation}, in which the prior covariance in the ensemble Kalman filter is (partly) represented by past ensemble members that closely resemble the prior model state, but the use is different as the dictionary-based method allows for nonlinear relations between observed and unobserved variables, while in the analog method these relations are assumed to be linear.
}

\section{Numerical experiments}\label{sec:exp}
We demonstrate the superior performance of the proposed EnSF with inpainting
by comparing to the original EnSF method introduced in \cite{ensf_cmame} and the state-of-the-art LETKF method. The experiment design is introduced in Section \ref{sec:ex_design}, the comprehensive comparison between the proposed method and the baseline methods is provided in Section \ref{sec:ex_result}, and an ablation study on a few special cases is given in Section \ref{sec:ex_ablation} with additional results given in Appendix \ref{sec:app}. 

\begin{remark}[Reproducibility]
   The proposed method is implemented in Python. All the numerical results presented in this section can be reproduced exactly using the code on GitHub. The source code is publicly available at \href{https://github.com/Siming-Liang/EnSFInpainting}{https://github.com/Siming-Liang/EnSFInpainting}.
\end{remark}

\subsection{Experimental design}\label{sec:ex_design}
\begin{figure}[h!]
\hspace{0.6cm}
    \begin{subfigure}[t]{0.23\textwidth}
        \centering
        \includegraphics[width=\textwidth]{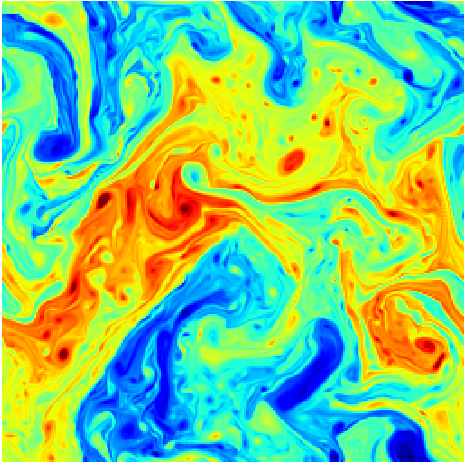}
        \caption{Truth}
        \label{fig: N256 12hour truth}
    \end{subfigure}
    \begin{subfigure}[t]{0.23\textwidth}
        \centering
        \includegraphics[width=\textwidth]{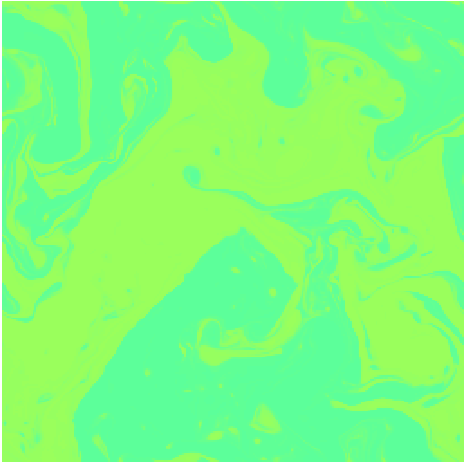}
        \caption{100\% observation}
        \label{fig:obs 100per}
    \end{subfigure}
    \begin{subfigure}[t]{0.23\textwidth}
        \centering
        \includegraphics[width=\textwidth]{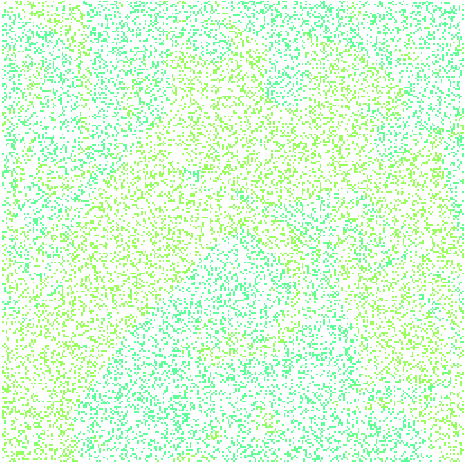}
        \caption{25\% observation}
        \label{fig:obs 25per}
    \end{subfigure}
    \begin{subfigure}[t]{0.23\textwidth}
        \centering
        \includegraphics[width=\textwidth]{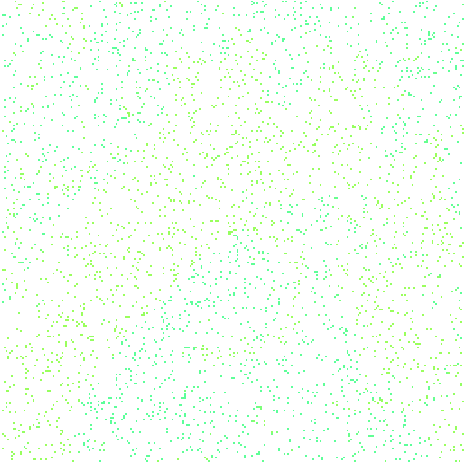}
        \caption{5\% observation}
        \label{fig:obs 5per}
    \end{subfigure}
    \vspace{-0.5cm}
    \caption{Illustration of nonlinear observation using arctangent operator. Subfigure (b) demonstrates that the observation information is significantly compressed compared to the ground truth. As the percentage of available observations decreases in (c) and (d), extracting information through the nonlinear operator becomes increasingly challenging.}
    \label{fig: Nonlinear observation}
    \vspace{-0.6cm}
\end{figure}

The proposed method is tested with 20 ensemble members using a typical Observing System Simulation Experiment (OSSE) setup. In this setup, synthetic observations are produced by adding random noise to the true dynamical trajectory. The generation of this true trajectory, or nature run closely adheres to the description provided in \cite{wang_et_al_2021}. 
We tested the performance of the proposed method in 16 different scenarios considering the following factors:
\vspace{0.15cm}
\begin{itemize}[leftmargin=15pt]\itemsep0.2cm
    \item \textbf{Spacial resolution}. Experiments are conducted at two different resolutions: $64 \times 64$ and $256 \times 256$. The coarse resolution of $64 \times 64$ represents the challenges of limited data availability in real-world scenarios, such as when observation instruments are sparsely distributed over large geographic regions (e.g., oceans, mountainous areas, or deserts). This coarser resolution reduces the correlation between nearby grid points in real-world models (e.g., weather models), resulting in ``pixels'' that are more independent and exhibit more nonlinearity. In contrast, the fine grid resolution of $256 \times 256$ represents scenarios where denser instrumentation provides greater detail. At this finer resolution, the ``pixels'' are more densely packed and tend to vary more linearly. Fine-tuning of the LETKF is performed only at the $64 \times 64$ resolution, as tuning for the $256 \times 256$ resolution is computationally prohibitive. This limitation underscores the LETKF's high sensitivity to parameter tuning and its reduced versatility across different resolutions.
\item \textbf{Observation operator}. We use fixed grid points for observations, reflecting real-world scenarios where instruments, such as surface weather stations, are geographically stationary. Additionally, we test both fully linear and fully nonlinear observation models to encompass a range of real-world application scenarios. In this study, we pick arctangent function as the nonlinear operator:
\begin{equation}
\label{arctan_op}
   Y_{n} = \arctan(X_n^{\rm obs}) + \varepsilon_n,
\end{equation}
 where $X_n^{\rm obs}$ denotes the observed states. In the case of linear observation, the observation error follows the standard normal distribution $\mathcal{N}(0,\mathbf{I})$. In the nonlinear case, the observation error variance is scaled relative to the linear observation experiments
 in order to reflect the narrower range of the arctangent function. In this case, we choose $\varepsilon_n \sim \mathcal{N}(0,0.01\mathbf{I})$. 
 \item \textbf{Data assimilation frequency}. We perform data assimilation at intervals of both 3 hours and 12 hours. The 3-hour frequency, as used in \cite{wang_et_al_2021}, results in smaller deviations from the true state, while the 12-hour frequency introduces larger differences, potentially leading to more pronounced departures from Gaussianity in the forecast ensemble.
\item 
\textbf{Observation sparsity}. The available observations are sparse, with two settings: 5\% and 25\%. The 5\% observation setting reflects the challenges of sparse data availability in real-world scenarios, while the 25\% setting represents idealized conditions with denser coverage, reducing the effects of observation sparsity. Figures \ref{fig:pde_inpaint} and \ref{fig:dct_inpaint} illustrate the 5\% and 25\% linear observation images. In Figure \ref{fig: Nonlinear observation}, the nonlinear arctangent operator demonstrates a highly compressed representation of the information. These settings enable an evaluation of our method's robustness under varying levels of data availability.
\end{itemize}
\vspace{-0.3cm}

\subsection{Comprehensive comparison}\label{sec:ex_result}
The above factors combine to create 16 experiments, enabling us to evaluate a wide range of scenarios and conditions. For simplicity, we summarize the 16 scenarios in Table \ref{table:scenario}. We conduct comparison of the methods listed in Table \ref{table:method}.

\subsubsection{LETKF fine tuning}\label{sec:letkf}
LETKF is an advanced data assimilation method widely used in the geosciences—particularly in weather forecasting and climate modeling. However, the performance of LETKF heavily relies on fine-tuning of its hyper-parameters, particularly the horizontal localization scale $L_h$ (in km) and inflation parameters. Inflation is a way to increase the spread in the ensemble, which is typically too small from to undersampling due to small ensemble sizes. The inflation method we use is the so-called Relaxation to Prior Spread (RTPS).

\begin{table}[h!]
    \centering
    \rowcolors{2}{gray!25}{white}
    \begin{tabular}{c|cccc}
    \toprule
        Label &  Resolution & Obs operator & DA frequency & Obs sparsity\\
    \midrule
      ($\mathbf{C}_1$)  & $64\times64$ & Linear & 3 hour & 5\% \\
      ($\mathbf{C}_2$)  & $64\times64$ & Linear & 3 hour & 25\% \\
      ($\mathbf{C}_3$)  & $64\times64$ & Linear & 12 hour & 5\% \\
      ($\mathbf{C}_4$)  & $64\times64$ & Linear & 12 hour & 25\% \\
      ($\mathbf{C}_5$)  & $64\times64$ & Nonlinear & 3 hour & 5\% \\
      ($\mathbf{C}_6$)  & $64\times64$ & Nonlinear & 3 hour & 25\% \\
      ($\mathbf{C}_7$)  & $64\times64$ & Nonlinear & 12 hour & 5\% \\
      ($\mathbf{C}_8$)  & $64\times64$ & Nonlinear & 12 hour & 25\% \\
      ($\mathbf{C}_9$)  & $256\times256$ & Linear & 3 hour & 5\% \\
      ($\mathbf{C}_{10}$)  & $256\times256$ & Linear & 3 hour & 25\% \\
      ($\mathbf{C}_{11}$)  & $256\times256$ & Linear & 12 hour & 5\% \\
      ($\mathbf{C}_{12}$)  & $256\times256$ & Linear & 12 hour & 25\% \\
      ($\mathbf{C}_{13}$)  & $256\times256$ & Nonlinear & 3 hour & 5\% \\
      ($\mathbf{C}_{14}$)  & $256\times256$ & Nonlinear & 3 hour & 25\% \\
      ($\mathbf{C}_{15}$)  & $256\times256$ & Nonlinear & 12 hour & 5\% \\
      ($\mathbf{C}_{16}$)  & $256\times256$ & Nonlinear & 12 hour & 25\% \\
      \bottomrule
    \end{tabular}
    \caption{The 16 different scenarios tested in this work. In the rest of the paper, we use the labels in the first column to refer to the test cases.}
    \label{table:scenario}
    \vspace{-0.5cm}
\end{table}
%
\begin{table}[h!]
    \centering
    \small
    \renewcommand{\arraystretch}{1.25}
    \begin{tabular}{c|c}
     \toprule
          Label & Description of the method\\
    \midrule
         {\bf LETKF} & The Local Ensemble Transform Kalman Filter method\\
         {\bf EnSF Only} & The original EnSF method proposed in \cite{ensf_cmame}\\
         {\bf EnSF+Bi} & EnSF with biharmonic inpaiting in Section \ref{sec:PDE_inpaint}\\
         {\bf EnSF+NS} & EnSF with Navier-Stokes inpaiting in Section \ref{sec:PDE_inpaint}\\
         {\bf EnSF+DL} & EnSF with dictionary-learning-based inpaiting in Section \ref{sec:DL_inpaint}\\
    \bottomrule
    \end{tabular}
    \caption{List of the methods tested under the 16 different scenarios given in Table \ref{table:scenario}.}
    \label{table:method}
    \vspace{-0.5cm}
\end{table}

\begin{figure}[h!]
    \centering
\includegraphics[width=0.99\textwidth]{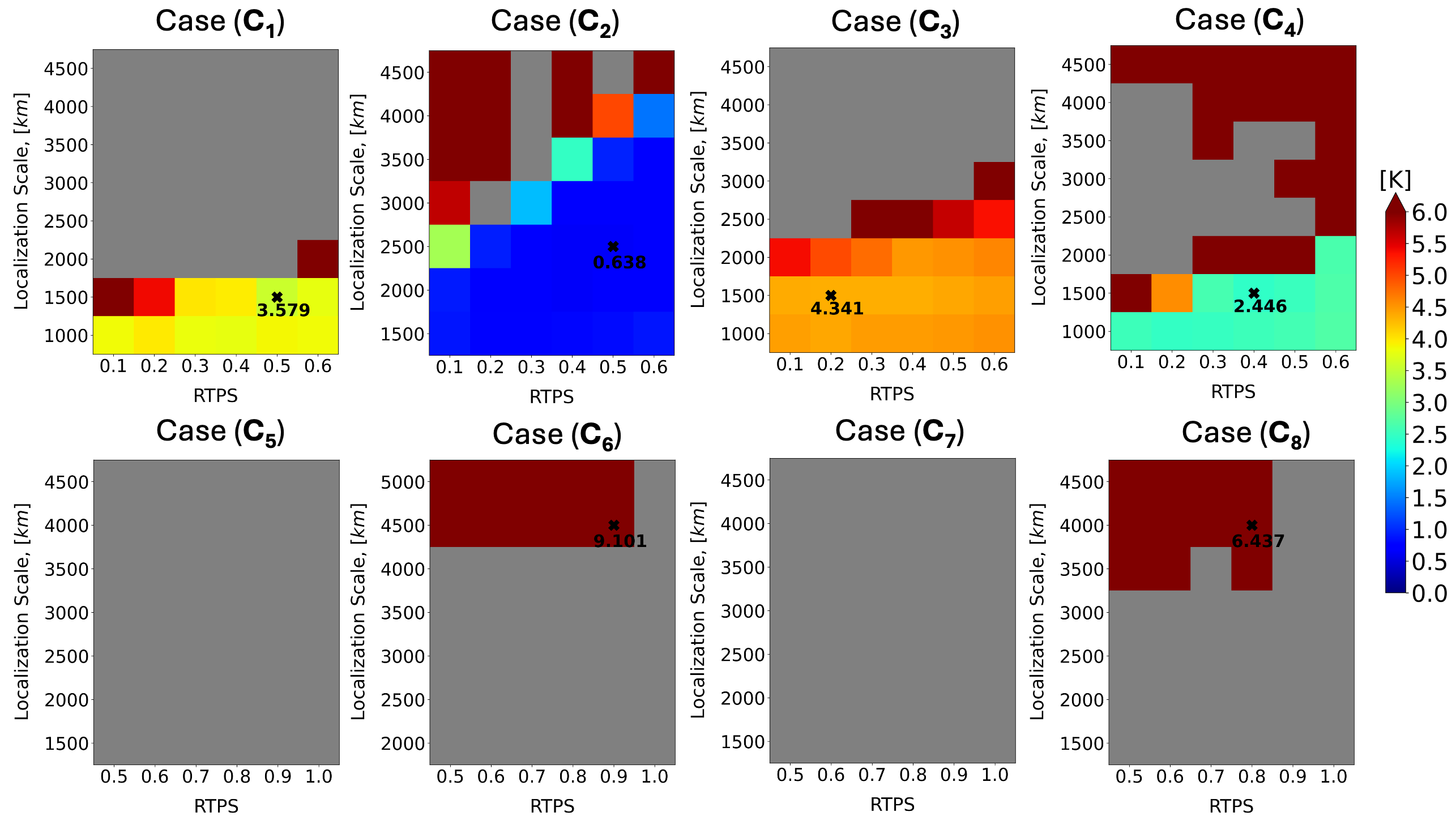}
\vspace{-0.1cm}
    \caption{The LETKF fine tuning chart at the $64 \times 64$ resolution, where the case labels are defined in Table \ref{table:scenario}. The optimal parameter pairs, if available, are marked on the chart. An RMSE greater than 6 indicates total failure of LETKF in tracking the SQG model. The grey regions represent cases where the RMSE diverges to undefined values (i.e., NAN). There is no optimal parameter pairs for cases ($\mathbf{C}_5$) and ($\mathbf{C}_7$) due to the nonlinearity and the high observation sparsity.}
    \label{fig:LETKF 64 tuning chart}
    \vspace{-0.8cm}
\end{figure}

Therefore, we first fine-tune the localization and RTPS parameters of LETKF at the $64 \times 64$ resolution by the standard grid search method. Specifically, we run LETKF at each of the $10\times12$ combinations of localization scale and RTPS parameter values ($L_h$ from 1000 to 5500 with an increment of 500; RTPS 0.1 to 1.2 with an increment of 0.1) and try to find the best combination. The fine-tuning at the $64 \times 64$ resolution for all 8 cases takes approximately 20 days on a workstation with a single CPU. The cost of fine-tuning at the $256 \times 256$ resolution for all 8 cases will be around 79 years, which is computationally prohibitive. As such, we transfer the best hyper-parameter values obtained at the $64 \times 64$ resolution to the corresponding cases at the $256 \times 256$ resolution. This approach can also help reveal the sensitivity of the LETKF's performance to the hyper-parameters. The fine-tuning results at the $64 \times 64$ resolution are given in Figure \ref{fig:LETKF 64 tuning chart}. Each colored block corresponds to the root mean square error (RMSE).

%
%

\subsubsection{RMSE comparison}\label{sec:ex_comp} 
\begin{figure}[h!]
    \centering
\includegraphics[width=0.99\linewidth]{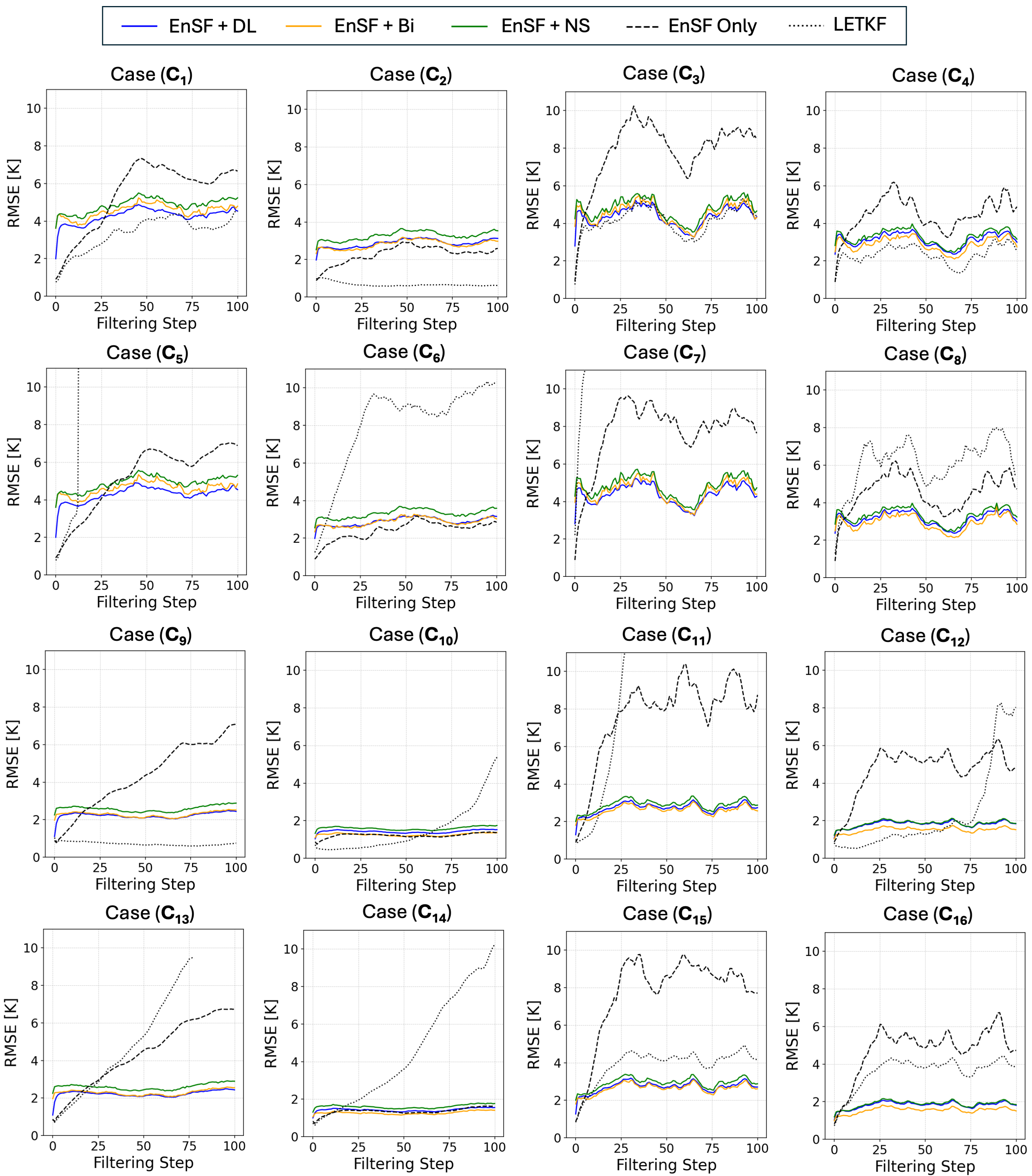}
   \caption{Comparison of the total RMSE for the 16 scenarios listed in Table \ref{table:scenario}. The horizontal axis indicates the number of filtering steps, i.e., the discrete time step $n$ in Eq.~\eqref{eq:filtering-state} and Eq.~\eqref{eq:state_obs}. When having linear observation operators, i.e.,  $(\mathbf{C}_1)-(\mathbf{C}_4)$, and $(\mathbf{C}_9)-(\mathbf{C}_{12})$, we observe that EnSF with inpainting achieves performance comparable to the fine-tuned LETKF, even without requiring any fine-tuning. When having nonlinear observations, i.e., $(\mathbf{C}_5)-(\mathbf{C}_8)$, and $(\mathbf{C}_{13})-(\mathbf{C}_{16})$, EnSF with inpainting significantly outperform LETKF and the original EnSF.}
    \label{fig:RMSE 64}
    \vspace{-0.5cm}
\end{figure}

The comparison of the total RMSE for the 16 cases listed in Table \ref{table:scenario} is given in Figure \ref{fig:RMSE 64}. In the cases of having linear observation operators, i.e., the cases $(\mathbf{C}_1)-(\mathbf{C}_4)$, and $(\mathbf{C}_9)-(\mathbf{C}_{12})$, we observe that EnSF with inpainting achieves performance comparable to the fine-tuned LETKF, even without requiring any fine-tuning. The lowest RMSE occurs in the ideal case with 3-hourly assimilation and 25\% observation coverage, while the highest RMSE is observed in the most challenging scenario with 12-hourly assimilation and 5\% observation coverage.
Moving beyond the idealized linear observation scenario, which favors LETKF, we evaluate performance under a nonlinear arctangent observation operator, i.e., the cases $(\mathbf{C}_5)-(\mathbf{C}_8)$, and $(\mathbf{C}_{13})-(\mathbf{C}_{16})$.  LETKF fails across all scenarios, as expected due to its reliance on Gaussian assumptions during the update step. In contrast, EnSF with inpainting demonstrates robust performance, with only a slight increase in RMSE compared to the linear observation cases.

As mentioned earlier, tuning for the $256 \times 256$ resolution is computationally prohibitive. Therefore, we use the tuned parameters from the $64 \times 64$ case for the $256 \times 256$ experiments. For linear observations, LETKF achieves good results only in the 3-hourly assimilation with 5\% observation case, despite theoretically being expected to produce low RMSE in all linear cases. This highlights LETKF's high sensitivity to parameter tuning and its limited versatility. In contrast, EnSF with inpainting performs consistently well. For nonlinear observations, LETKF performs poorly, as expected, due to its inherent limitations. Meanwhile, EnSF shows strong performance, with only a slight RMSE increase compared to the linear cases, further underscoring its robustness across all scenarios.

\subsubsection{Ensemble uncertainty estimation}\label{uq}
\begin{figure}[h!]
    \centering
\includegraphics[width=0.99\linewidth]{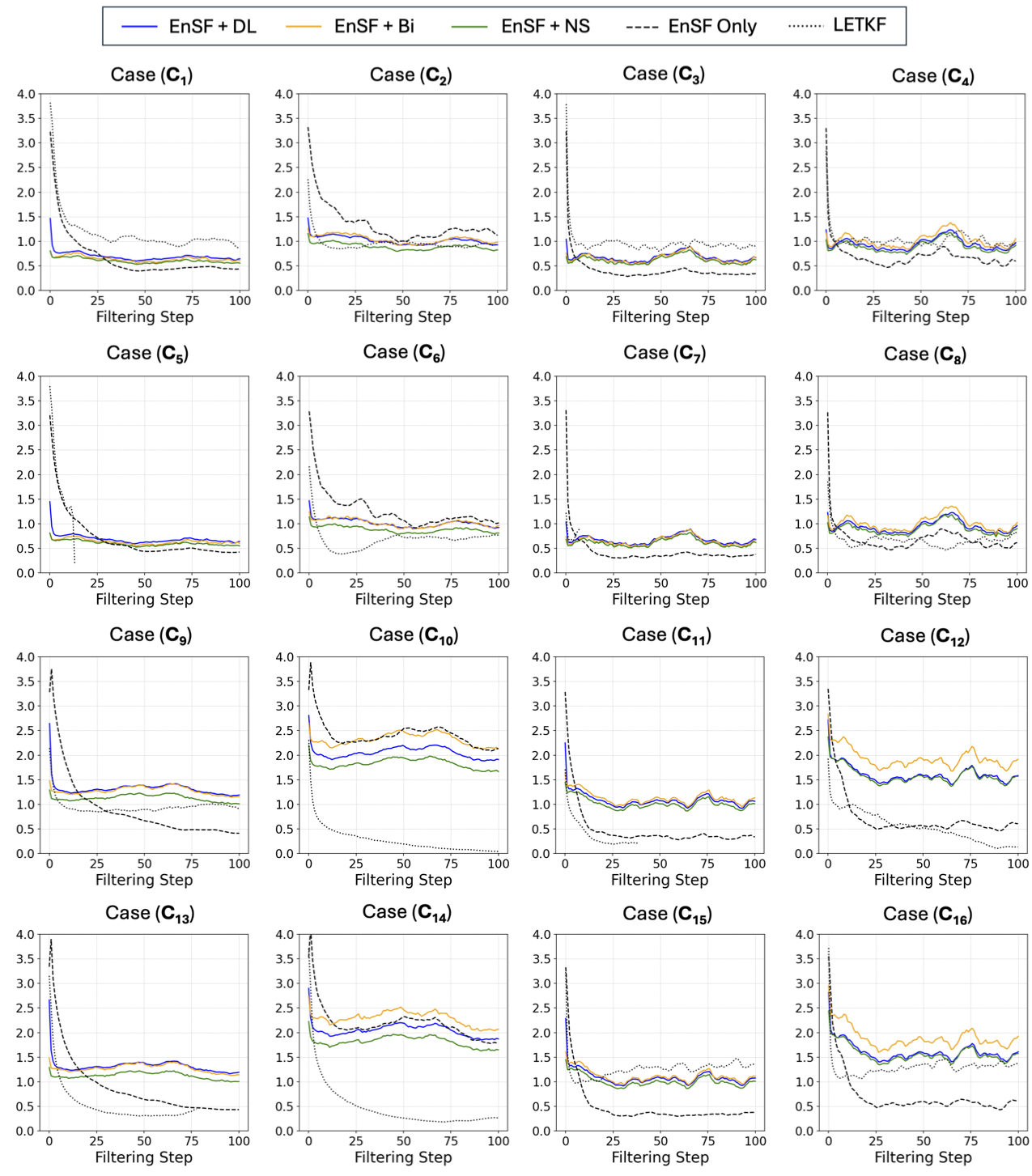}
   \caption{Comparison of the uncertainty estimation for the 16 scenarios listed in Table \ref{table:scenario}. The horizontal axis is the same as Figure \ref{fig:RMSE 64}. The vertical axis is the ratio of the ensemble spread (ensemble standard deviation) over RMSE.}
    \label{fig:uq}
\end{figure}

The comparison of the uncertainty estimation for the 16 cases listed in Table \ref{table:scenario} is given in Figure \ref{fig:uq}. At each filtering step, the ratio of ensemble spread over RMSE is computed using the formula:
\begin{equation*}
\frac{\text{Root Mean Square Spread}}{\text{RMSE}} = \frac{\sqrt{\frac{1}{N} \sum_{i=1}^N \left( \sigma_{i} \right)^2}}{\sqrt{\frac{1}{N} \sum_{i=1}^N \left( X_i - \bar{X}_i \right)^2}}, \sigma_i = \sqrt{\frac{1}{20} \sum_{k=1}^{20} \left( X_i^{ens_k} - \bar{X}_i \right)^2}
\end{equation*}
 where \(X_i\) represents the true values, \(\bar{X}_i\) denotes the mean of 20 ensemble members, \(N\) is the total number of grid points, and \(\sigma_i\) is the ensemble spread at the grid point.

A minimal requirement for the ensemble to be a good representation of the width of the posterior pdf is that the ensemble standard deviation is similar to the difference between the ensemble mean and the true state. Hence, we expect a value close to 1 for their ratio. Fig. 7 shows that all the inpainting methods perform well in this measure. The higher ratio values of close to 2 in cases C10, C12, and C14, correspond to high observation densities, contrast with the too low values for the LETKF in these cases. Since a too high uncertainty estimate leads to more robust estimates the inpainting methods are preferable over the LETKF even in those cases. This can be seen clearly in Fig. 6, which shows that the LETKF diverges from the truth in these cases. 

\subsection{Ablation study}\label{sec:ex_ablation}
Because the proposed EnSF with inpainting shows a significantly better performance over the LETKF and the original EnSF methods, we explore the results in greater depth
for select cases with nonlinear observations, i.e., $(\mathbf{C}_6)$, $(\mathbf{C}_7)$, $(\mathbf{C}_{14})$, and $(\mathbf{C}_{15})$. Case$(\mathbf{C}_7)$ represents the most challenging scenarios due to its coarser grid, high observation sparsity, and low data assimilation frequency, meaning we have the least amount of data to exploit. On the other hand, case $(\mathbf{C}_{14})$ is the easiest case due to the fine grid, less sparse observation, and higher data assimilation frequency. Cases $(\mathbf{C}6)$ and $(\mathbf{C}{15})$ fall between these extremes.

\subsubsection{\texorpdfstring{The case study on $(\mathbf{C}_{7})$}{The case study on (C7)}}
\begin{figure}[h!]
    \centering
    \begin{subfigure}[t]{0.9\textwidth}
        \centering
        \includegraphics[width=\textwidth]{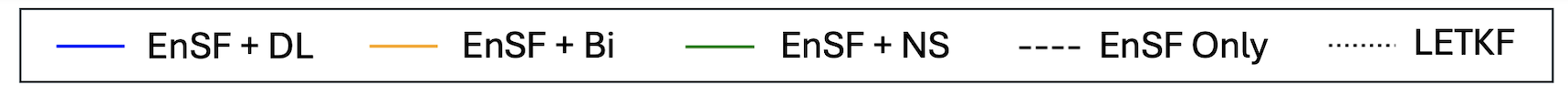}
    \end{subfigure} 
    \begin{subfigure}[t]{0.3\textwidth}
        \centering
        \includegraphics[width=\textwidth]{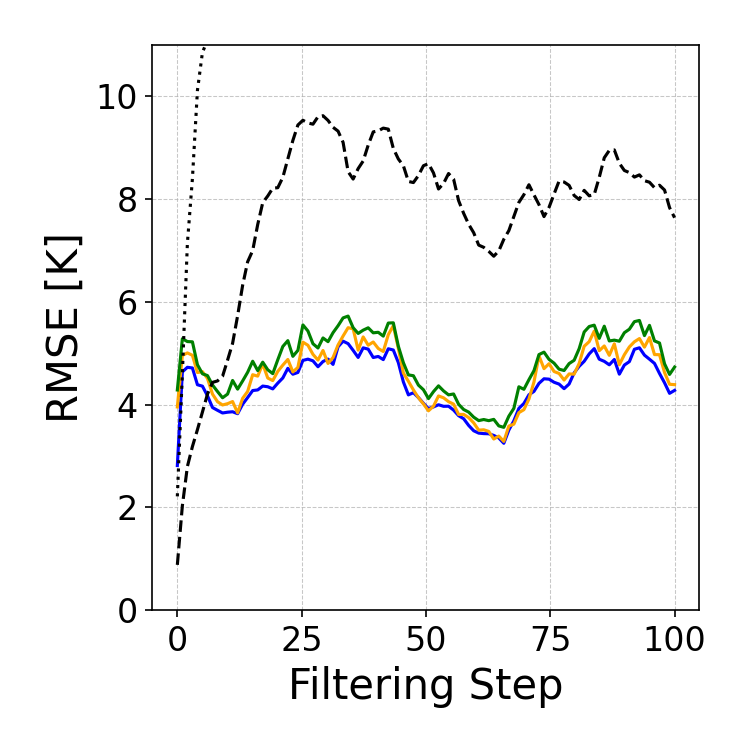}
        \caption{Total RMSE}
        \label{fig:total rmse arctan 64 12 5per}
    \end{subfigure}
    \begin{subfigure}[t]{0.3\textwidth}
        \centering
        \includegraphics[width=\textwidth]{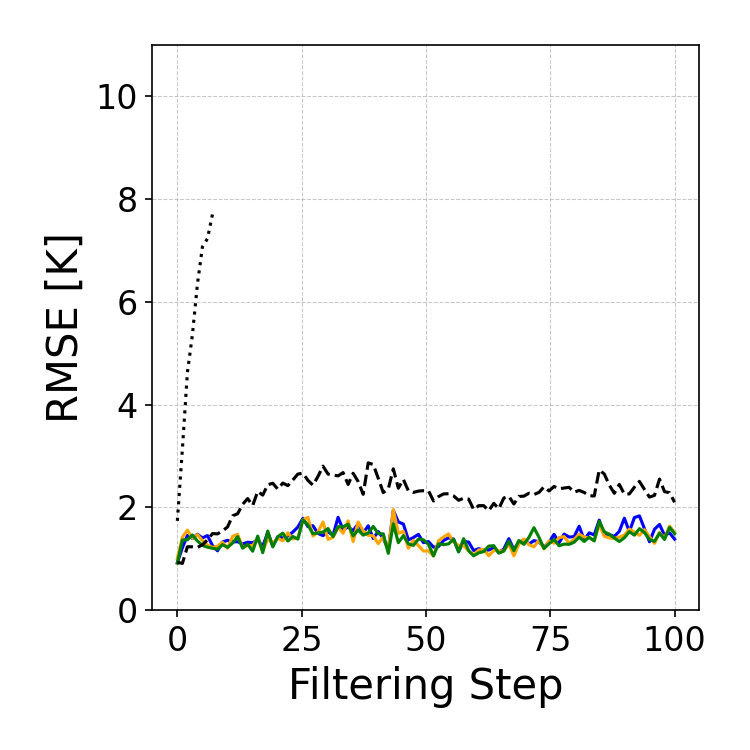}
        \caption{RMSE {observed}}
        \label{fig:obs rmse arctan 64 12 5per}
    \end{subfigure}
    \begin{subfigure}[t]{0.3\textwidth}
        \centering
        \includegraphics[width=\textwidth]{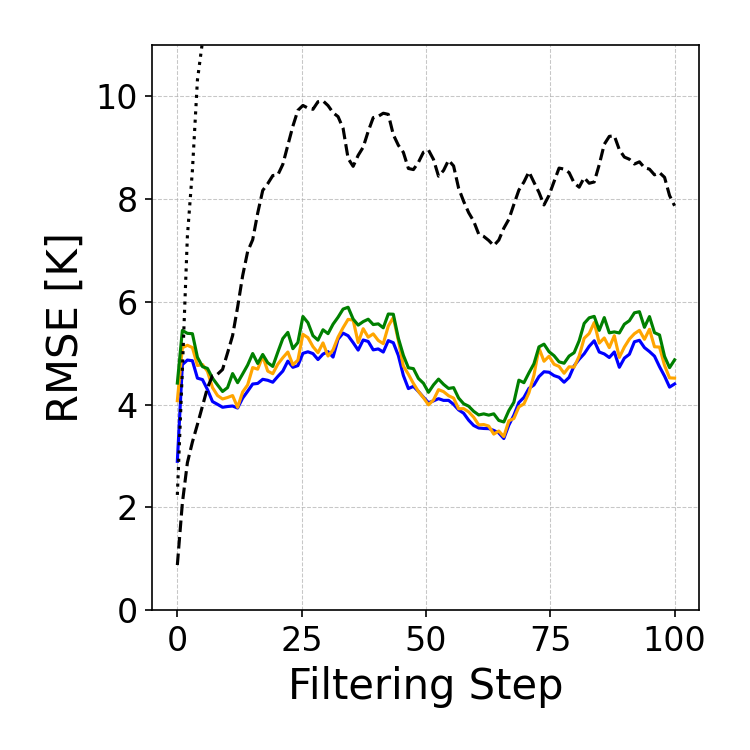}
        \caption{RMSE {unobserved}}
        \label{fig:unob rmse arctan 64 12 5per}
    \end{subfigure}
    \vspace{-0.5cm}
    \caption{Results of $(\mathbf{C}_7)$: {(a) The total RMSE including all state variables, (b) the RMSE of the observed variables (c) RMSE of only unobserved variables.}}
    \label{fig:rmse arctan 64 12hourly 5}
\end{figure}

\begin{figure}[h!]
    \centering
    \begin{subfigure}[t]{0.25\textwidth}
        \centering
        \includegraphics[width=\textwidth]{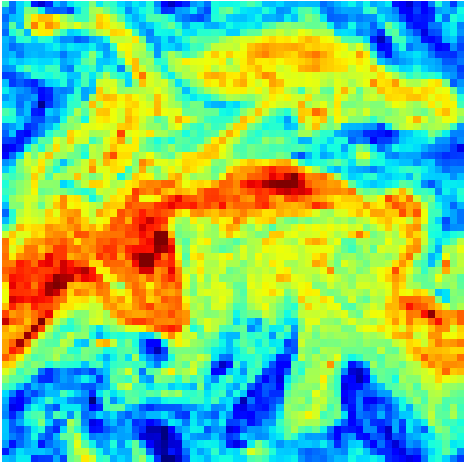}
        \caption{Truth}
        \label{Arctan_N64_12hrly_5per Truth}
    \end{subfigure}
    \begin{subfigure}[t]{0.25\textwidth}
        \centering
        \includegraphics[width=\textwidth]{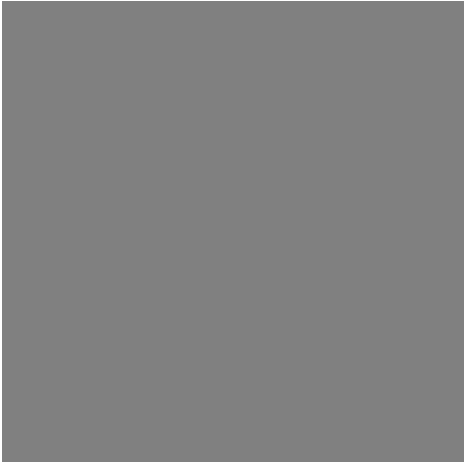}
        \caption{LETKF}
        \label{Arctan_N64_12hrly_5per LETKF}
    \end{subfigure}
    \begin{subfigure}[t]{0.25\textwidth}
        \centering
        \includegraphics[width=\textwidth]{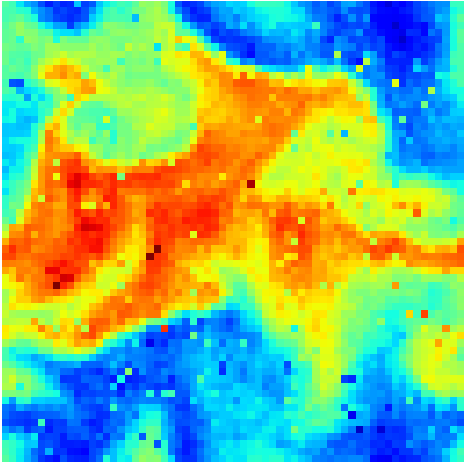}
        \caption{EnSF Only}
        \label{Arctan_N64_12hrly_5per EnSF}
    \end{subfigure}
    \begin{subfigure}[t]{0.25\textwidth}
        \centering
        \includegraphics[width=\textwidth]{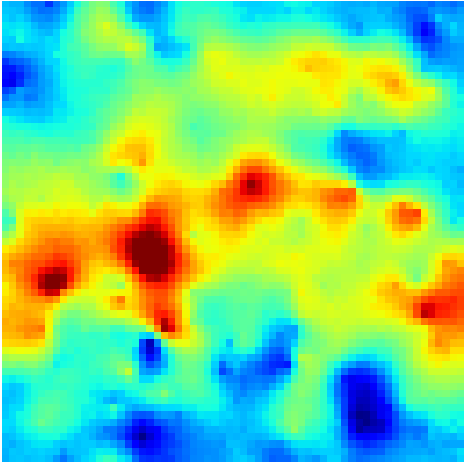}
        \caption{EnSF+Bi}
        \label{Arctan_N64_12hrly_5per SciKit}
    \end{subfigure}
    \begin{subfigure}[t]{0.25\textwidth}
        \centering
        \includegraphics[width=\textwidth]{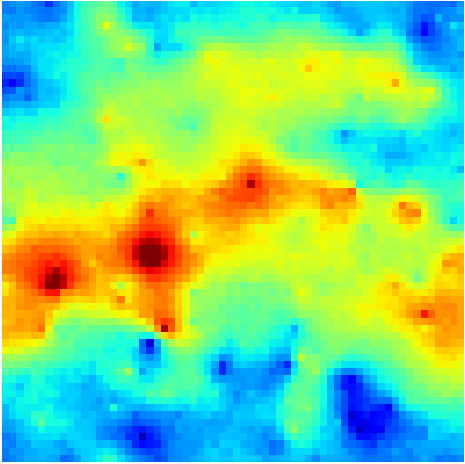}
        \caption{EnSF+DL}
        \label{Arctan_N64_12hrly_5per DCT}
    \end{subfigure}
    \begin{subfigure}[t]{0.25\textwidth}
        \centering
        \includegraphics[width=\textwidth]{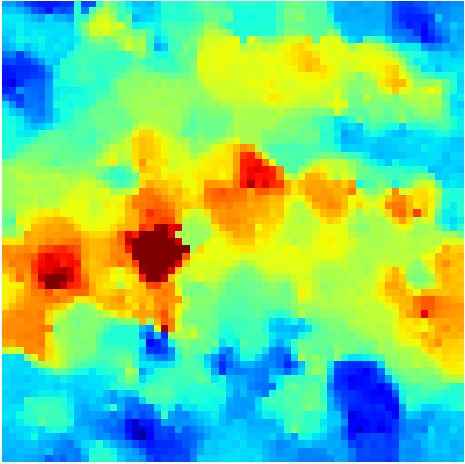}
        \caption{EnSF+NS}
        \label{fig:Arctan_N64_12hrly_5per cv2}
    \end{subfigure}
    \vspace{-0.4cm}
    \caption{Results of $(\mathbf{C}_7)$: snapshot at filtering step 100. LETKF (b) diverges to undefined values, EnSF without inpainting shows a total failure, EnSF inpainting can track the major dynamics.}
    \label{snap:Arctan_N64_12hrly_5per}
\end{figure}

We compare the performance on the $64 \times 64$ grid where the state of the SQG model is 5\% observed at fixed grid points through the nonlinear arctangent operator, with a 12-hour assimilation interval. As shown in the $64 \times 64$ tuning chart in Figure \ref{fig:LETKF 64 tuning chart}, LETKF cannot handle the combination of long assimilation intervals, low observation percentages, and nonlinear observations, resulting in no viable tuning parameter pairs. In Figure \ref{fig:total rmse arctan 64 12 5per}, the RMSE of LETKF increases rapidly, indicating a failure in tracking the SQG dynamics. Similarly, EnSF without inpainting also fails under these extremely challenging conditions. However, integrating the inpainting method with EnSF reduces the RMSE of the three EnSF inpainting variants to approximately 4.6. While this value is not low, Figure \ref{snap:Arctan_N64_12hrly_5per} demonstrates that, compared to the failure, these methods still capture the major dynamics of the SQG model.

Delving deeper into Figure \ref{fig:obs rmse arctan 64 12 5per}, we observe the strength of EnSF: even when the total RMSE indicates failure in tracking the SQG model, the RMSE at observed points remains low. This suggests that EnSF provides accurate estimates at observed points, offering a reliable foundation for the inpainting method to reconstruct the unobserved regions. The RMSE of observed points across the three EnSF inpainting methods is consistently low and stable. In Figure \ref{fig:unob rmse arctan 64 12 5per}, we see that most errors originate from the unobserved state variables. Given that $(\mathbf{C}_7)$ represents the most challenging scenario—characterized by a coarse $64 \times 64$ grid, the longest assimilation interval, the lowest observation availability, and a nonlinear arctangent observation operator—the EnSF inpainting methods demonstrate remarkable robustness and effectiveness in this extreme case.

\subsubsection{\texorpdfstring{The case study on $(\mathbf{C}_{15})$}{The case study on (C15)}}
In this experiment, we increase the resolution from the coarse $64 \times 64$ grid to the fine $256 \times 256$ grid while keeping all other factors the same as in $(\mathbf{C}_{7})$. The finer grid provides more available information for the model and more ``pixels'' for the EnSF inpainting methods, which exhibit more linear changes in their neighborhood regions. For LETKF, since no viable parameter pairs were found in the $64 \times 64$ case, we use the pair that results in the slowest RMSE explosion: $L_h = 4500$ km and RTPS = 0.9.

\begin{figure}[h!]
    \centering
    \begin{subfigure}[t]{0.9\textwidth}
        \centering
        \includegraphics[width=\textwidth]{separate_legend.png}
    \end{subfigure} 
    \begin{subfigure}[t]{0.3\textwidth}
        \centering
        \includegraphics[width=\textwidth]{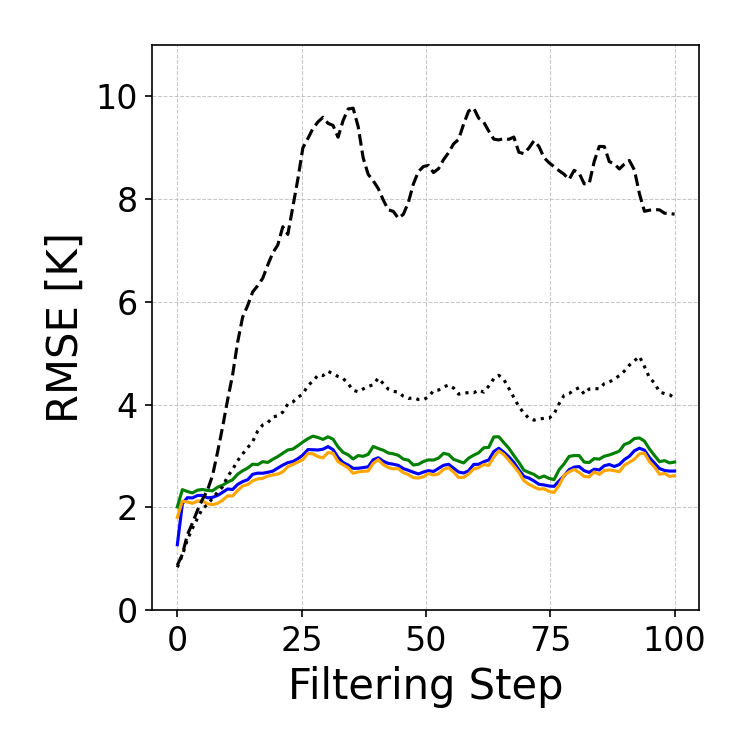}
        \caption{Total RMSE}
        \label{fig:total rmse arctan 256 12 5per}
    \end{subfigure}
    \begin{subfigure}[t]{0.3\textwidth}
        \centering
        \includegraphics[width=\textwidth]{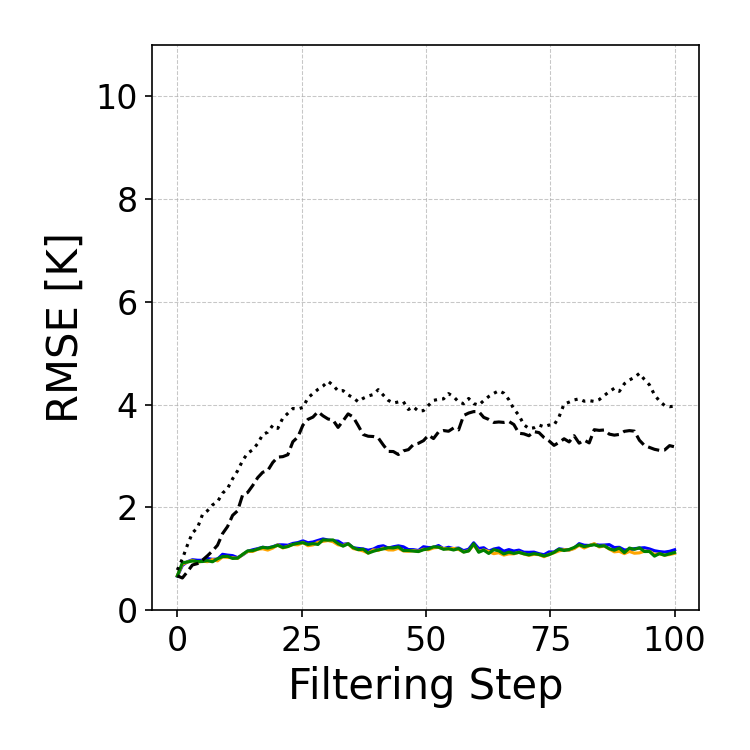}
        \caption{observed RMSE}
        \label{fig:obs rmse arctan 256 12 5per}
    \end{subfigure}
    \begin{subfigure}[t]{0.3\textwidth}
        \centering
        \includegraphics[width=\textwidth]{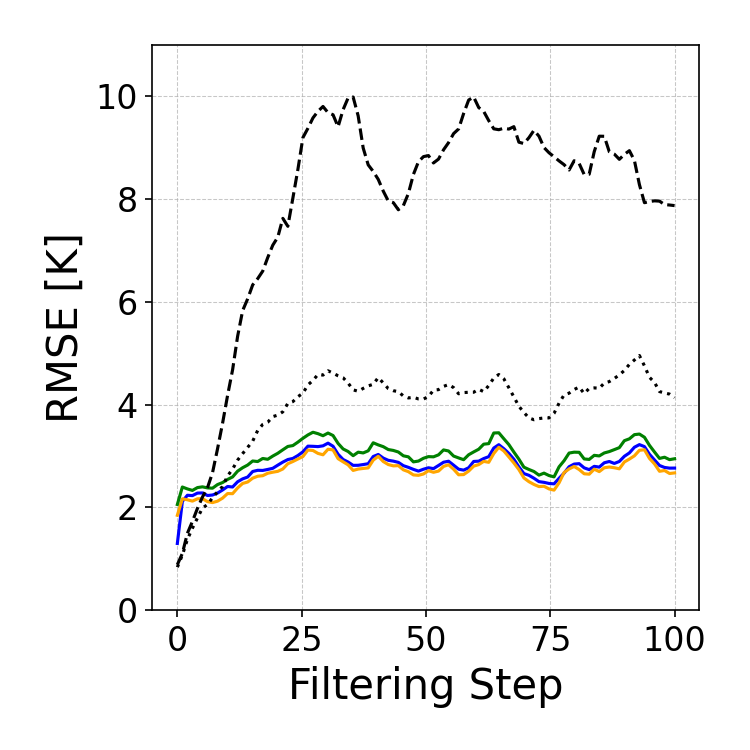}
        \caption{unobserved RMSE}
        \label{fig:unob rmse arctan 256 12 5per}
    \end{subfigure}
    \vspace{-0.5cm}
    \caption{Results of $(\mathbf{C}_{15})$: The total RMSE (a) includes all state points, the observed RMSE (b) only includes observed state points, and the unobserved RMSE (c) only includes unobserved state points.}
    \label{fig:rmse arctan 256 12hourly 5}
\end{figure}

\begin{figure}[h!]
    \centering
    \begin{subfigure}[t]{0.25\textwidth}
        \centering
        \includegraphics[width=\textwidth]{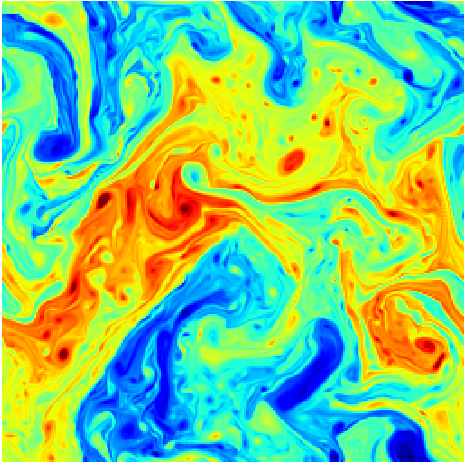}
        \caption{Truth}
        \label{Truth 256 12 5per}
    \end{subfigure}
    \begin{subfigure}[t]{0.25\textwidth}
        \centering
        \includegraphics[width=\textwidth]{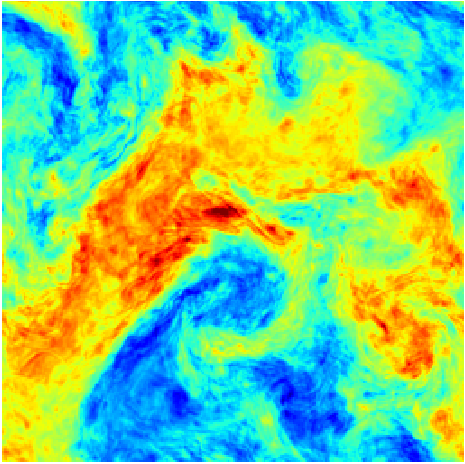}
        \caption{LETKF}
        \label{LETKF 256 12 5per}
    \end{subfigure}
    \begin{subfigure}[t]{0.25\textwidth}
        \centering
        \includegraphics[width=\textwidth]{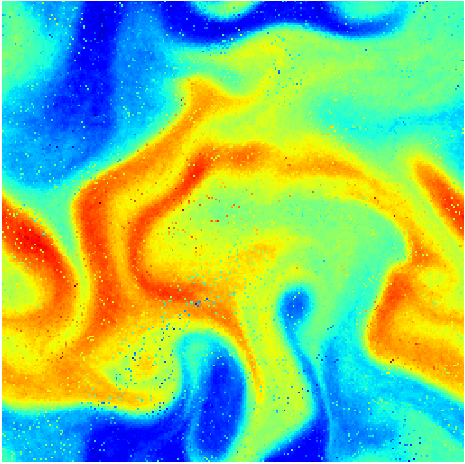}
        \caption{EnSF Only}
        \label{EnSF 256 12 5per}
    \end{subfigure}
    \\
    \begin{subfigure}[t]{0.25\textwidth}
        \centering
        \includegraphics[width=\textwidth]{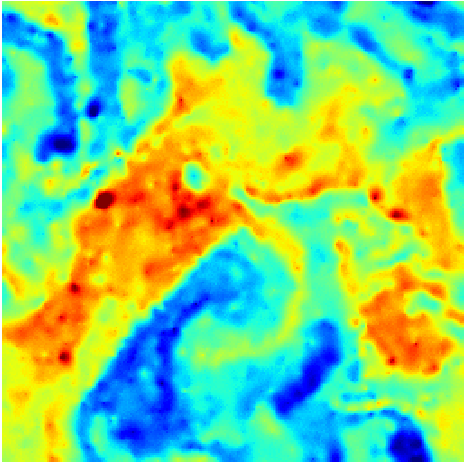}
        \caption{EnSF + Bi}
        \label{SciKit 256 12 5per}
    \end{subfigure}
    \begin{subfigure}[t]{0.25\textwidth}
        \centering
        \includegraphics[width=\textwidth]{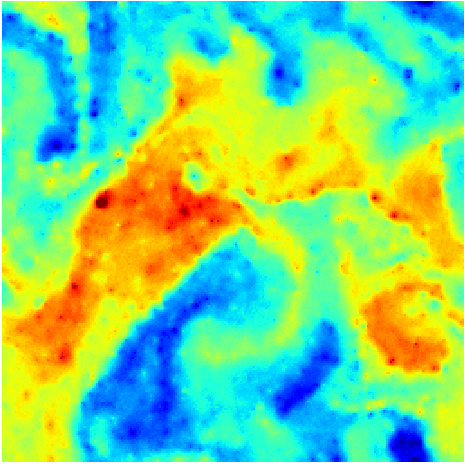}
        \caption{EnSF + DL}
        \label{DCT 256 12 5per}
    \end{subfigure}
    \begin{subfigure}[t]{0.25\textwidth}
        \centering
        \includegraphics[width=\textwidth]{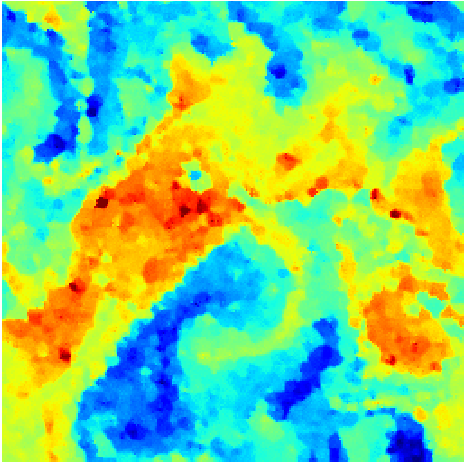}
        \caption{EnSF + NS}
        \label{cv2  256 12 5per}
    \end{subfigure}
    \caption{Results of $(\mathbf{C}_{15})$: Snapshot at filtering step 100. LETKF (b) captures the overall shape of the major dynamics, missing most extreme events (dark red and blue regions) and mismatching small-scale details. EnSF without inpainting shows a total failure, EnSF inpainting can track the major dynamics but lose some small-scale details.}
    \label{snap:Arctan_N256_12hrly_5per}
\end{figure}

In Figures \ref{fig:total rmse arctan 256 12 5per} and \ref{LETKF 256 12 5per}, we observe that with more observational information, LETKF's RMSE no longer explodes. While LETKF appears to track both major dynamics and small-scale details, closer inspection reveals that it only captures the overall shape of the major dynamics, missing most extreme events (dark red and blue regions) and mismatching small-scale details.
In contrast, the EnSF inpainting methods achieve a lower total RMSE compared to LETKF. Figures \ref{SciKit 256 12 5per}, \ref{DCT 256 12 5per}, and \ref{cv2 256 12 5per} show that EnSF inpainting methods successfully track all major dynamics, although they lose some small-scale details. However, compared to $(\mathbf{C}_{7})$, the finer $256 \times 256$ grid result captures more details overall. Recall that the total RMSE for the three EnSF inpainting methods was around 4.6 in $(\mathbf{C}_{7})$. In $(\mathbf{C}_{15})$, the total RMSE decreases to approximately 2.8.

Furthermore, comparing the RMSE of observed states between $(\mathbf{C}_{7})$ in Figure \ref{fig:obs rmse arctan 64 12 5per} and $(\mathbf{C}_{15})$ in Figure \ref{fig:obs rmse arctan 256 12 5per}, we see that the RMSE is more stable in $(\mathbf{C}_{15})$ due to the increased availability of observational information. These results highlight the ability of EnSF inpainting methods to effectively utilize additional observation information to reduce assimilation error.

\subsubsection{\texorpdfstring{The case study on $(\mathbf{C}_{6})$}{The case study on (C6)}}
In this experiment, we again evaluate performance in the $64 \times 64$ case, but compared to $(\mathbf{C}_{7})$, we increase the percentage of observed grid points to 25\% using the arctangent nonlinear operator and reduce the assimilation frequency to 3 hours. The shorter 3-hour assimilation interval decreases deviations from the true state compared to the 12-hour interval. Additionally, the 5 times increase in observations compared to $(\mathbf{C}_{7})$ significantly reduces the uncertainty in tracking the SQG model.
Despite these improvements, as shown in the tuning chart for $64 \times 64$ in Figure \ref{fig:LETKF 64 tuning chart}, the best-tuned parameter pair for LETKF results in a total failure, with an RMSE of 9.1. Figure \ref{Arctan_N64_3hrly_25per LETKF} illustrates LETKF's inability to track the SQG model under these conditions.

\begin{figure}[h!]
    \centering
    \begin{subfigure}[t]{0.9\textwidth}
        \centering
        \includegraphics[width=\textwidth]{separate_legend.png}
    \end{subfigure} 
    \begin{subfigure}[t]{0.3\textwidth}
        \centering
        \includegraphics[width=\textwidth]{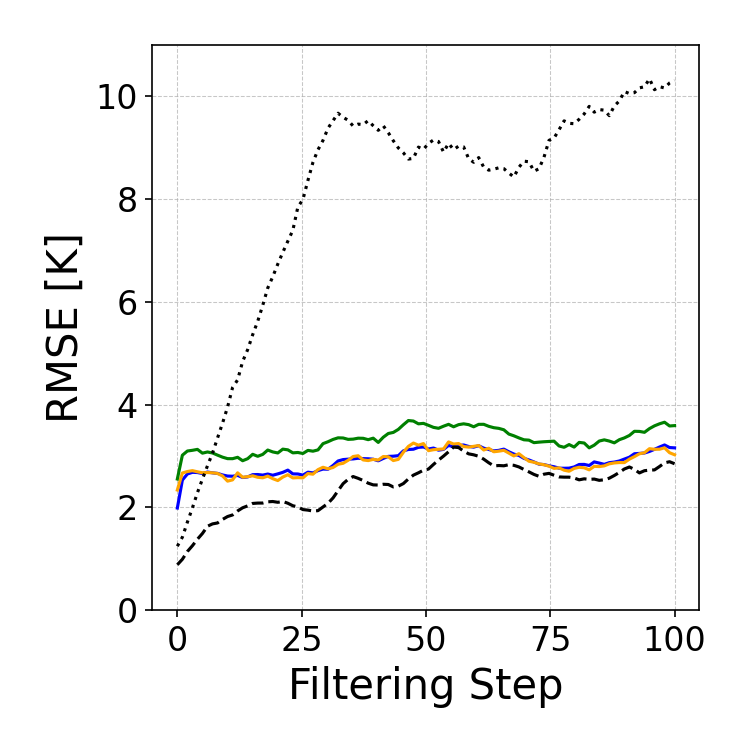}
        \caption{Total RMSE}
        \label{fig:total rmse arctan 64 3 25per}
    \end{subfigure}
    \begin{subfigure}[t]{0.3\textwidth}
        \centering
        \includegraphics[width=\textwidth]{  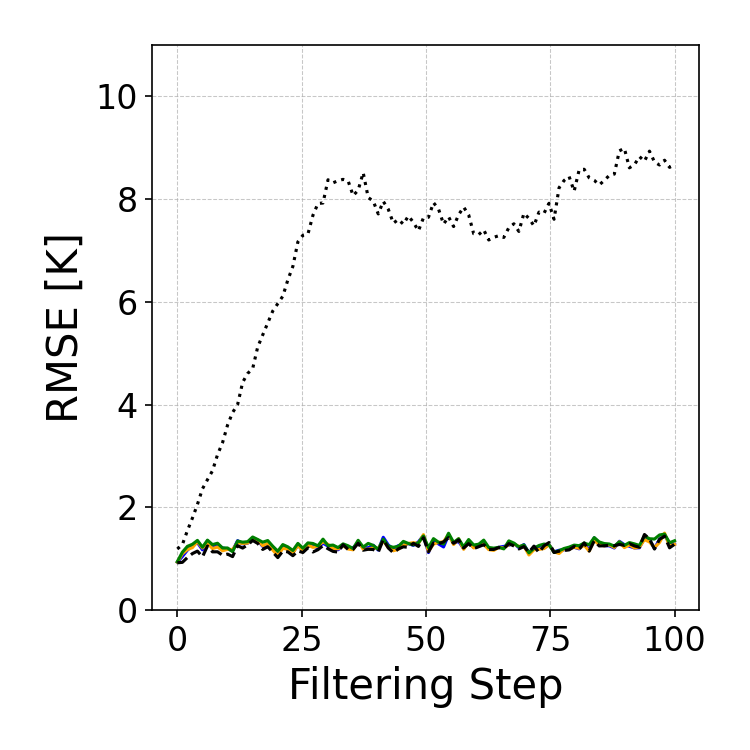}
        \caption{observed RMSE}
        \label{fig:obs rmse arctan 64 3 25per}
    \end{subfigure}
    \begin{subfigure}[t]{0.3\textwidth}
        \centering
        \includegraphics[width=\textwidth]{  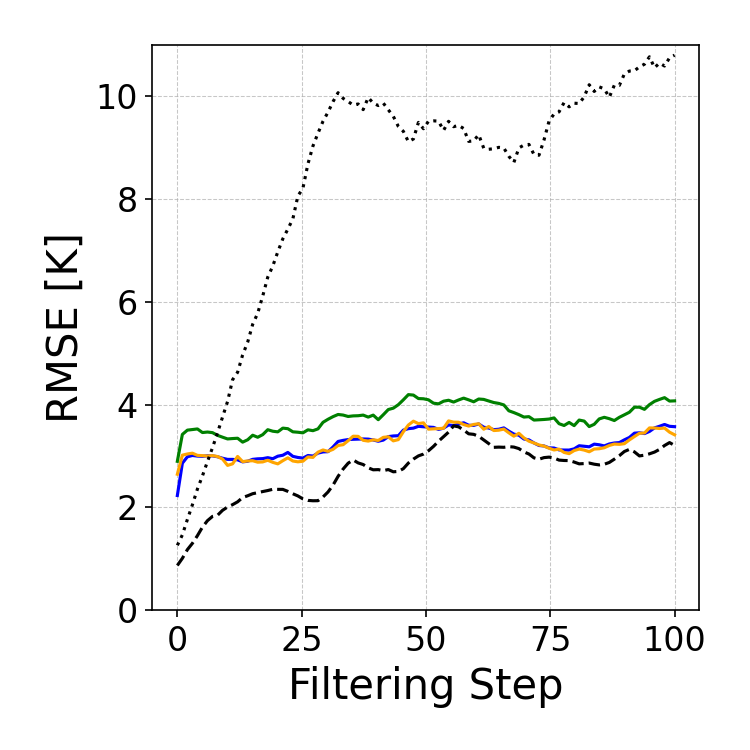}
        \caption{unobserved RMSE}
        \label{fig:unob rmse arctan 64 3 25per}
    \end{subfigure}
    \vspace{-0.5cm}
    \caption{Results of $(\mathbf{C}_{6})$: The total RMSE (a) includes all state points, the observed RMSE (b) only includes observed state points, and the unobserved RMSE (c) only includes unobserved state points.}
    \label{fig:rmse arctan 64 3hourly 25}
\end{figure}

\begin{figure}[h!]
    \centering
    \begin{subfigure}[t]{0.25\textwidth}
        \centering
        \includegraphics[width=\textwidth]{  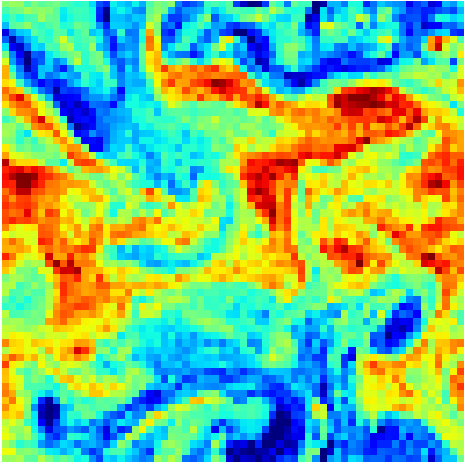}
        \caption{Truth}
        \label{Arctan_N64_3hrly_25per Truth}
    \end{subfigure}
    \begin{subfigure}[t]{0.25\textwidth}
        \centering
        \includegraphics[width=\textwidth]{  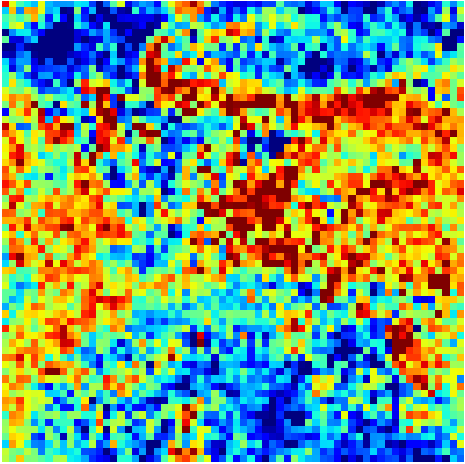}
        \caption{LETKF}
        \label{Arctan_N64_3hrly_25per LETKF}
    \end{subfigure}
    \begin{subfigure}[t]{0.25\textwidth}
        \centering
        \includegraphics[width=\textwidth]{  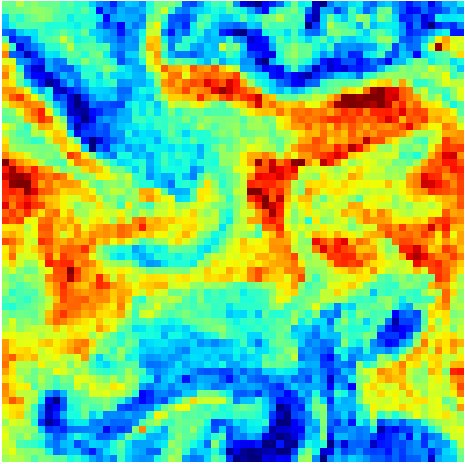}
        \caption{EnSF Only}
        \label{Arctan_N64_3hrly_25per EnSF}
    \end{subfigure}
    \begin{subfigure}[t]{0.25\textwidth}
        \centering
        \includegraphics[width=\textwidth]{  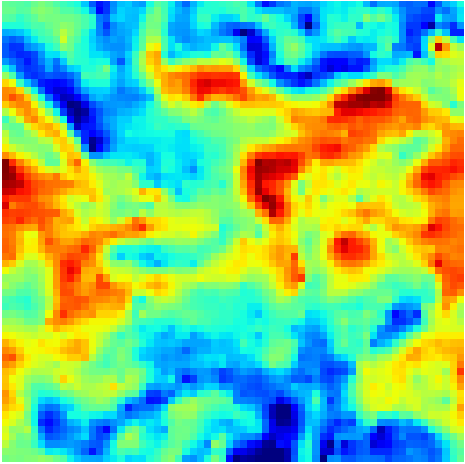}
        \caption{EnSF+Bi}
        \label{Arctan_N64_3hrly_25per SciKit}
    \end{subfigure}
    \begin{subfigure}[t]{0.25\textwidth}
        \centering
        \includegraphics[width=\textwidth]{  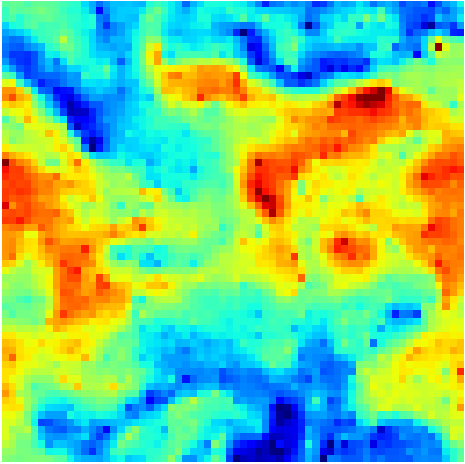}
        \caption{EnSF+DL}
        \label{Arctan_N64_3hrly_25per DCT}
    \end{subfigure}
    \begin{subfigure}[t]{0.25\textwidth}
        \centering
        \includegraphics[width=\textwidth]{  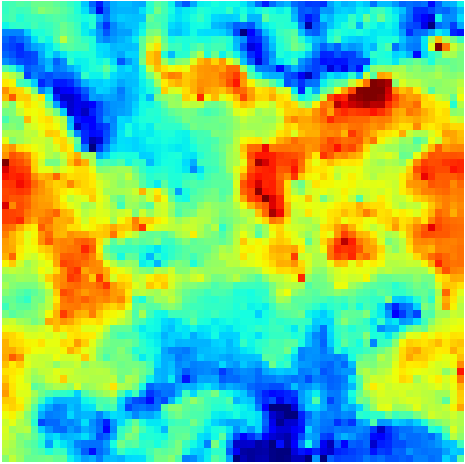}
        \caption{EnSF+NS}
        \label{fig:Arctan_N64_3hrly_25per cv2}
    \end{subfigure}
    \vspace{-0.4cm}
    \caption{Results of $(\mathbf{C}_{6})$: Snapshot at filtering step 100. LETKF (b) shows a total failure. EnSF captures both major dynamics and small-scale detail. EnSF inpainting can track the major dynamics but lose some small-scale details.}
    \label{snap:Arctan_N64_3hrly_25per}
\end{figure}

In Figure \ref{fig:total rmse arctan 64 3 25per}, EnSF without inpainting achieves the best performance, while EnSF inpainting methods exhibit higher RMSE. From Figures \ref{fig:obs rmse arctan 64 3 25per} and \ref{fig:unob rmse arctan 64 3 25per}, we observe that the RMSE of the observed state is nearly identical across methods, with most errors originating from the unobserved state. This indicates a limitation of inpainting methods in accurately reconstructing the unobserved state, as the inpainting process itself can introduce additional errors.
Figure \ref{snap:Arctan_N64_3hrly_25per} provides a clear visual comparison: EnSF captures both major dynamics and small-scale details, while EnSF inpainting methods successfully capture the major dynamics but lose some small-scale details. We will further discuss these findings at the end of this section.

\subsubsection{\texorpdfstring{The case study on $(\mathbf{C}_{14})$}{The case study on (C14)}} We increase the resolution from the coarse $64 \times 64$ grid to the fine $256 \times 256$ grid while keeping all other factors the same as in $(\mathbf{C}_{6})$. Compared to $(\mathbf{C}_{7})$, this represents the most ideal scenario under the nonlinear arctangent observation operator, with a fine $256 \times 256$ grid, the shortest assimilation frequency, and the highest percentage of available observations. For LETKF, we use the same fine-tuned parameter pairs as in the $64 \times 64$ case from $(\mathbf{C}_{6})$.

\begin{figure}[h!]
    \centering
    \begin{subfigure}[t]{0.9\textwidth}
        \centering
        \includegraphics[width=\textwidth]{ separate_legend.png}
    \end{subfigure} 
    \begin{subfigure}[t]{0.3\textwidth}
        \centering
        \includegraphics[width=\textwidth]{ 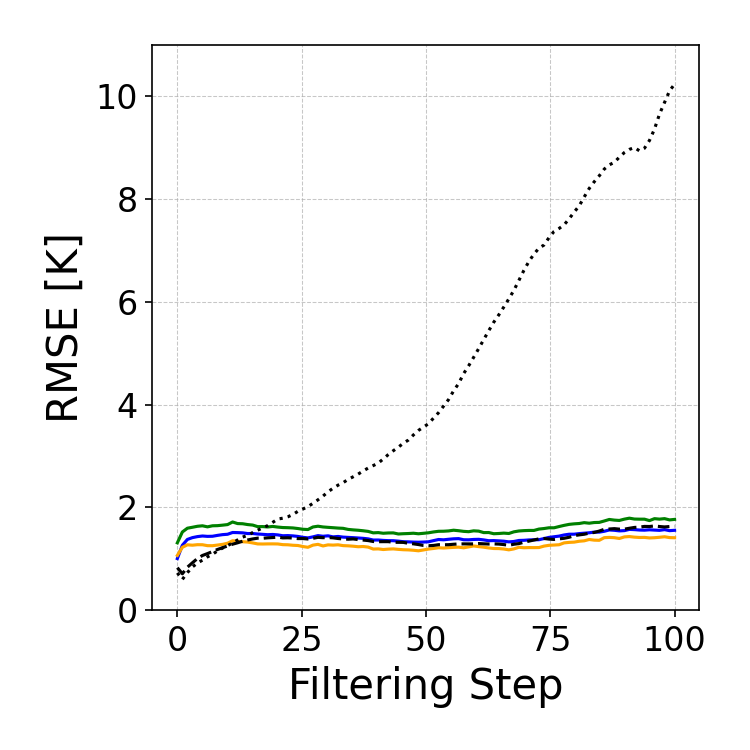}
        \caption{Total RMSE}
        \label{fig:total rmse arctan 256 3 25per}
    \end{subfigure}
    \begin{subfigure}[t]{0.3\textwidth}
        \centering
        \includegraphics[width=\textwidth]{ 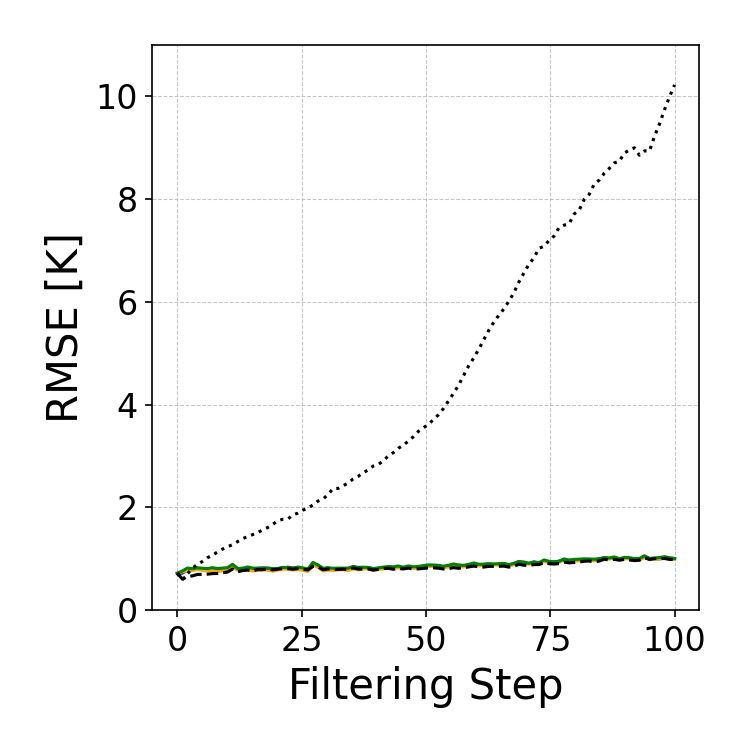}
        \caption{observed RMSE}
        \label{fig:obs rmse arctan 256 3 25per}
    \end{subfigure}
    \begin{subfigure}[t]{0.3\textwidth}
        \centering
        \includegraphics[width=\textwidth]{ 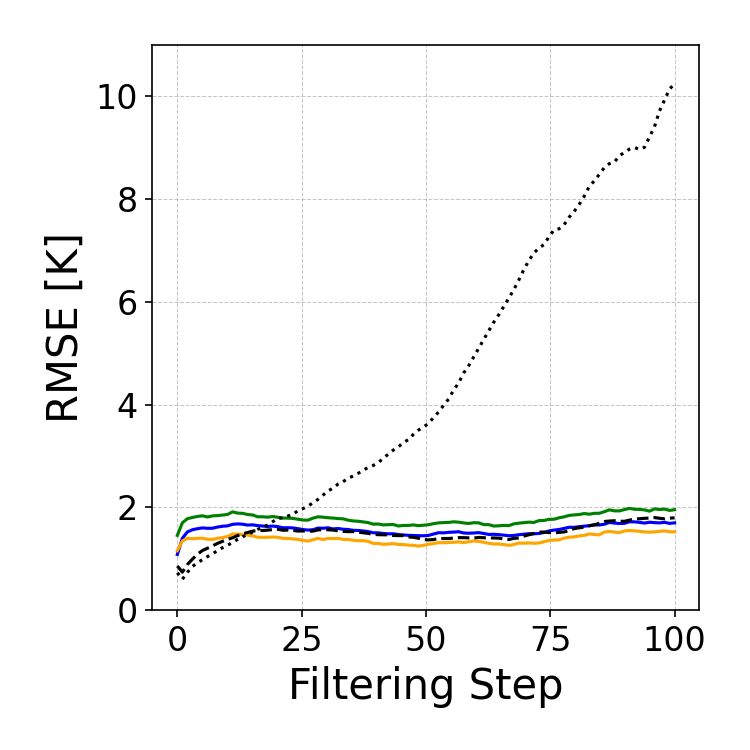}
        \caption{unobserved RMSE}
        \label{fig:unob rmse arctan 256 3 25per}
    \end{subfigure}
    \vspace{-0.4cm}
    \caption{Results of $(\mathbf{C}_{14})$: The total RMSE (a) includes all state points, the observed RMSE (b) only includes observed state points, and the unobserved RMSE (c) only includes unobserved state points.}
    \label{fig:rmse arctan 256 3hourly 25}
\end{figure}

\begin{figure}[h!]
    \centering
    \begin{subfigure}[t]{0.25\textwidth}
        \centering
        \includegraphics[width=\textwidth]{ 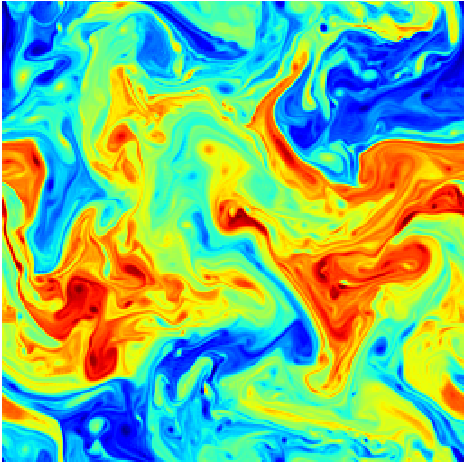}
        \caption{Truth}
        \label{Truth 256 3 25per}
    \end{subfigure}
    \begin{subfigure}[t]{0.25\textwidth}
        \centering
        \includegraphics[width=\textwidth]{ 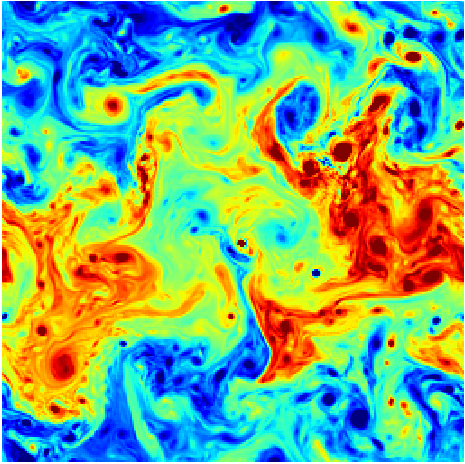}
        \caption{LETKF}
        \label{LETKF 256 3 25per}
    \end{subfigure}
    \begin{subfigure}[t]{0.25\textwidth}
        \centering
        \includegraphics[width=\textwidth]{ 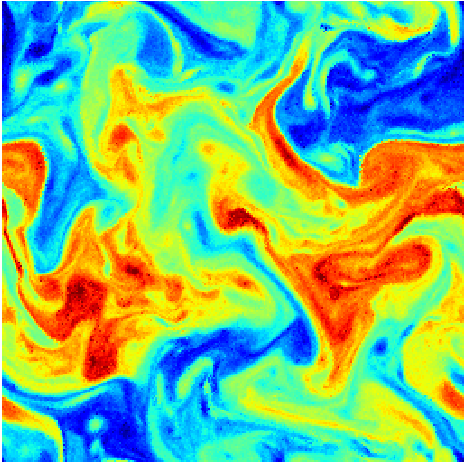}
        \caption{EnSF Only}
        \label{EnSF 256 3 25per}
    \end{subfigure}
    \\
    \begin{subfigure}[t]{0.25\textwidth}
        \centering
        \includegraphics[width=\textwidth]{ 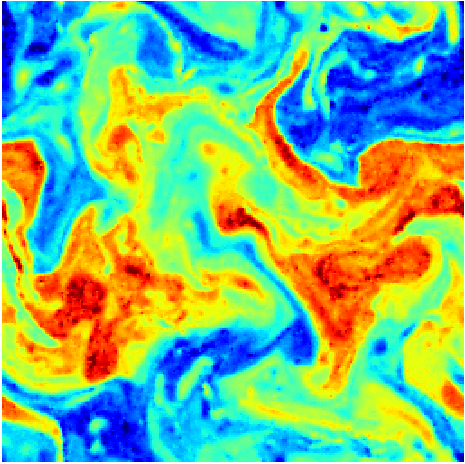}
        \caption{EnSF + Bi}
        \label{SciKit 256 3 25per}
    \end{subfigure}
    \begin{subfigure}[t]{0.25\textwidth}
        \centering
        \includegraphics[width=\textwidth]{ 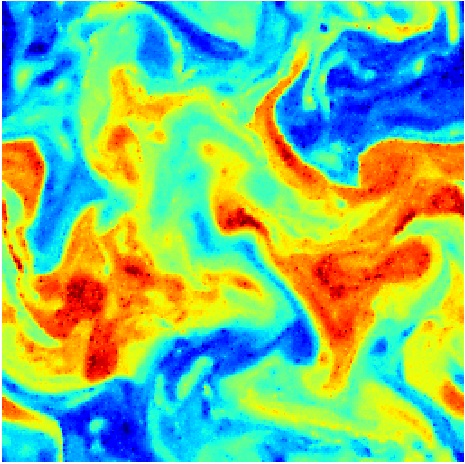}
        \caption{EnSF + DL}
        \label{DCT 256 3 25per}
    \end{subfigure}
    \begin{subfigure}[t]{0.25\textwidth}
        \centering
        \includegraphics[width=\textwidth]{ 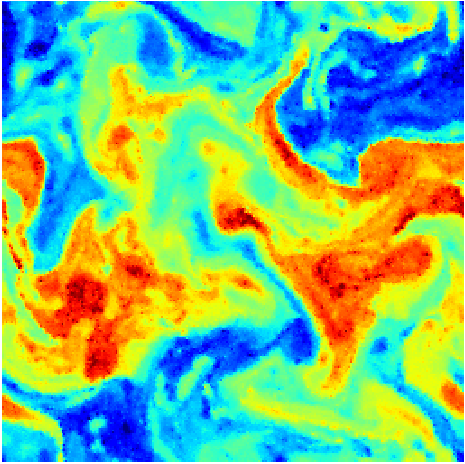}
        \caption{EnSF + NS}
        \label{cv2 256 3 25per}
    \end{subfigure}
    \vspace{-0.4cm}
    \caption{Results of $(\mathbf{C}_{14})$: snapshot at filtering step 100. LETKF (b) shows a total failure. EnSF and EnSF inpainting both capture major dynamics and small-scale detail.}
    \label{snap:Arctan_N256_3hrly_25per}
\end{figure}

Figures \ref{fig:total rmse arctan 256 3 25per} and \ref{LETKF 256 3 25per} show that even under the most ideal settings, the RMSE of LETKF still explodes. LETKF fails to capture both major dynamics and small-scale details.
In contrast, EnSF inpainting methods demonstrate excellent accuracy under these ideal conditions. The RMSE is highly stable, as shown in Figures \ref{fig:obs rmse arctan 256 3 25per} and \ref{fig:unob rmse arctan 256 3 25per}. Compared to $(\mathbf{C}_{6})$, the differences in total RMSE between EnSF without inpainting and EnSF with inpainting are minimal. The lowest total RMSE, as reported in Table \ref{tab:rmse_table}, is achieved by EnSF combined with Biharmonic PDE-based inpainting, at 1.27, while EnSF without inpainting achieves 1.4. Further discussion on these results will be provided in the next section.

\subsection{Additional discussion} In this work, we conducted 16 experiments to explore the interplay of assimilation frequency, resolution, observation network, and observation sparsity. Table \ref{tab:rmse_table} summarizes the average total RMSE across all experiments for the three EnSF inpainting methods, EnSF without inpainting, and the reference LETKF.
\begin{table}[h!]
    \centering
    \begin{tabular}{c|ccccc}
    \toprule
        Label &  EnSF+DL & EnSF+Bi & EnSF+NS & EnSF Only & LETKF\\
    \midrule
      ($\mathbf{C}_1$)  & 4.47 & 4.71 & 5.05 & 6.49 & \cellcolor{gray!25} 3.86 \\
      ($\mathbf{C}_2$)  & 2.86 & 2.81 & 3.27 & \cellcolor{gray!25} 2.29 & 0.67\\
      ($\mathbf{C}_3$)  & 4.43 & 4.57 & 4.82 & 8.44 & \cellcolor{gray!25} 4.27 \\
      ($\mathbf{C}_4$)  & 3.14 & 2.90 & 3.30 & 4.66 & \cellcolor{gray!25} 2.39 \\
      ($\mathbf{C}_5$)  & \cellcolor{gray!25} 4.29 & 4.60 & 4.90 & 6.38 & NaN \\
      ($\mathbf{C}_6$)  & 2.98 & 2.96 & 3.41 & \cellcolor{gray!25} 2.63 & 9.27 \\
      ($\mathbf{C}_7$)  & \cellcolor{gray!25} 4.46 & 4.61 & 4.86 & 8.40 & NaN \\
      ($\mathbf{C}_8$)  & 3.17 & \cellcolor{gray!25} 2.98 & 3.35 & 4.77 & 6.52 \\
      ($\mathbf{C}_9$)  & \cellcolor{gray!25} 2.22 & 2.28 & 2.59 & 4.32 & \cellcolor{gray!25} 0.71 \\
      ($\mathbf{C}_{10}$)  & 1.42 & 1.26 & 1.59 & \cellcolor{gray!25}1.24 & 4.48 \\
      ($\mathbf{C}_{11}$)  & 2.77 & \cellcolor{gray!25} 2.66 & 2.95 & 8.85 & NaN \\
      ($\mathbf{C}_{12}$)  & 1.86 & \cellcolor{gray!25} 1.55 & 1.90 & 5.20 & 7.85 \\
      ($\mathbf{C}_{13}$)  & \cellcolor{gray!25} 2.22 & 2.29 & 2.60 & 6.51 & NaN \\
      ($\mathbf{C}_{14}$)  & 1.43 & \cellcolor{gray!25} 1.27 & 1.61 & 1.40 & 8.66 \\
      ($\mathbf{C}_{15}$)  & 2.82 & \cellcolor{gray!25} 2.74 & 3.02 & 8.67 & 4.28 \\
      ($\mathbf{C}_{16}$)  & 1.89 & \cellcolor{gray!25} 1.61 & 1.95 & 5.32 & 3.94 \\
      \bottomrule
    \end{tabular}
    \caption{This table summarizes the average total RMSE across all experiments. The average is calculated by excluding the initial spin-up steps and then averaging the remaining steps to accurately represent the RMSE levels. The best-performing method in each case is highlighted}
    \label{tab:rmse_table}
\end{table}


The LETKF, as expected, performs well under linear settings after fine-tuning. This is evident in the $64 \times 64$ grid case, where computational costs remain manageable. However, when transitioning to the $256 \times 256$ grid and reusing the fine-tuned parameters from the $64 \times 64$ case, the results reveal LETKF's high sensitivity to these parameters. This sensitivity highlights the lack of versatility in adapting to different problem settings. Additionally, the study demonstrates LETKF's limitations in assimilating dynamic systems with nonlinear observation operators. It is possible to use so-called adaptive inflation techniques which could result in better performance of the LETKF. We did not pursue that extension in this paper, since, as Fig. 5 shows, even in the low-resolution runs the localization and inflation tuning failed. The dynamics and observation operators are simply too nonlinear for the LETKF to work properly.

For EnSF without inpainting, Figures \ref{fig:obs rmse arctan 64 12 5per}, \ref{fig:obs rmse arctan 256 12 5per}, \ref{fig:obs rmse arctan 64 3 25per},  \ref{fig:obs rmse arctan 256 3 25per}, and additional figures in the appendix illustrate its effectiveness in tracking the observed state. The RMSE is consistently low and stable. For the SQG model, when the assimilation frequency is 3 hours with 25\% observation coverage, EnSF alone performs well. In this setting, the 3-hour forward prediction introduces minimal deviations from the ground truth, and the higher observation density reduces uncertainty. These results are highlighted in the ``EnSF only'' column of Table \ref{tab:rmse_table}. Inpainting introduces additional error due to the inpainting process itself. This is evident in the 3-hour assimilation with 25\% observation settings, i.e., $(\mathbf{C}_2)$, $(\mathbf{C}_6)$, $(\mathbf{C}_{10})$, and $(\mathbf{C}_{14})$, which represent the most ideal conditions in terms of assimilation frequency and observation sparsity. Under such ideal conditions, inpainting methods may not be necessary. However, even with the added error, the overall performance of EnSF with inpainting remains strong, with minimal additional RMSE.

In real-world applications, ideal scenarios are rare, and most cases are highly challenging. Table \ref{tab:rmse_table} shows that EnSF inpainting methods maintain robust and accurate performance across all settings. Given EnSF's advantage in handling nonlinear observation operators, the difference in total RMSE between linear and nonlinear cases under the same assimilation frequency, resolution, and observation sparsity is negligible. This demonstrates the efficiency and flexibility of EnSF with inpainting in adapting to diverse settings without requiring parameter fine-tuning.

This study introduces one dictionary learning-based inpainting method and two PDE-based inpainting methods. The primary distinction lies in their approach: dictionary learning-based inpainting leverages prior information to construct a basis, while PDE-based inpainting relies solely on the observed state points updated by EnSF. Table \ref{tab:rmse_table} reveals that dictionary learning-based inpainting outperforms PDE-based methods in nearly all 5\% observation settings, except for the $256 \times 256$, 12-hour assimilation case—the most challenging scenario in both linear and nonlinear settings. When observation data is scarce, PDE-based methods lack sufficient ``pixels'' to fill the unobserved state, whereas dictionary learning-based methods utilize prior basis information effectively. However, in extreme cases, the prior basis may become less accurate, making the limited observational data more reliable. Even in such cases, the performance gap between PDE-based and dictionary learning-based inpainting remains small.

Examining the total RMSE curves in Figures \ref{fig:total rmse arctan 64 12 5per}, \ref{fig:total rmse arctan 256 12 5per}, \ref{fig:total rmse arctan 64 3 25per},   and \ref{fig:total rmse arctan 256 3 25per}, along with Table \ref{tab:rmse_table}, we observe that increasing observation coverage from 5\% to 25\% consistently improves the performance of PDE-based inpainting. When sufficient ``pixels" are available, PDE-based methods effectively fill the unobserved state. Among the two PDE-based methods, the biharmonic equation-based method consistently outperforms the Navier-Stokes equation-based method. However, since this study focuses on the SQG model, we refrain from concluding that the Navier-Stokes-based method is inferior—it may perform better for other dynamic systems.

\section{Conclusion}\label{sec:con}
This work has demonstrated the successful integration of image inpainting techniques into ensemble score filtering for data assimilation with partial observations. The proposed framework effectively addresses three major challenges in data assimilation: high dimensionality, nonlinear observations, and partial observations. Our comprehensive experimental results across 16 different scenarios show that EnSF with inpainting consistently outperforms traditional LETKF, particularly in handling nonlinear observations and sparse data scenarios. The dictionary learning-based inpainting approach shows particular promise in scenarios with highly sparse observations (5\% coverage), while PDE-based methods excel when more observational data is available (25\% coverage).

Looking ahead, several promising research directions emerge from this work. First, the current dictionary learning approach could be enhanced by incorporating physical constraints and conservation laws into the learning process, potentially improving the accuracy of state reconstruction in unobserved regions. Second, the framework could be extended to handle time-varying observation operators and adaptive observation networks, which are common in operational weather forecasting. Third, investigating hybrid approaches that dynamically select between PDE-based and dictionary learning-based inpainting methods based on local observation density and flow characteristics could optimize performance across different scenarios.
A particularly important direction for future research is the application of this framework to more complex geophysical systems beyond the SQG model, such as primitive equation models and coupled atmosphere-ocean systems. Additionally, exploring the potential of modern deep learning-based inpainting techniques, while maintaining the training-free advantage of the current approach, could further improve reconstruction accuracy in challenging scenarios. Finally, developing efficient parallel implementations of the proposed methods, particularly for the dictionary learning component, would be crucial for operational applications in weather forecasting and climate prediction.
These future developments would further advance the field of data assimilation, potentially leading to improved weather forecasts and climate predictions, especially in regions where observations are sparse or when dealing with nonlinear measurement systems.

\section*{Acknowledgments}
This material is based upon work supported by the U.S. Department of Energy, Office of Science, Office of Advanced Scientific Computing Research, Applied Mathematics program under the contract ERKJ443 at the Oak Ridge National Laboratory, which is operated by UT-Battelle, LLC, for the U.S. Department of Energy under Contract DE-AC05-00OR22725. The third author (FB) would also like to acknowledge the support from U.S. National Science Foundation through project DMS-2142672 and the support from the U.S. Department of Energy, Office of Science, Office of Advanced Scientific Computing Research, Applied Mathematics program under Grant DE-SC0025412.

\newpage

\newpage
\appendix

\section{Additional ablation study results}\label{sec:app}
This section provides additional ablation study results for the cases (in Table \ref{table:scenario}) that are not included in Section \ref{sec:ex_ablation}. The these results provide additional evidence of the superior performance of the proposed EnSF with inpainting techniques. 
\begin{figure}[h!]
    \centering
    \begin{subfigure}[t]{0.8\textwidth}
        \centering
        \includegraphics[width=\textwidth]{ separate_legend.png}
    \end{subfigure} 
    \begin{subfigure}[t]{0.25\textwidth}
        \centering
        \includegraphics[width=\textwidth]{  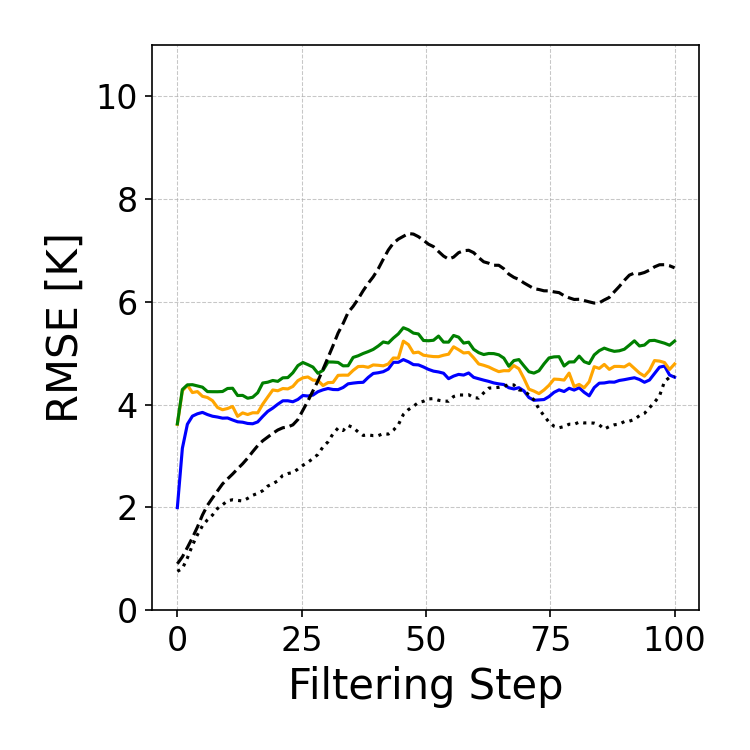}
        \caption{Total RMSE}
    \end{subfigure}
    \begin{subfigure}[t]{0.25\textwidth}
        \centering
        \includegraphics[width=\textwidth]{  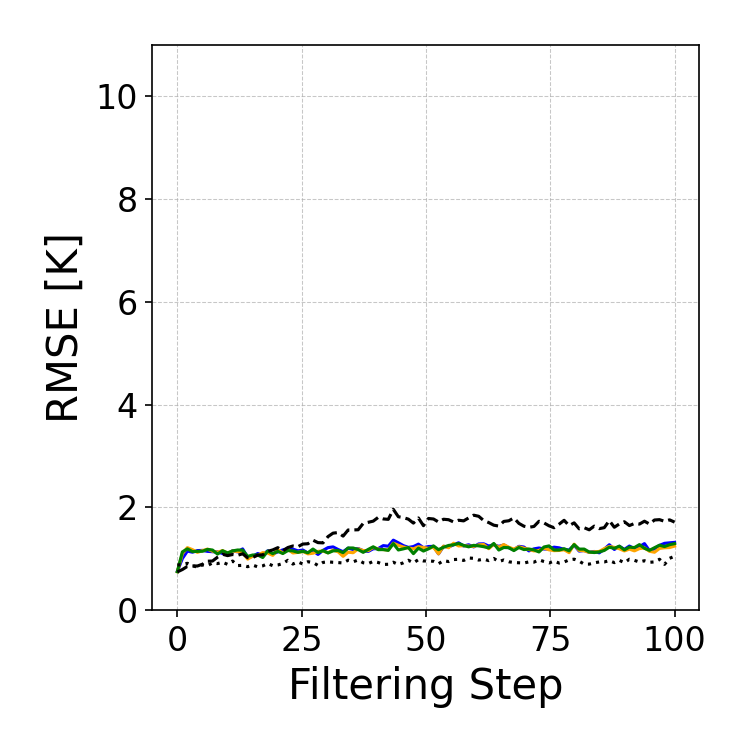}
        \caption{observed RMSE}
    \end{subfigure}
    \begin{subfigure}[t]{0.25\textwidth}
        \centering
        \includegraphics[width=\textwidth]{  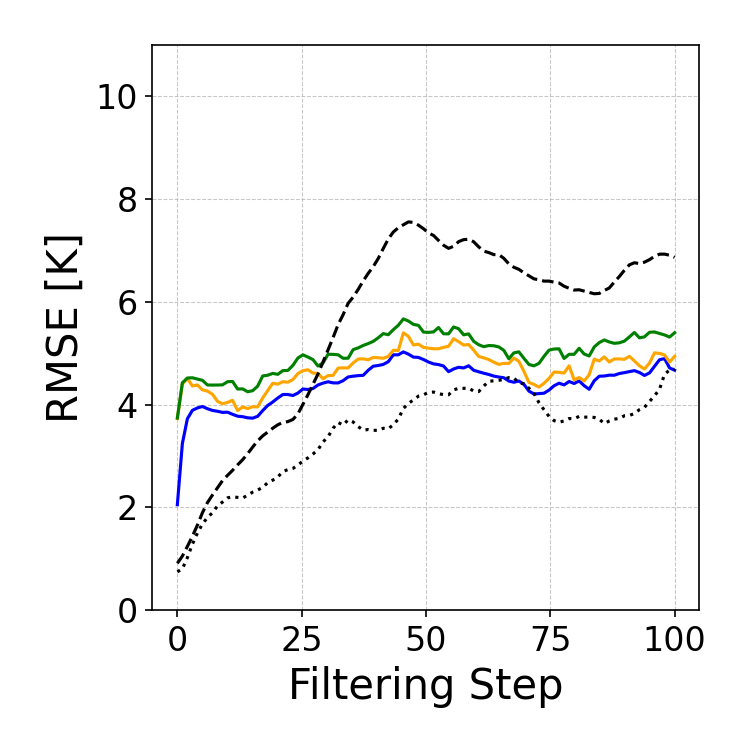}
        \caption{unobserved RMSE}
    \end{subfigure}
    \vspace{-0.5cm}
    \caption{RMSE of $(\mathbf{C}_1)$}
    \label{fig:rmse linear 64 3hourly 5}
    \vspace{-0.4cm}
\end{figure}
\begin{figure}[h!]
    \centering
    \begin{subfigure}[t]{0.2\textwidth}
        \centering
        \includegraphics[width=\textwidth]{  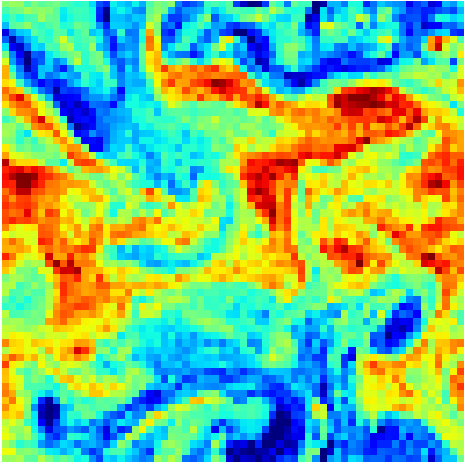}
        \caption{Truth}
    \end{subfigure}
    \begin{subfigure}[t]{0.2\textwidth}
        \centering
        \includegraphics[width=\textwidth]{  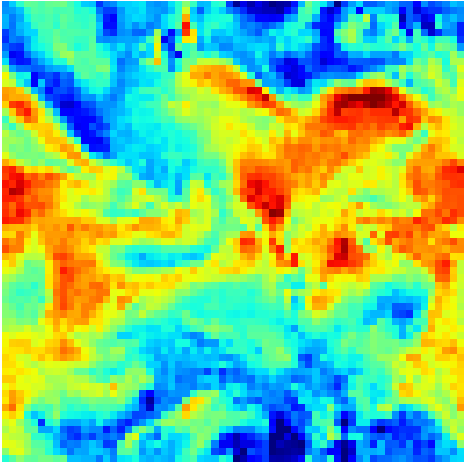}
        \caption{LETKF}
    \end{subfigure}
    \begin{subfigure}[t]{0.2\textwidth}
        \centering
        \includegraphics[width=\textwidth]{  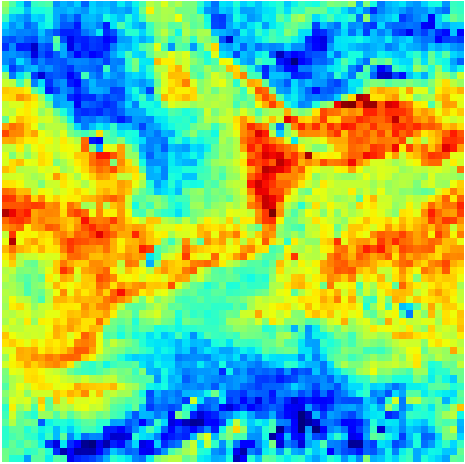}
        \caption{EnSF Only}
    \end{subfigure}\\
    \begin{subfigure}[t]{0.2\textwidth}
        \centering
        \includegraphics[width=\textwidth]{  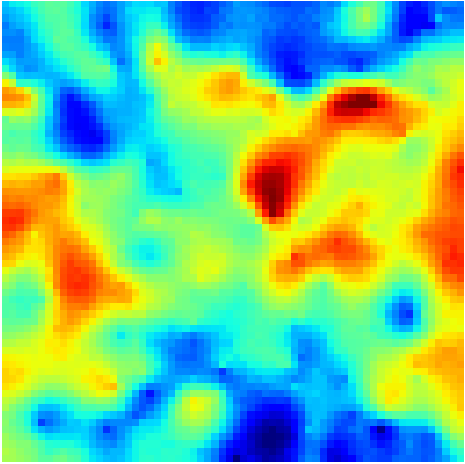}
        \caption{EnSF+Bi}
    \end{subfigure}
    \begin{subfigure}[t]{0.2\textwidth}
        \centering
        \includegraphics[width=\textwidth]{  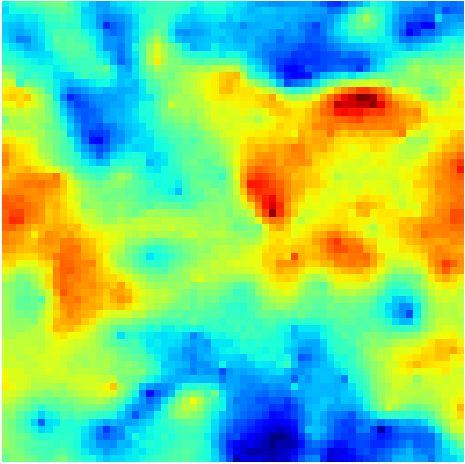}
        \caption{EnSF+DL}
    \end{subfigure}
    \begin{subfigure}[t]{0.2\textwidth}
        \centering
        \includegraphics[width=\textwidth]{  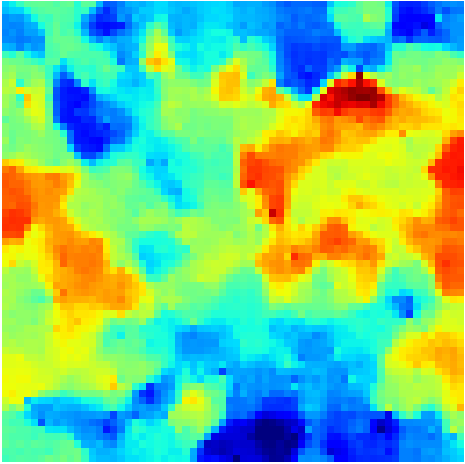}
        \caption{EnSF+NS}
    \end{subfigure}
    \vspace{-0.4cm}
    \caption{Snapshot at filtering step 100 of $(\mathbf{C}_{1})$}
    \label{snap:linear_N64_3hrly_5per}
    \vspace{-0.4cm}
\end{figure}
\begin{figure}[h!]
    \centering
    \begin{subfigure}[h]{0.8\textwidth}
        \centering
        \includegraphics[width=\textwidth]{ separate_legend.png}
    \end{subfigure} 
    \begin{subfigure}[h]{0.25\textwidth}
        \centering
        \includegraphics[width=\textwidth]{ 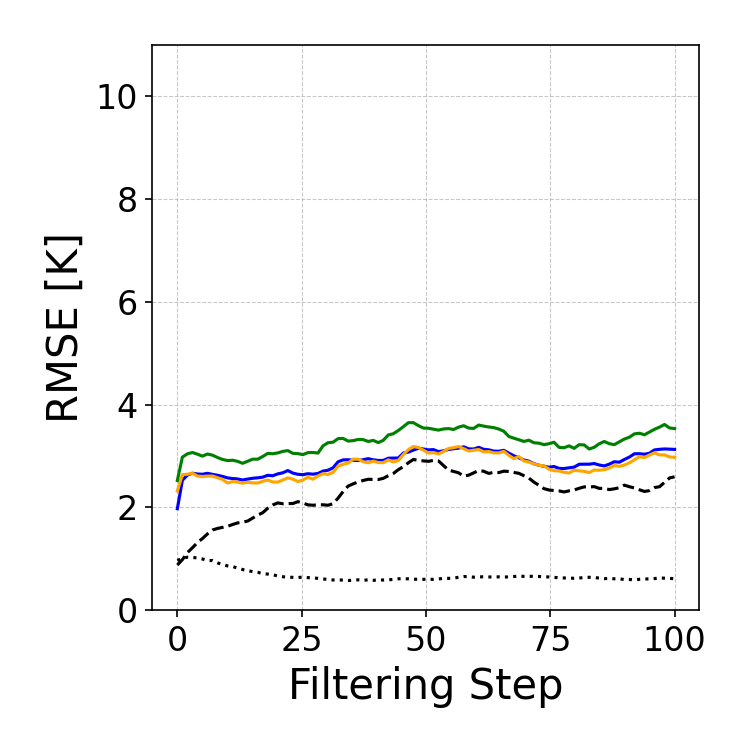}
        \caption{Total RMSE}
    \end{subfigure}
    \begin{subfigure}[h]{0.25\textwidth}
        \centering
        \includegraphics[width=\textwidth]{ 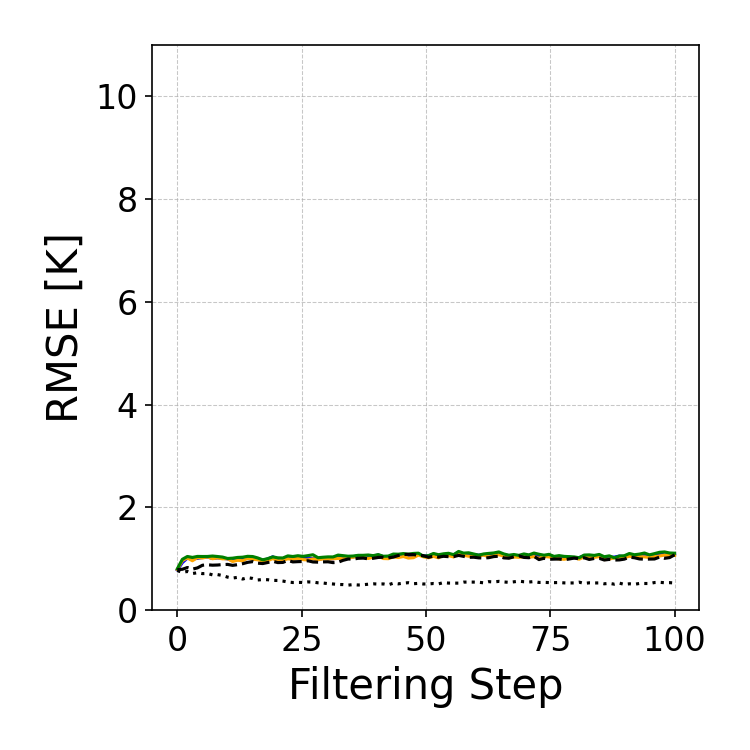}
        \caption{observed RMSE}
    \end{subfigure}
    \begin{subfigure}[h]{0.25\textwidth}
        \centering
        \includegraphics[width=\textwidth]{ 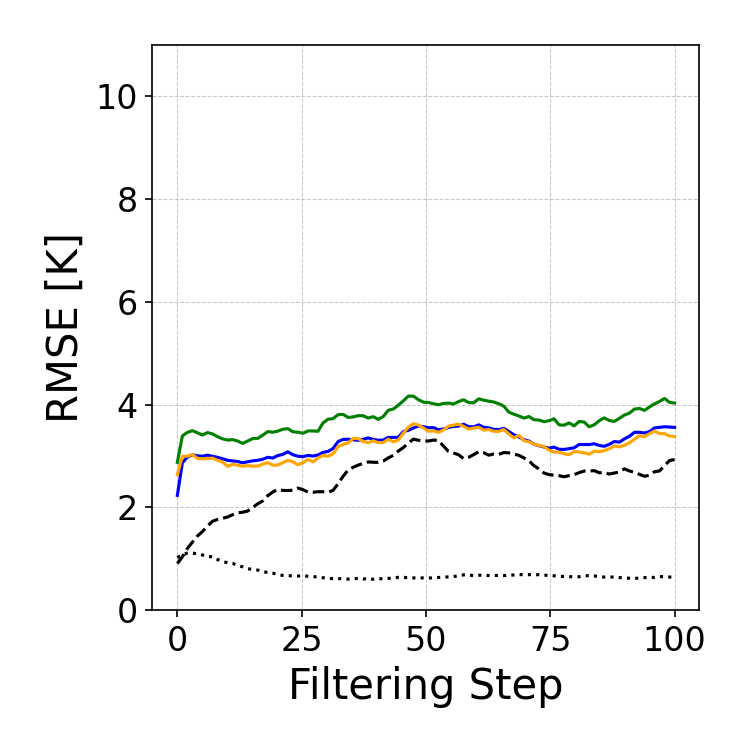}
        \caption{unobserved RMSE}
    \end{subfigure}
    \vspace{-0.5cm}
    \caption{RMSE of $(\mathbf{C}_2)$}
    \label{fig:rmse linear 64 3hourly 25}
    \vspace{-2cm}
\end{figure}
\begin{figure}[h!]
    \centering
    \begin{subfigure}[t]{0.2\textwidth}
        \centering
        \includegraphics[width=\textwidth]{ 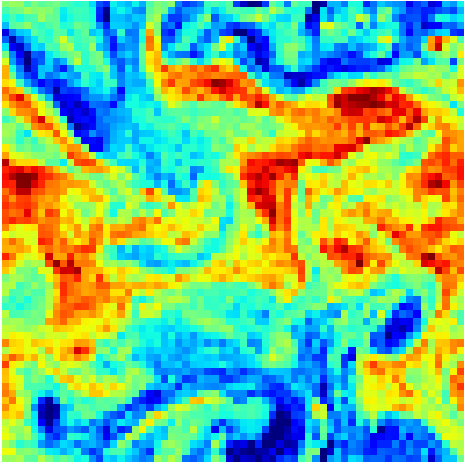}
        \caption{Truth}
    \end{subfigure}
    \begin{subfigure}[t]{0.2\textwidth}
        \centering
        \includegraphics[width=\textwidth]{ 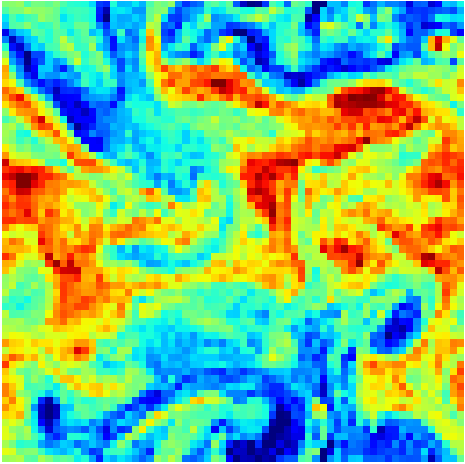}
        \caption{LETKF}
    \end{subfigure}
    \begin{subfigure}[t]{0.2\textwidth}
        \centering
        \includegraphics[width=\textwidth]{ 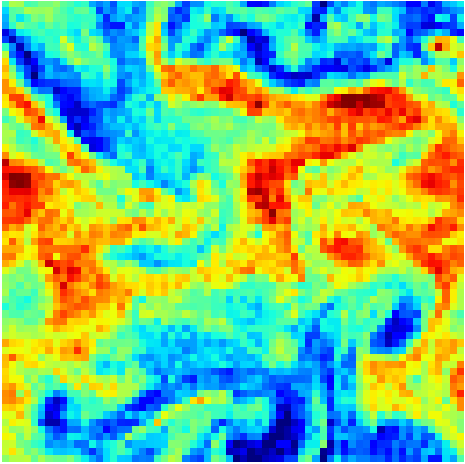}
        \caption{EnSF Only}
    \end{subfigure}\\
    \begin{subfigure}[t]{0.2\textwidth}
        \centering
        \includegraphics[width=\textwidth]{ 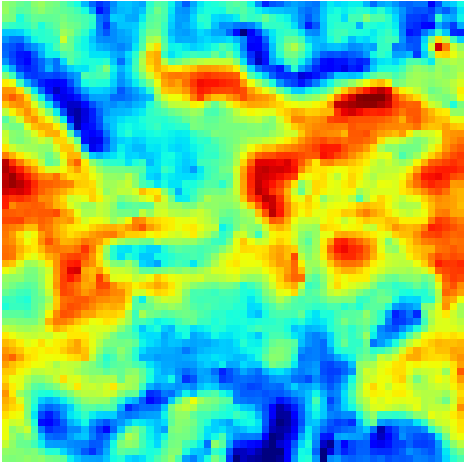}
        \caption{EnSF+Bi}
    \end{subfigure}
    \begin{subfigure}[t]{0.2\textwidth}
        \centering
        \includegraphics[width=\textwidth]{ 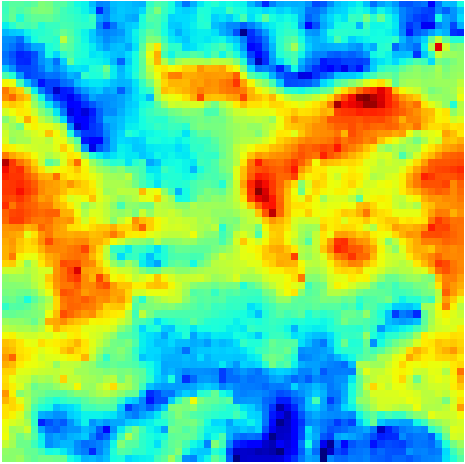}
        \caption{EnSF+DL}
    \end{subfigure}
    \begin{subfigure}[t]{0.2\textwidth}
        \centering
        \includegraphics[width=\textwidth]{ 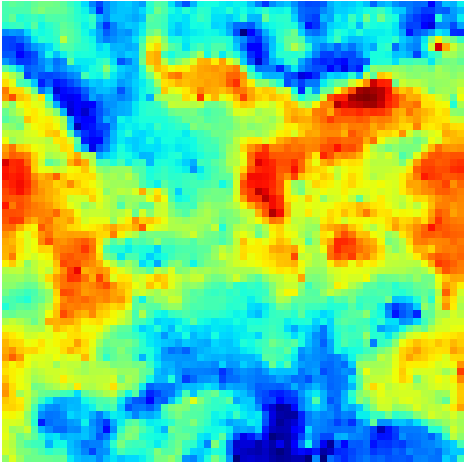}
        \caption{EnSF+NS}
    \end{subfigure}
    \vspace{-0.6cm}
    \caption{Snapshot at filtering step 100 of $(\mathbf{C}_{2})$}
    \label{snap:linear_N64_3hrly_25per}
    \vspace{-0.2cm}
\end{figure}
\begin{figure}[h!]
    \centering
    \begin{subfigure}[t]{0.8\textwidth}
        \centering
        \includegraphics[width=\textwidth]{ separate_legend.png}
    \end{subfigure} 
    \begin{subfigure}[t]{0.25\textwidth}
        \centering
        \includegraphics[width=\textwidth]{  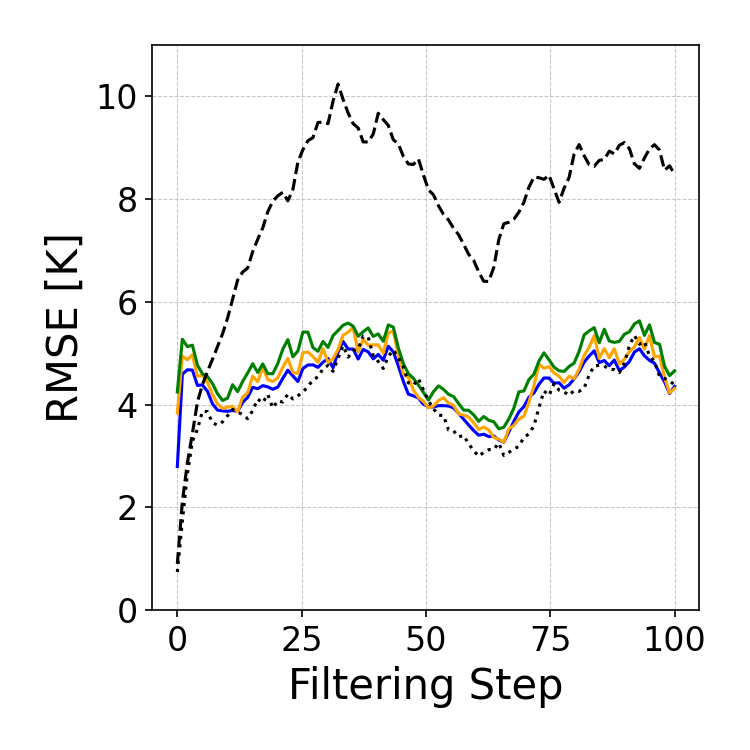}
        \caption{Total RMSE}
    \end{subfigure}
    \begin{subfigure}[t]{0.25\textwidth}
        \centering
        \includegraphics[width=\textwidth]{  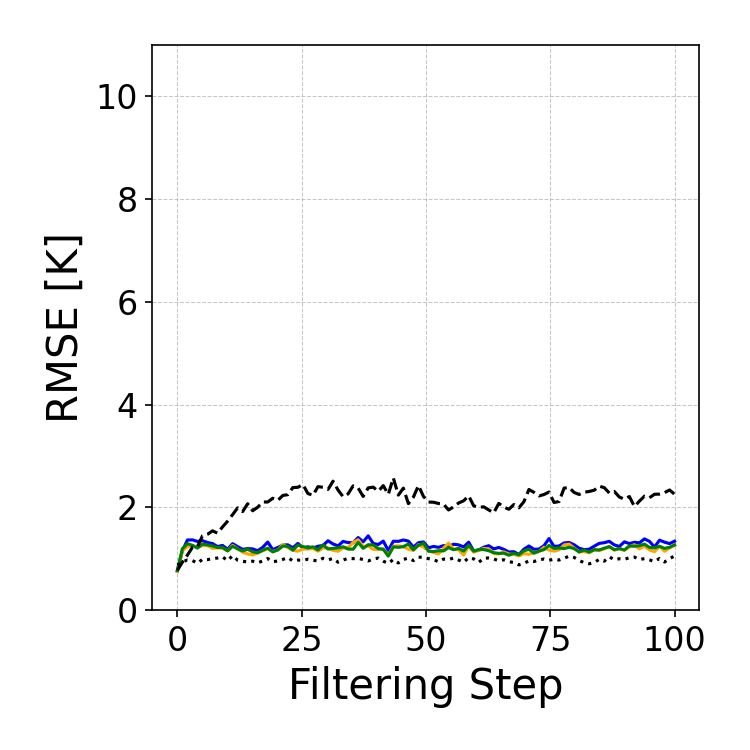}
        \caption{observed RMSE}
    \end{subfigure}
    \begin{subfigure}[t]{0.25\textwidth}
        \centering
        \includegraphics[width=\textwidth]{  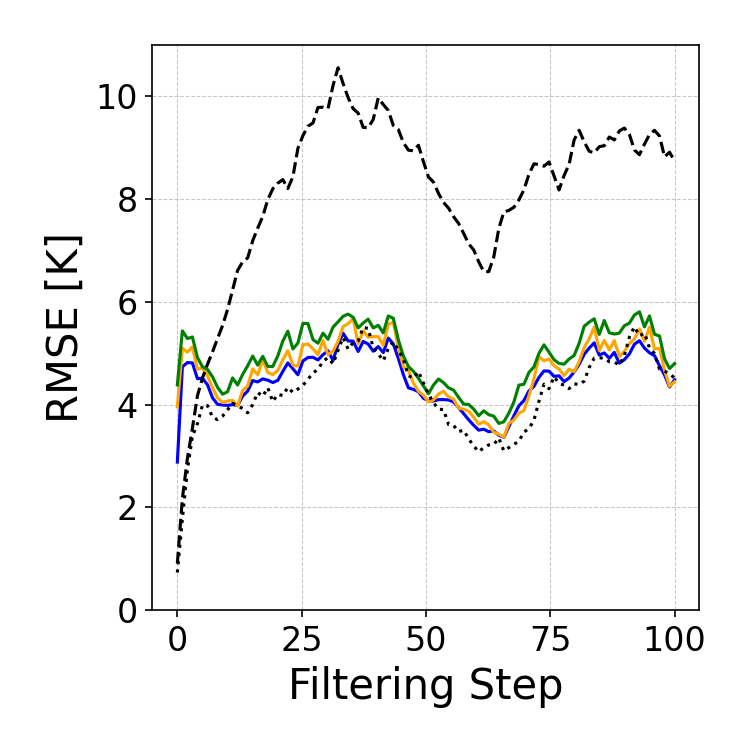}
        \caption{unobserved RMSE}
    \end{subfigure}
    \vspace{-0.5cm}
    \caption{RMSE of $(\mathbf{C}_3)$}
    \label{fig:rmse linear 64 12hourly 5}
    \vspace{-0.3cm}
\end{figure}
\begin{figure}[h!]
    \centering
    \begin{subfigure}[t]{0.2\textwidth}
        \centering
        \includegraphics[width=\textwidth]{  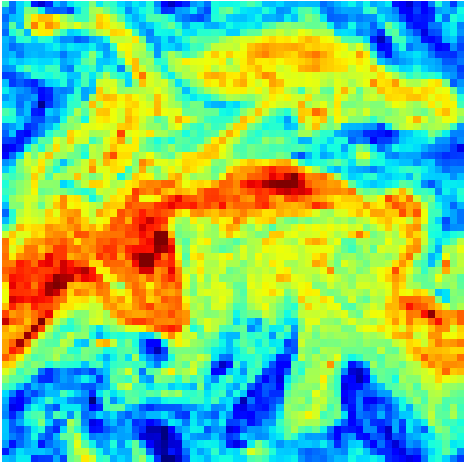}
        \caption{Truth}
    \end{subfigure}
    \begin{subfigure}[t]{0.2\textwidth}
        \centering
        \includegraphics[width=\textwidth]{  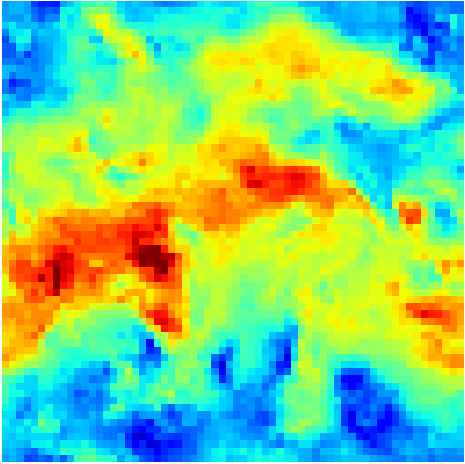}
        \caption{LETKF}
    \end{subfigure}
    \begin{subfigure}[t]{0.2\textwidth}
        \centering
        \includegraphics[width=\textwidth]{  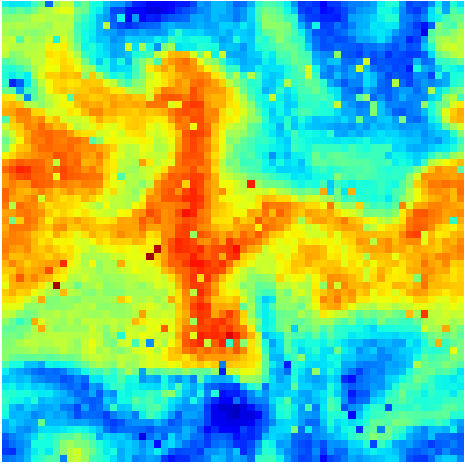}
        \caption{EnSF Only}
    \end{subfigure}\\
    \begin{subfigure}[t]{0.2\textwidth}
        \centering
        \includegraphics[width=\textwidth]{  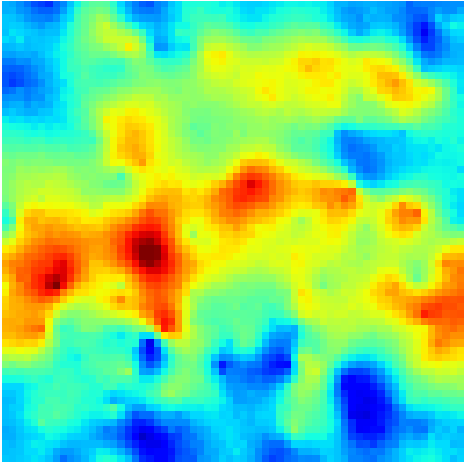}
        \caption{EnSF+Bi}
    \end{subfigure}
    \begin{subfigure}[t]{0.2\textwidth}
        \centering
        \includegraphics[width=\textwidth]{  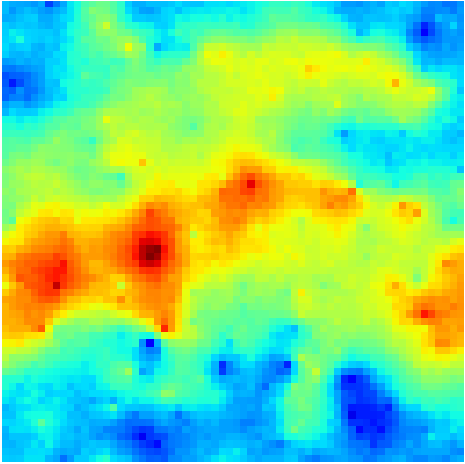}
        \caption{EnSF+DL}
    \end{subfigure}
    \begin{subfigure}[t]{0.2\textwidth}
        \centering
        \includegraphics[width=\textwidth]{  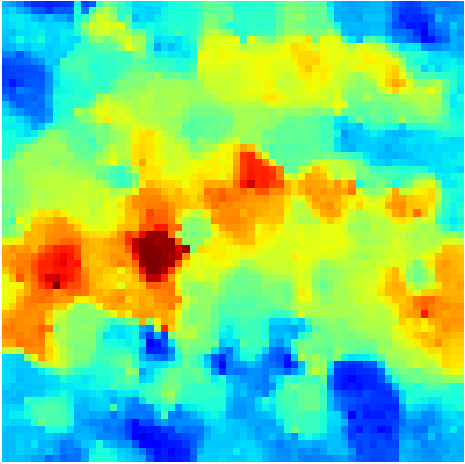}
        \caption{EnSF+NS}
    \end{subfigure}
    \vspace{-0.6cm}
    \caption{Snapshot at filtering step 100 of $(\mathbf{C}_{3})$}
    \label{snap:linear_N64_12hrly_5per}
\end{figure}
\begin{figure}[h!]
    \centering
    \begin{subfigure}[t]{0.8\textwidth}
        \centering
        \includegraphics[width=\textwidth]{ separate_legend.png}
    \end{subfigure} 
    \begin{subfigure}[t]{0.25\textwidth}
        \centering
        \includegraphics[width=\textwidth]{ 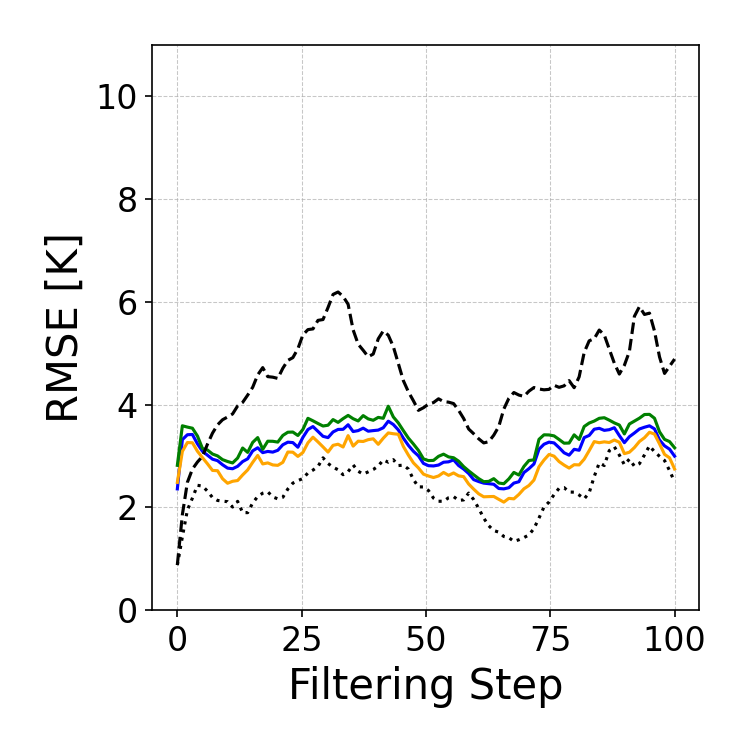}
        \caption{Total RMSE}
    \end{subfigure}
    \begin{subfigure}[t]{0.25\textwidth}
        \centering
        \includegraphics[width=\textwidth]{ 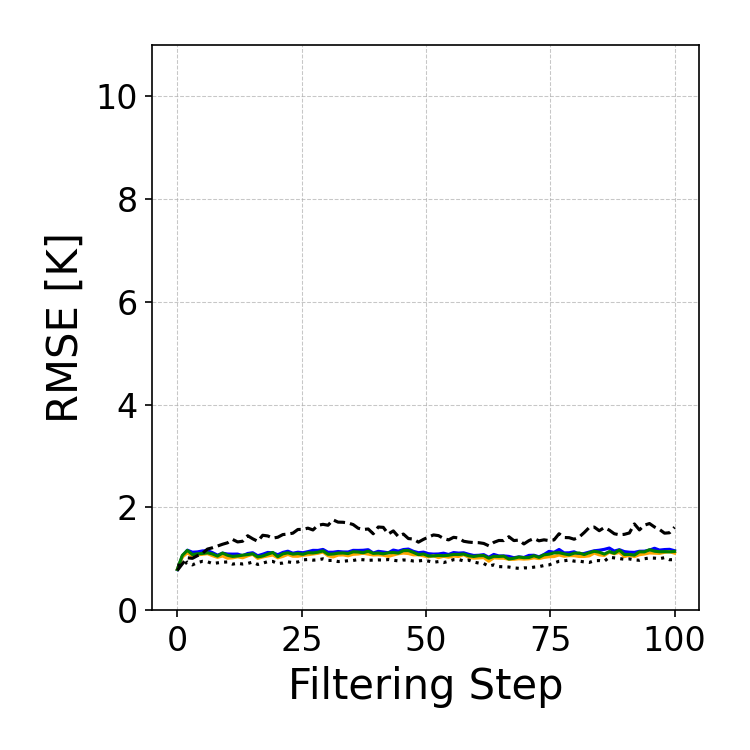}
        \caption{observed RMSE}
    \end{subfigure}
    \begin{subfigure}[t]{0.25\textwidth}
        \centering
        \includegraphics[width=\textwidth]{ 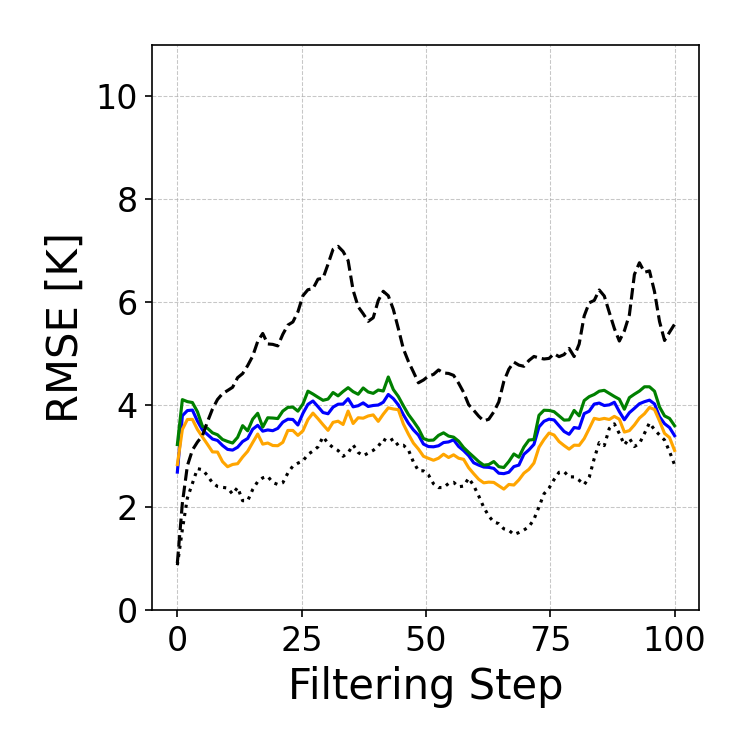}
        \caption{unobserved RMSE}
    \end{subfigure}
    \vspace{-0.5cm}
    \caption{RMSE of $(\mathbf{C}_4)$}
    \label{fig:rmse linear 64 12hourly 25}
    \vspace{-0.3cm}
\end{figure}
\begin{figure}[h!]
    \centering
    \begin{subfigure}[t]{0.2\textwidth}
        \centering
        \includegraphics[width=\textwidth]{ 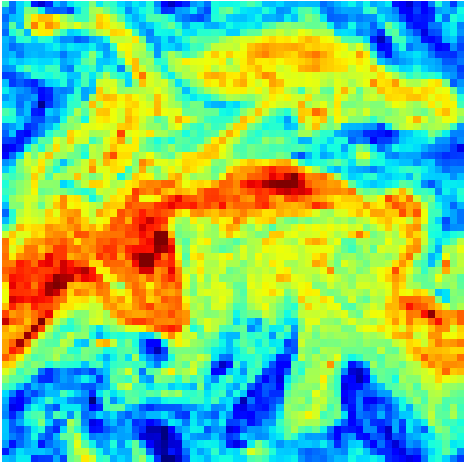}
        \caption{Truth}
    \end{subfigure}
    \begin{subfigure}[t]{0.2\textwidth}
        \centering
        \includegraphics[width=\textwidth]{ 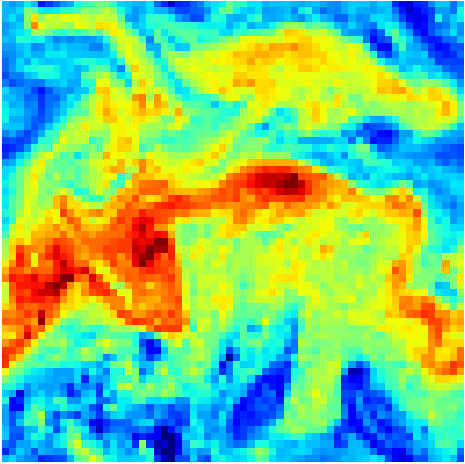}
        \caption{LETKF}
    \end{subfigure}
    \begin{subfigure}[t]{0.2\textwidth}
        \centering
        \includegraphics[width=\textwidth]{ 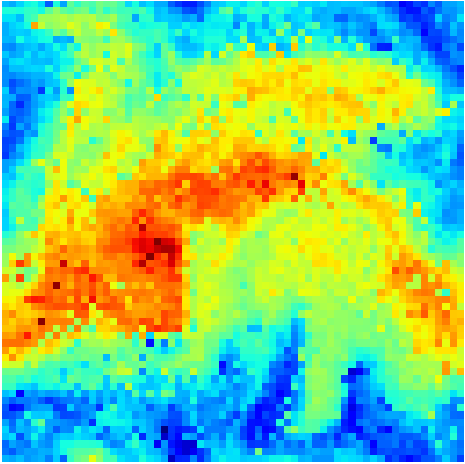}
        \caption{EnSF Only}
    \end{subfigure}\\
    \begin{subfigure}[t]{0.2\textwidth}
        \centering
        \includegraphics[width=\textwidth]{ 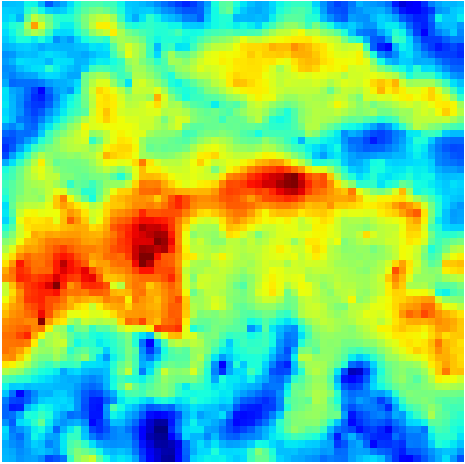}
        \caption{EnSF+Bi}
    \end{subfigure}
    \begin{subfigure}[t]{0.2\textwidth}
        \centering
        \includegraphics[width=\textwidth]{ 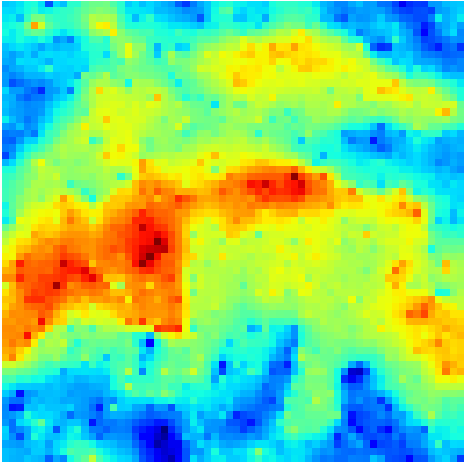}
        \caption{EnSF+DL}
    \end{subfigure}
    \begin{subfigure}[t]{0.2\textwidth}
        \centering
        \includegraphics[width=\textwidth]{ 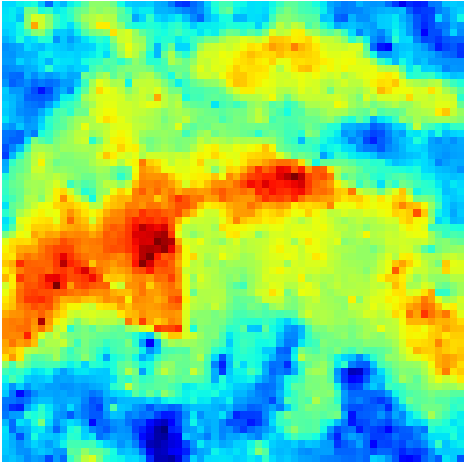}
        \caption{EnSF+NS}
    \end{subfigure}
    \vspace{-0.4cm}
    \caption{Snapshot at filtering step 100 of $(\mathbf{C}_{4})$}
    \label{snap:linear_N64_12hrly_25per}
    \vspace{-0.3cm}
\end{figure}
\begin{figure}[h!]
    \centering
    \begin{subfigure}[t]{0.8\textwidth}
        \centering
        \includegraphics[width=\textwidth]{ separate_legend.png}
    \end{subfigure} 
    \begin{subfigure}[t]{0.25\textwidth}
        \centering
        \includegraphics[width=\textwidth]{  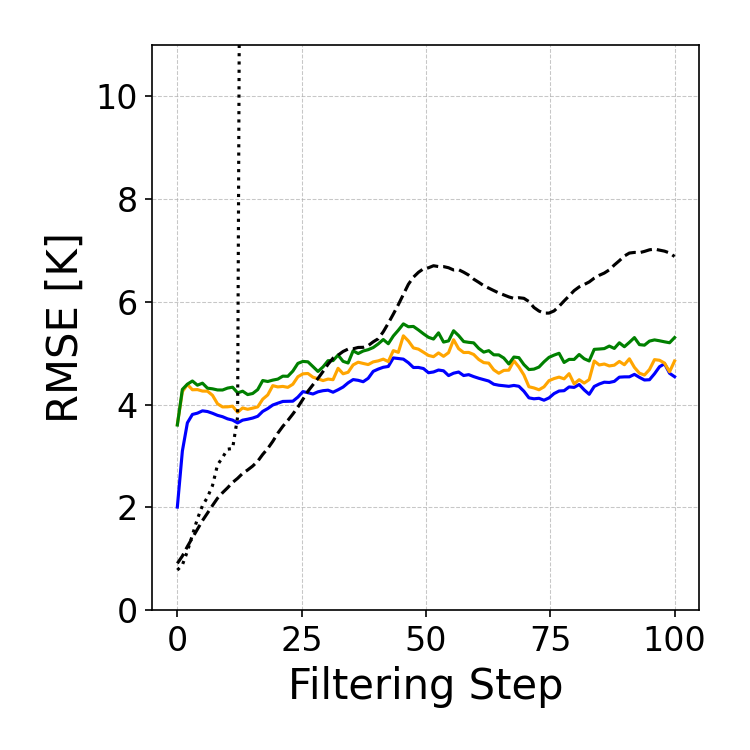}
        \caption{Total RMSE}
    \end{subfigure}
    \begin{subfigure}[t]{0.25\textwidth}
        \centering
        \includegraphics[width=\textwidth]{  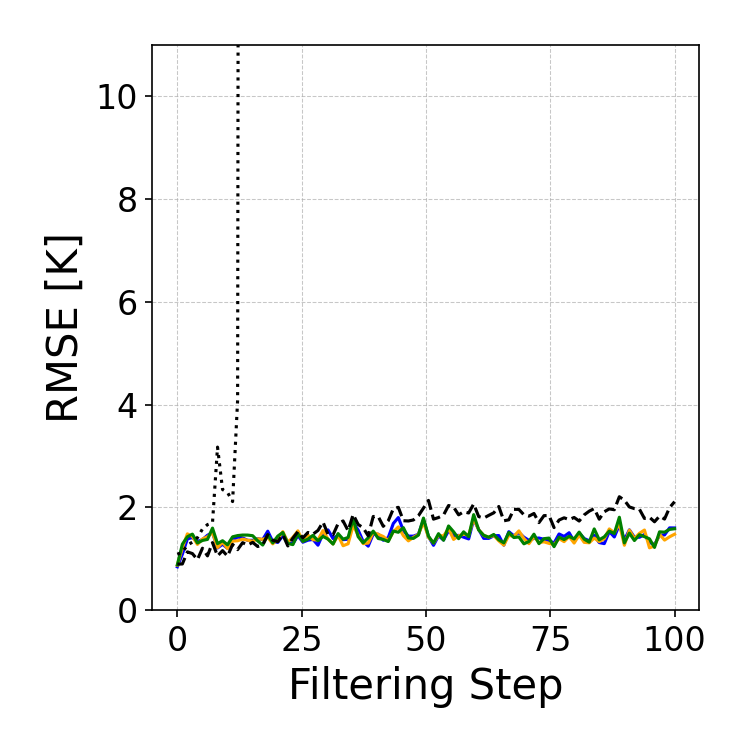}
        \caption{observed RMSE}
    \end{subfigure}
    \begin{subfigure}[t]{0.25\textwidth}
        \centering
        \includegraphics[width=\textwidth]{  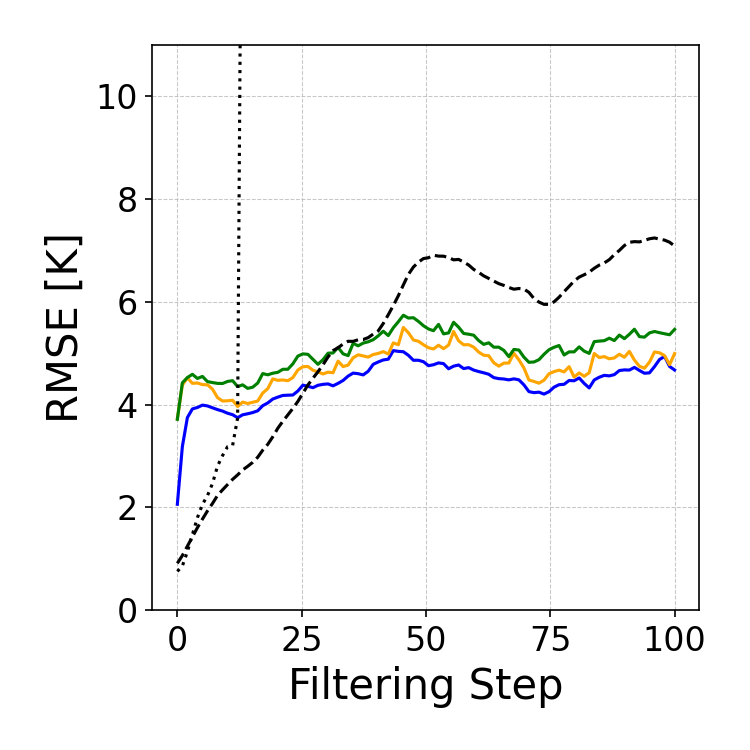}
        \caption{unobserved RMSE}
    \end{subfigure}
    \vspace{-0.5cm}
    \caption{RMSE of $(\mathbf{C}_5)$}
    \label{fig:rmse arctan 64 3hourly 5}
\end{figure}
\begin{figure}[h!]
    \centering
    \begin{subfigure}[t]{0.2\textwidth}
        \centering
        \includegraphics[width=\textwidth]{  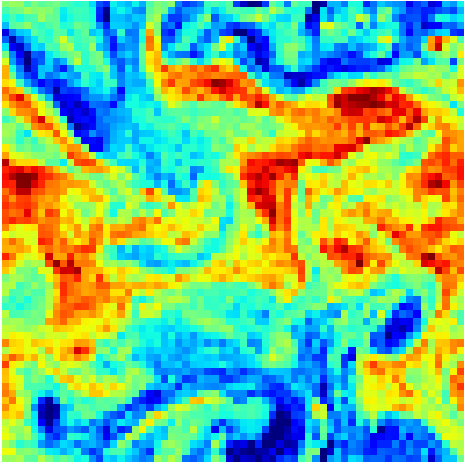}
        \caption{Truth}
    \end{subfigure}
    \begin{subfigure}[t]{0.2\textwidth}
        \centering
        \includegraphics[width=\textwidth]{  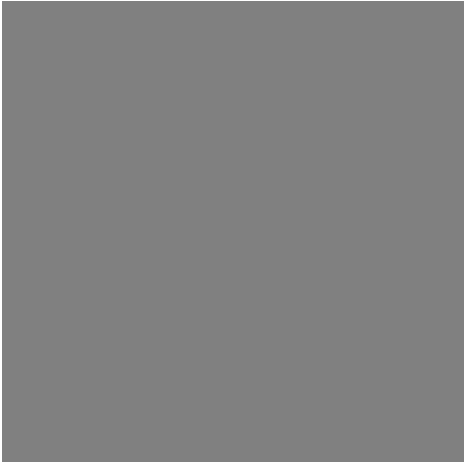}
        \caption{LETKF}
    \end{subfigure}
    \begin{subfigure}[t]{0.2\textwidth}
        \centering
        \includegraphics[width=\textwidth]{  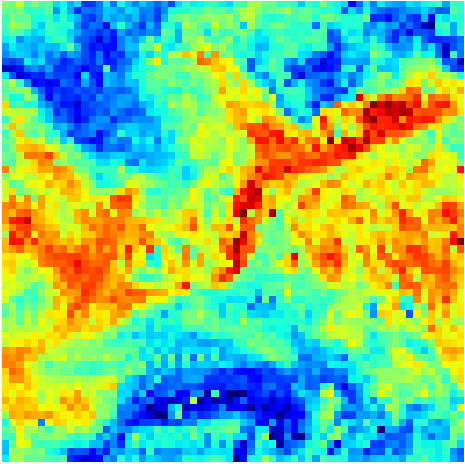}
        \caption{EnSF Only}
    \end{subfigure}\\
    \begin{subfigure}[t]{0.2\textwidth}
        \centering
        \includegraphics[width=\textwidth]{  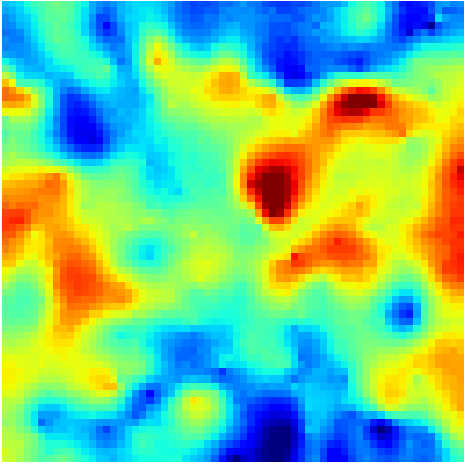}
        \caption{EnSF+Bi}
    \end{subfigure}
    \begin{subfigure}[t]{0.2\textwidth}
        \centering
        \includegraphics[width=\textwidth]{  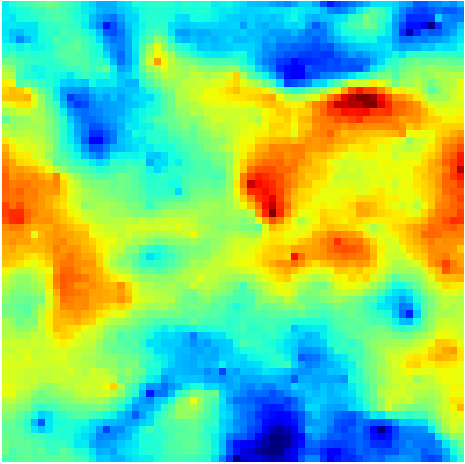}
        \caption{EnSF+DL}
    \end{subfigure}
    \begin{subfigure}[t]{0.2\textwidth}
        \centering
        \includegraphics[width=\textwidth]{  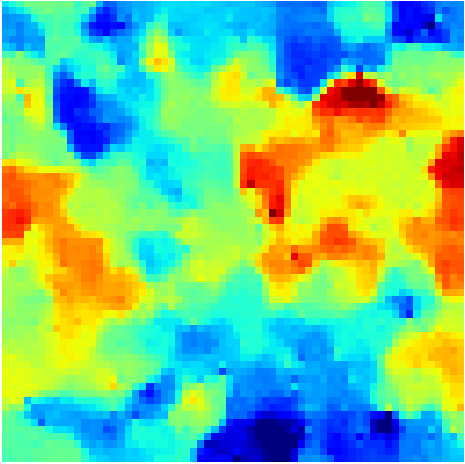}
        \caption{EnSF+NS}
    \end{subfigure}
    \vspace{-0.4cm}
    \caption{Snapshot at filtering step 100 of $(\mathbf{C}_{5})$}
    \label{snap:Arctan_N64_3hrly_5per}
    \vspace{-0.4cm}
\end{figure}
\begin{figure}[h!]
    \centering
    \begin{subfigure}[t]{0.8\textwidth}
        \centering
        \includegraphics[width=\textwidth]{ separate_legend.png}
    \end{subfigure} 
    \begin{subfigure}[t]{0.25\textwidth}
        \centering
        \includegraphics[width=\textwidth]{  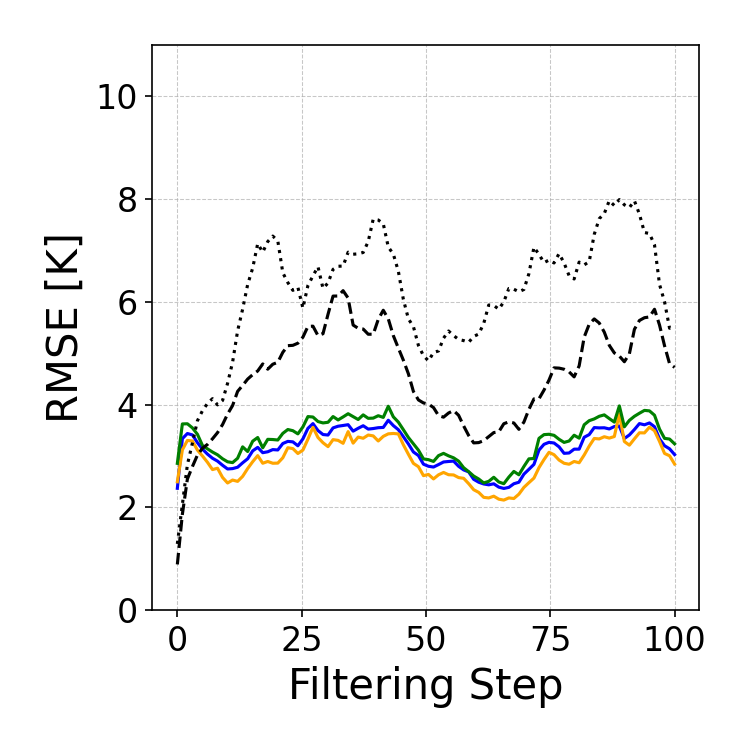}
        \caption{Total RMSE}
    \end{subfigure}
    \begin{subfigure}[t]{0.25\textwidth}
        \centering
        \includegraphics[width=\textwidth]{  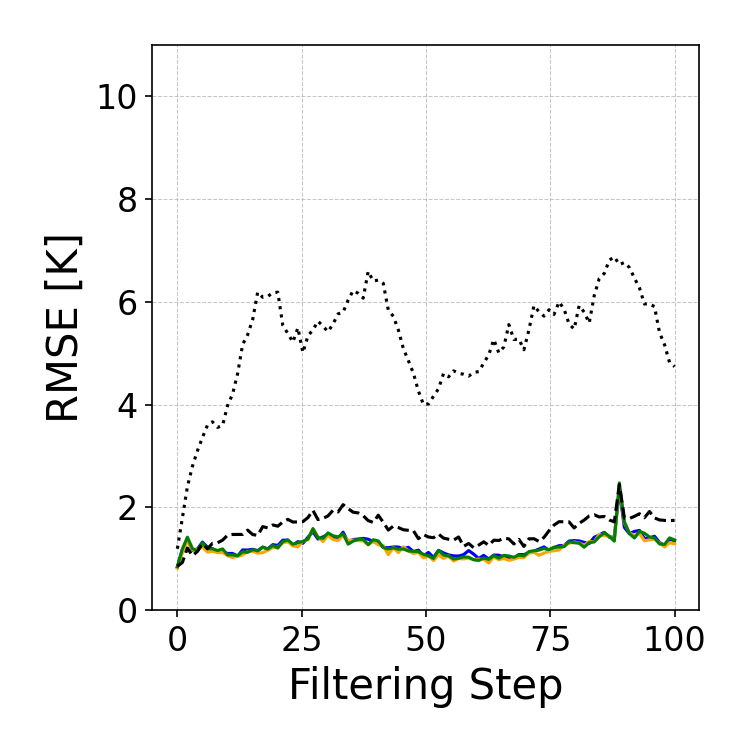}
        \caption{observed RMSE}
    \end{subfigure}
    \begin{subfigure}[t]{0.25\textwidth}
        \centering
        \includegraphics[width=\textwidth]{  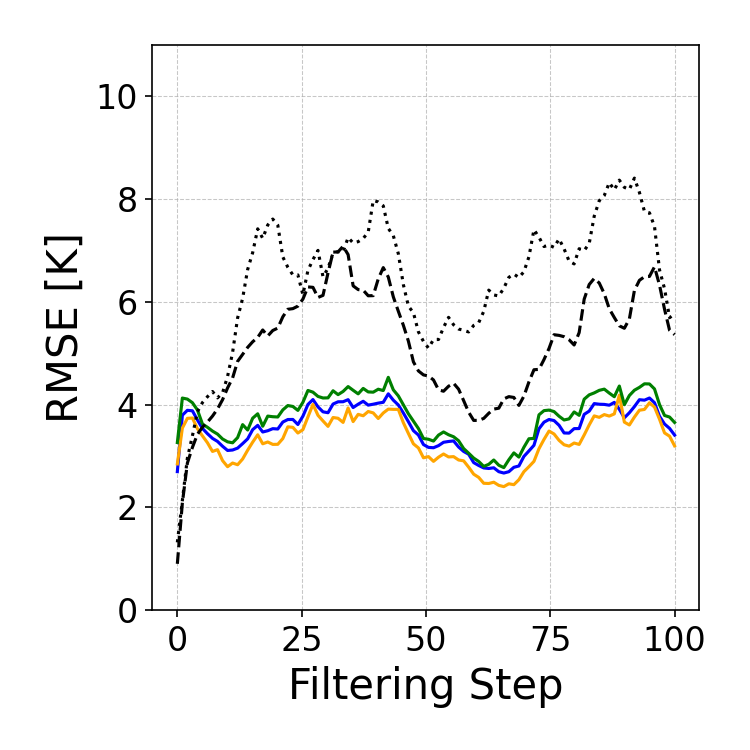}
        \caption{unobserved RMSE}
    \end{subfigure}
    \vspace{-0.5cm}
    \caption{RMSE of $(\mathbf{C}_8)$}
    \label{fig:rmse arctan 64 12hourly 25}
    \vspace{-0.5cm}
\end{figure}
\begin{figure}[h!]
    \centering
    \begin{subfigure}[t]{0.2\textwidth}
        \centering
        \includegraphics[width=\textwidth]{  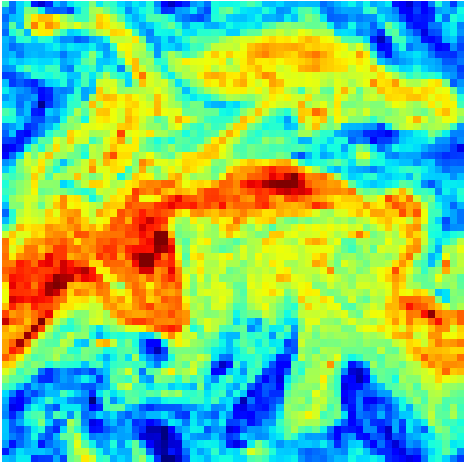}
        \caption{Truth}
    \end{subfigure}
    \begin{subfigure}[t]{0.2\textwidth}
        \centering
        \includegraphics[width=\textwidth]{  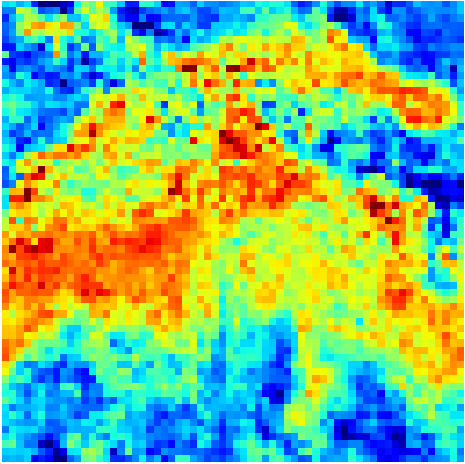}
        \caption{LETKF}
    \end{subfigure}
    \begin{subfigure}[t]{0.2\textwidth}
        \centering
        \includegraphics[width=\textwidth]{  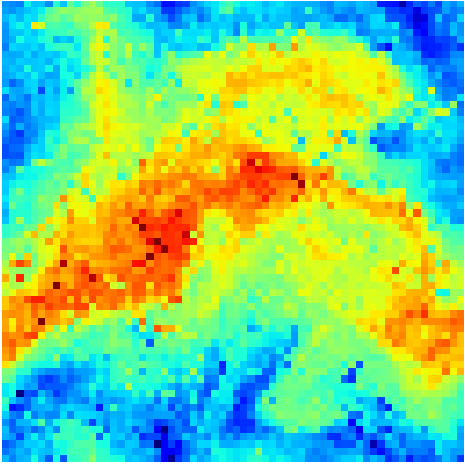}
        \caption{EnSF Only}
    \end{subfigure}\\
    \begin{subfigure}[t]{0.2\textwidth}
        \centering
        \includegraphics[width=\textwidth]{  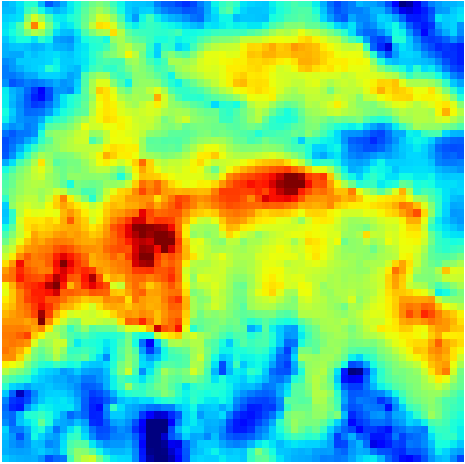}
        \caption{EnSF+Bi}
    \end{subfigure}
    \begin{subfigure}[t]{0.2\textwidth}
        \centering
        \includegraphics[width=\textwidth]{  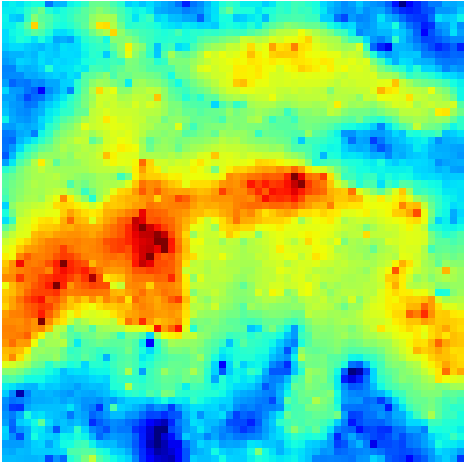}
        \caption{EnSF+DL}
    \end{subfigure}
    \begin{subfigure}[t]{0.2\textwidth}
        \centering
        \includegraphics[width=\textwidth]{  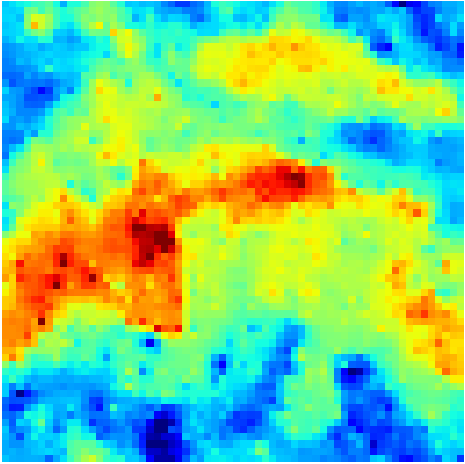}
        \caption{EnSF+NS}
    \end{subfigure}
    \vspace{-0.4cm}
    \caption{Snapshot at filtering step 100 of $(\mathbf{C}_{8})$}
    \label{snap:Arctan_N64_12hrly_25per}
    \vspace{-0.4cm}
\end{figure}
\begin{figure}[h!]
    \centering
    \begin{subfigure}[t]{0.8\textwidth}
        \centering
        \includegraphics[width=\textwidth]{ separate_legend.png}
    \end{subfigure} 
    \begin{subfigure}[t]{0.25\textwidth}
        \centering
        \includegraphics[width=\textwidth]{  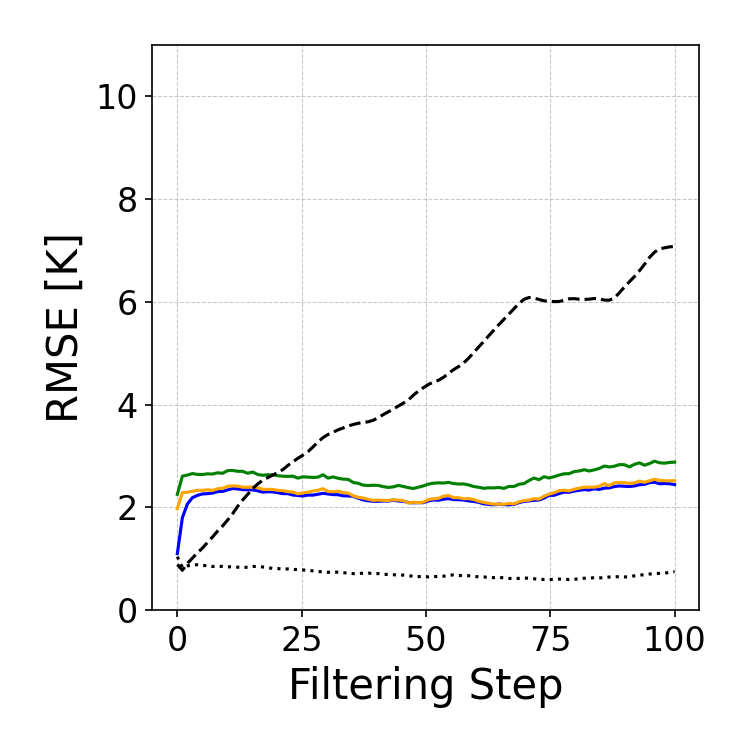}
        \caption{Total RMSE}
    \end{subfigure}
    \begin{subfigure}[t]{0.25\textwidth}
        \centering
        \includegraphics[width=\textwidth]{  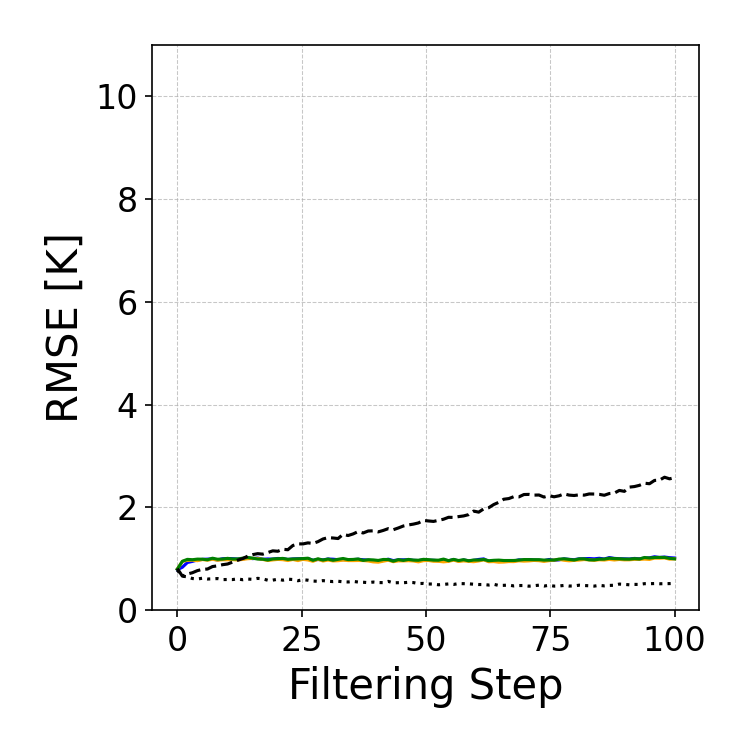}
        \caption{observed RMSE}
    \end{subfigure}
    \begin{subfigure}[t]{0.25\textwidth}
        \centering
        \includegraphics[width=\textwidth]{  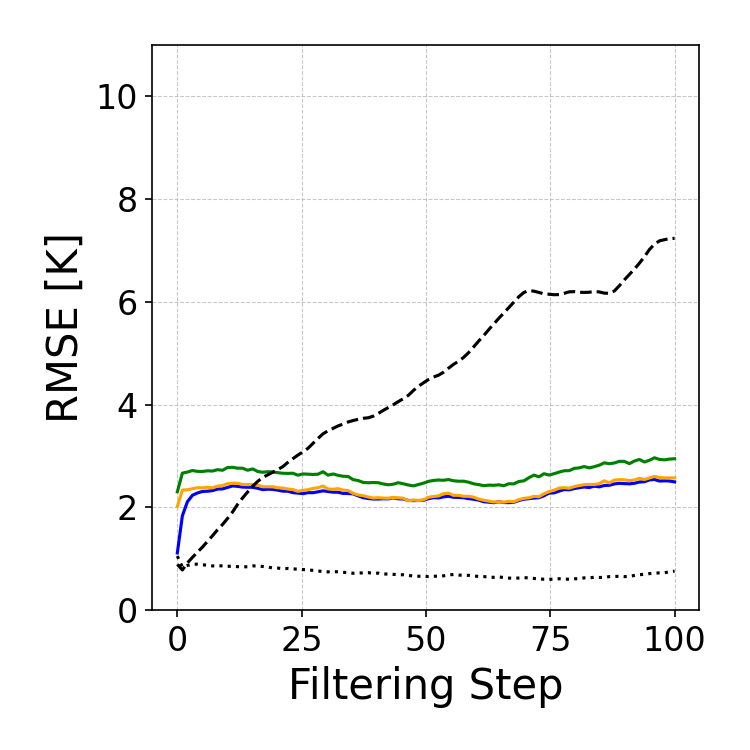}
        \caption{unobserved RMSE}
    \end{subfigure}
    \vspace{-0.5cm}
    \caption{RMSE of $(\mathbf{C}_9)$}
    \label{fig:rmse linear 256 3hourly 5}
    \vspace{-0.3cm}
\end{figure}
\begin{figure}[h!]
    \centering
    \begin{subfigure}[t]{0.2\textwidth}
        \centering
        \includegraphics[width=\textwidth]{  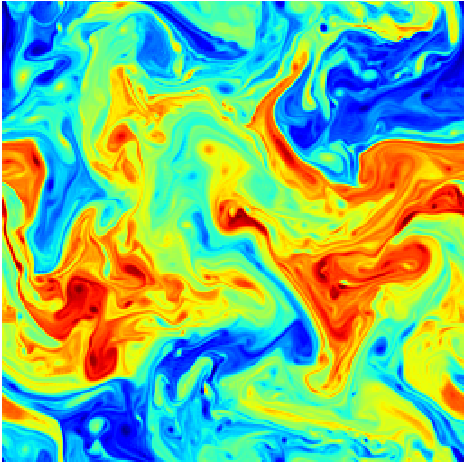}
        \caption{Truth}
    \end{subfigure}
    \begin{subfigure}[t]{0.2\textwidth}
        \centering
        \includegraphics[width=\textwidth]{  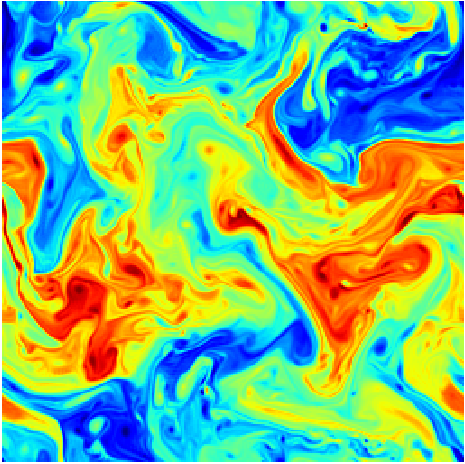}
        \caption{LETKF}
    \end{subfigure}
    \begin{subfigure}[t]{0.2\textwidth}
        \centering
        \includegraphics[width=\textwidth]{  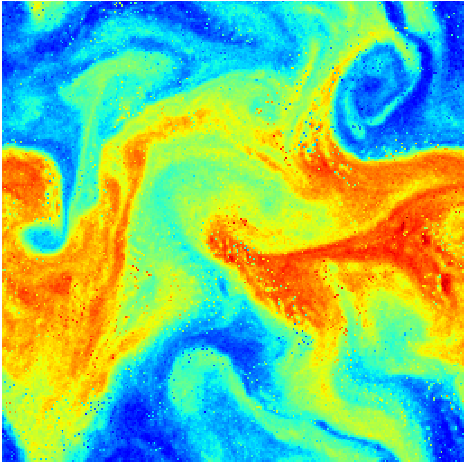}
        \caption{EnSF Only}
    \end{subfigure}\\
    \begin{subfigure}[t]{0.2\textwidth}
        \centering
        \includegraphics[width=\textwidth]{  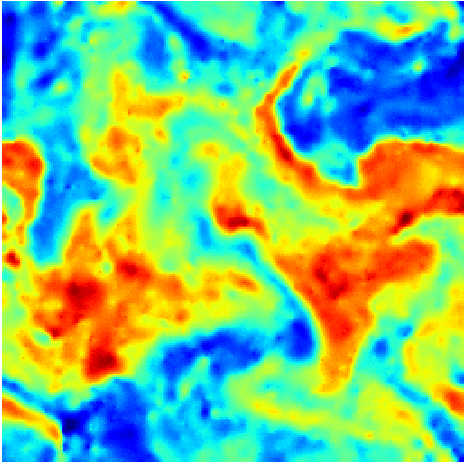}
        \caption{EnSF+Bi}
    \end{subfigure}
    \begin{subfigure}[t]{0.2\textwidth}
        \centering
        \includegraphics[width=\textwidth]{  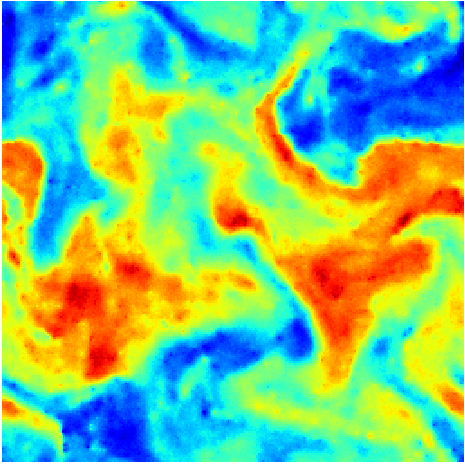}
        \caption{EnSF+DL}
    \end{subfigure}
    \begin{subfigure}[t]{0.2\textwidth}
        \centering
        \includegraphics[width=\textwidth]{  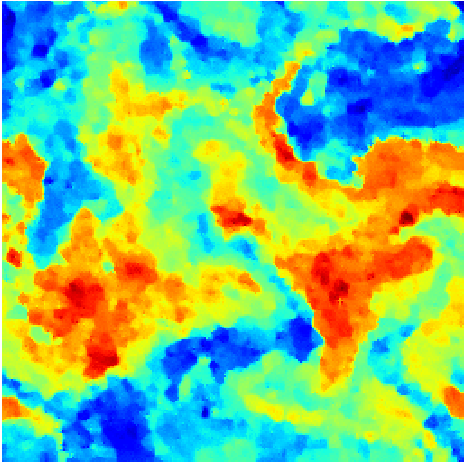}
        \caption{EnSF+NS}
    \end{subfigure}
    \vspace{-0.4cm}
    \caption{Snapshot at filtering step 100 of $(\mathbf{C}_{9})$}
    \label{snap:linear_N256_3hrly_5per}
    \vspace{-0.3cm}
\end{figure}
\begin{figure}[h!]
    \centering
    \begin{subfigure}[t]{0.8\textwidth}
        \centering
        \includegraphics[width=\textwidth]{ separate_legend.png}
    \end{subfigure} 
    \begin{subfigure}[t]{0.25\textwidth}
        \centering
        \includegraphics[width=\textwidth]{  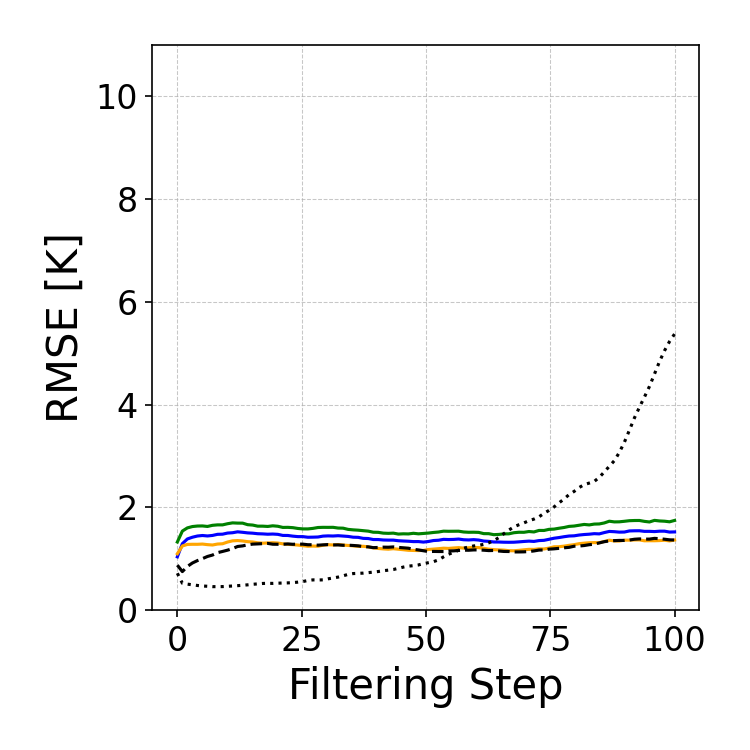}
        \caption{Total RMSE}
    \end{subfigure}
    \begin{subfigure}[t]{0.25\textwidth}
        \centering
        \includegraphics[width=\textwidth]{  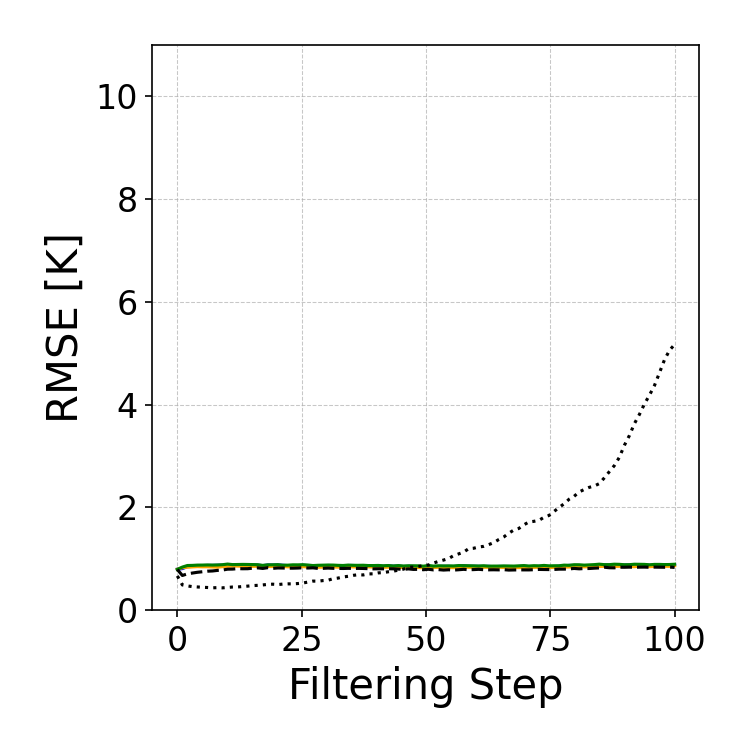}
        \caption{observed RMSE}
    \end{subfigure}
    \begin{subfigure}[t]{0.25\textwidth}
        \centering
        \includegraphics[width=\textwidth]{  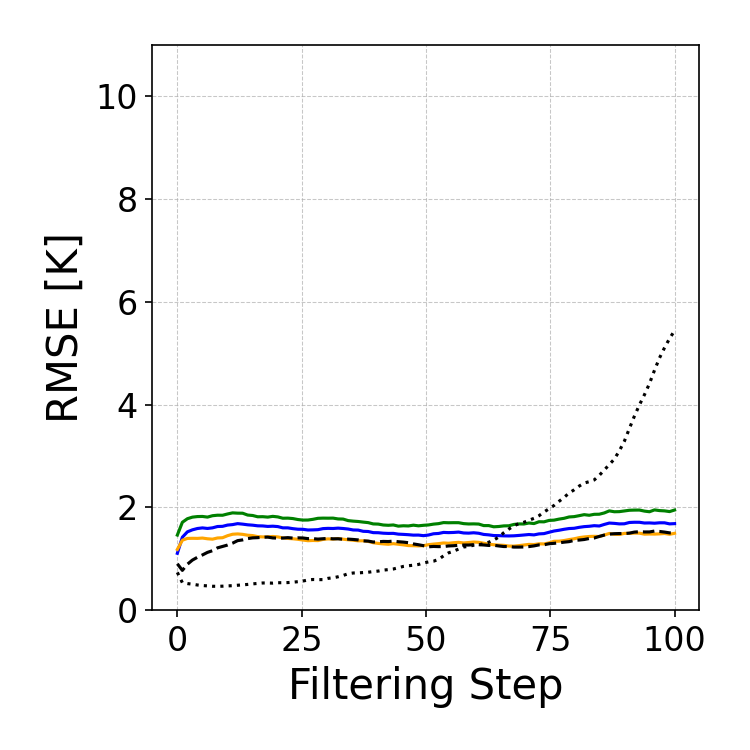}
        \caption{unobserved RMSE}
    \end{subfigure}
    \vspace{-0.5cm}
    \caption{RMSE of $(\mathbf{C}_{10})$}
    \label{fig:rmse linear 256 3hourly 25}
    \vspace{-0.3cm}
\end{figure}
\begin{figure}[h!]
    \centering
    \begin{subfigure}[t]{0.2\textwidth}
        \centering
        \includegraphics[width=\textwidth]{  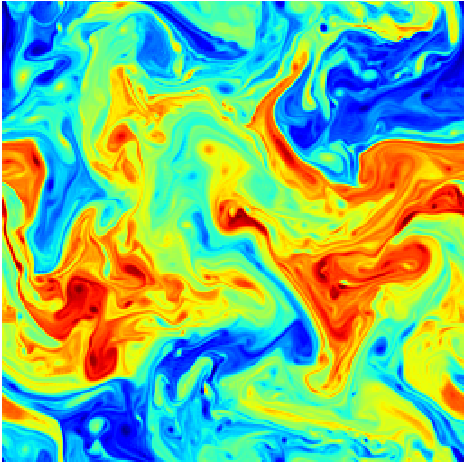}
        \caption{Truth}
    \end{subfigure}
    \begin{subfigure}[t]{0.2\textwidth}
        \centering
        \includegraphics[width=\textwidth]{  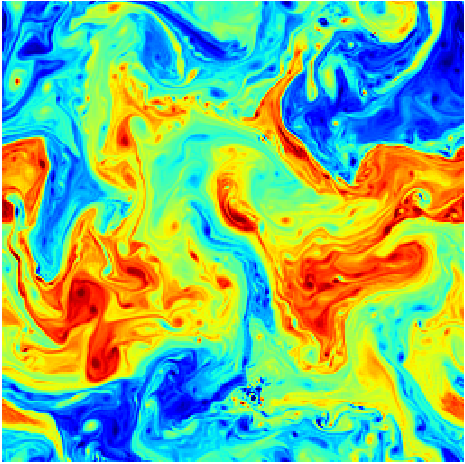}
        \caption{LETKF}
    \end{subfigure}
    \begin{subfigure}[t]{0.2\textwidth}
        \centering
        \includegraphics[width=\textwidth]{  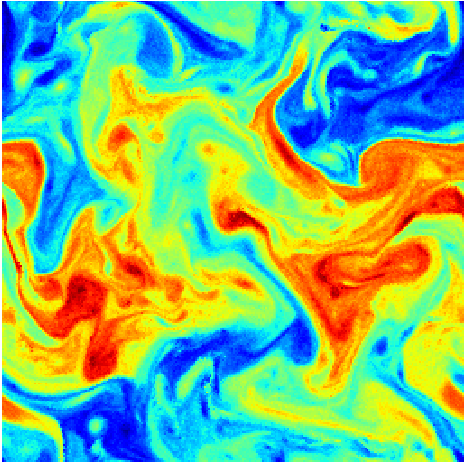}
        \caption{EnSF Only}
    \end{subfigure}\\
    \begin{subfigure}[t]{0.2\textwidth}
        \centering
        \includegraphics[width=\textwidth]{  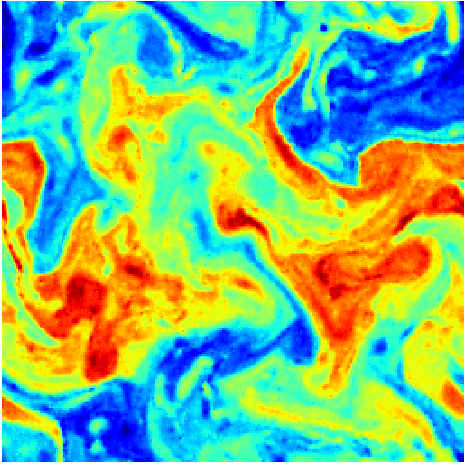}
        \caption{EnSF+Bi}
    \end{subfigure}
    \begin{subfigure}[t]{0.2\textwidth}
        \centering
        \includegraphics[width=\textwidth]{  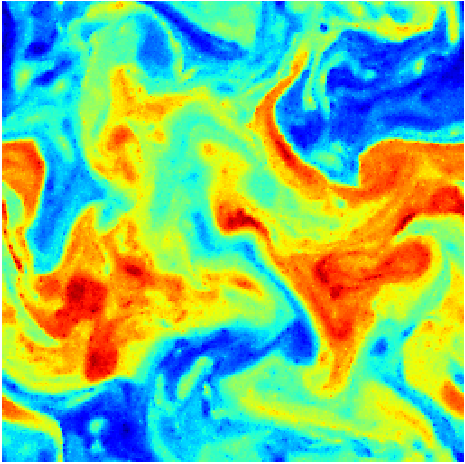}
        \caption{EnSF+DL}
    \end{subfigure}
    \begin{subfigure}[t]{0.2\textwidth}
        \centering
        \includegraphics[width=\textwidth]{  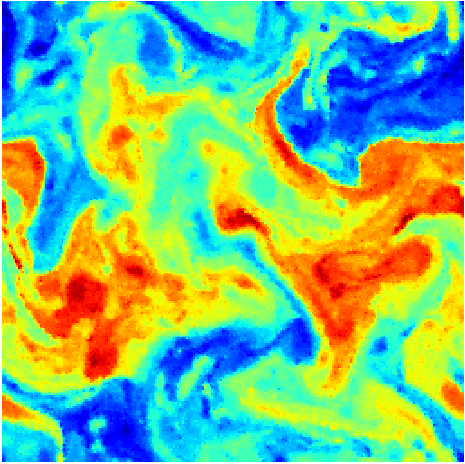}
        \caption{EnSF+NS}
    \end{subfigure}
    \vspace{-0.4cm}
    \caption{Snapshot at filtering step 100 of $(\mathbf{C}_{10})$}
    \label{snap:linear_N256_3hrly_25per}
    \vspace{-0.3cm}
\end{figure}
\begin{figure}[h!]
    \centering
    \begin{subfigure}[t]{0.8\textwidth}
        \centering
        \includegraphics[width=\textwidth]{ separate_legend.png}
    \end{subfigure} 
    \begin{subfigure}[t]{0.25\textwidth}
        \centering
        \includegraphics[width=\textwidth]{  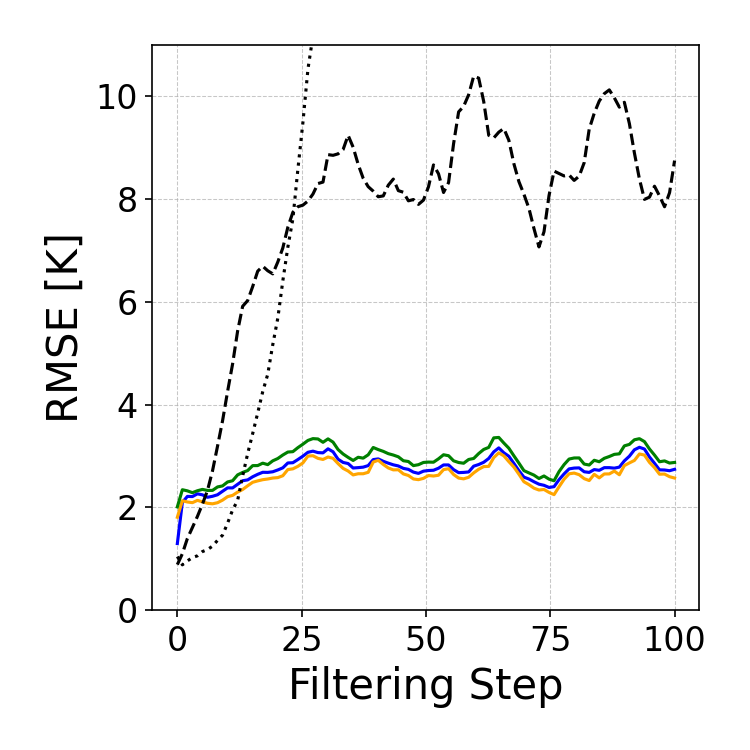}
        \caption{Total RMSE}
    \end{subfigure}
    \begin{subfigure}[t]{0.25\textwidth}
        \centering
        \includegraphics[width=\textwidth]{  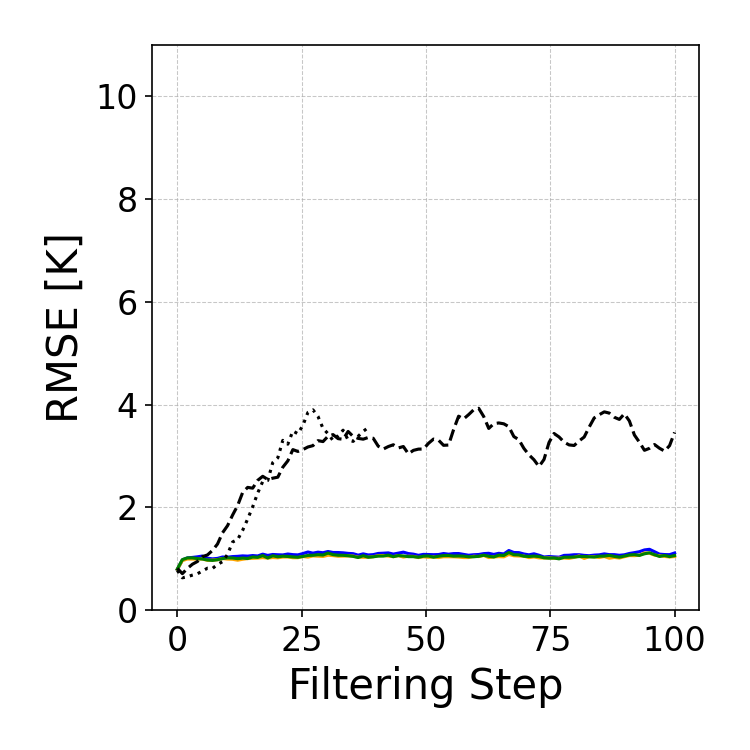}
        \caption{observed RMSE}
    \end{subfigure}
    \begin{subfigure}[t]{0.25\textwidth}
        \centering
        \includegraphics[width=\textwidth]{  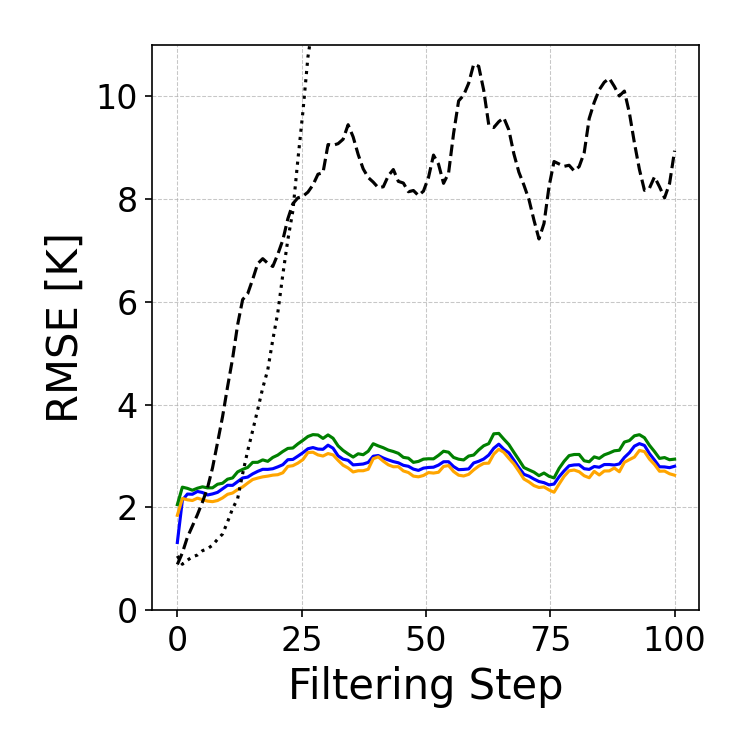}
        \caption{unobserved RMSE}
    \end{subfigure}
    \vspace{-0.5cm}
    \caption{RMSE of $(\mathbf{C}_{11})$}
    \label{fig:rmse linear 256 12hourly 5}
    \vspace{-0.3cm}
\end{figure}
\begin{figure}[h!]
    \centering
    \begin{subfigure}[t]{0.2\textwidth}
        \centering
        \includegraphics[width=\textwidth]{  truth_linear_N256_12hrly_5per.png}
        \caption{Truth}
    \end{subfigure}
    \begin{subfigure}[t]{0.2\textwidth}
        \centering
        \includegraphics[width=\textwidth]{  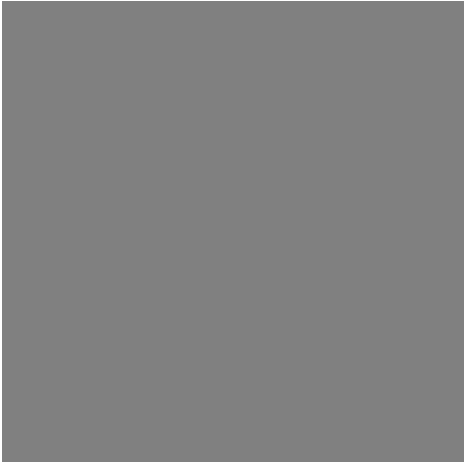}
        \caption{LETKF}
    \end{subfigure}
    \begin{subfigure}[t]{0.2\textwidth}
        \centering
        \includegraphics[width=\textwidth]{  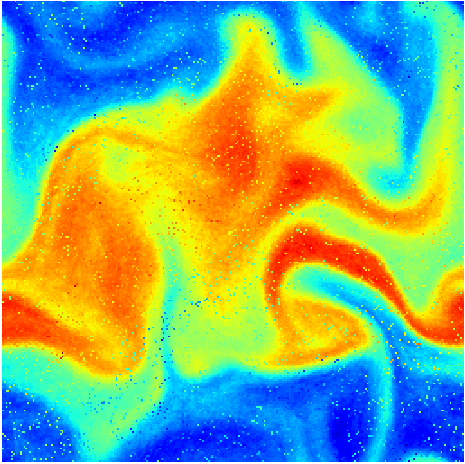}
        \caption{EnSF Only}
    \end{subfigure}\\
    \begin{subfigure}[t]{0.2\textwidth}
        \centering
        \includegraphics[width=\textwidth]{  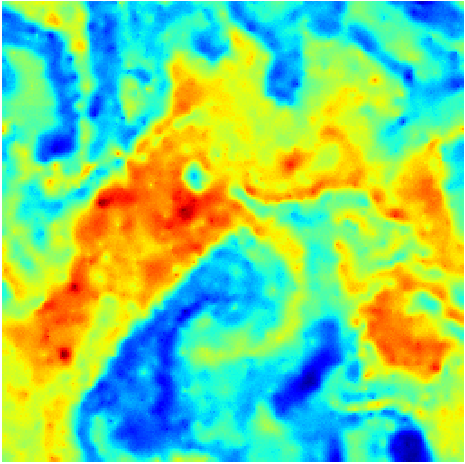}
        \caption{EnSF+Bi}
    \end{subfigure}
    \begin{subfigure}[t]{0.2\textwidth}
        \centering
        \includegraphics[width=\textwidth]{  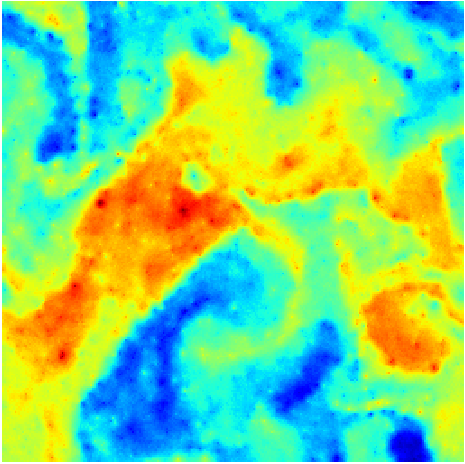}
        \caption{EnSF+DL}
    \end{subfigure}
    \begin{subfigure}[t]{0.2\textwidth}
        \centering
        \includegraphics[width=\textwidth]{  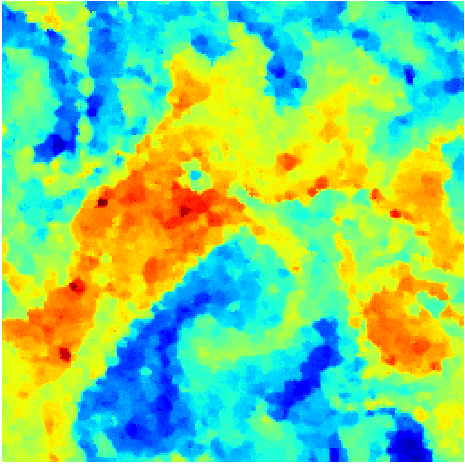}
        \caption{EnSF+NS}
    \end{subfigure}
    \vspace{-0.4cm}
    \caption{Snapshot at filtering step 100 of $(\mathbf{C}_{11})$}
    \label{snap:linear_N256_12hrly_5per}
    \vspace{-0.3cm}
\end{figure}
\begin{figure}[h!]
    \centering
    \begin{subfigure}[t]{0.8\textwidth}
        \centering
        \includegraphics[width=\textwidth]{ separate_legend.png}
    \end{subfigure} 
    \begin{subfigure}[t]{0.25\textwidth}
        \centering
        \includegraphics[width=\textwidth]{  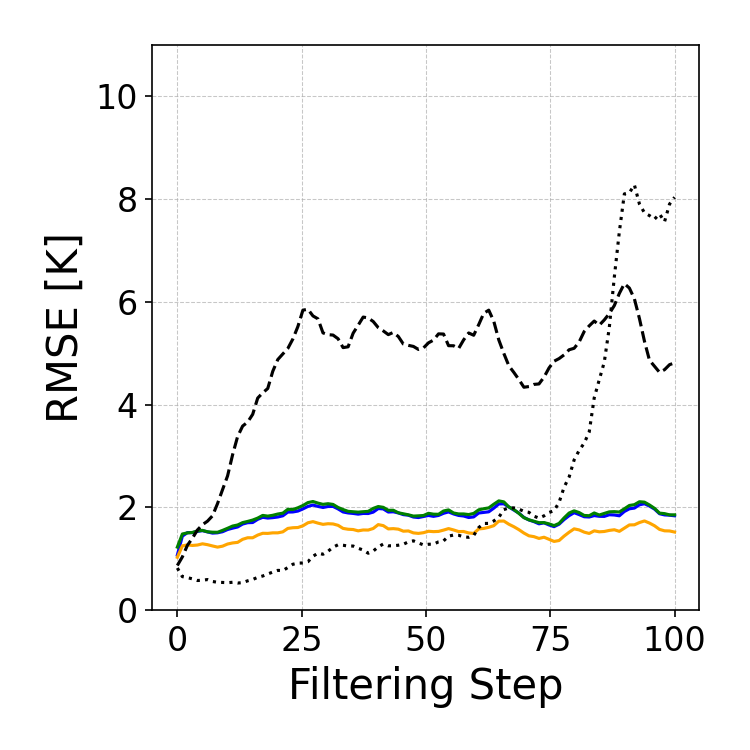}
        \caption{Total RMSE}
    \end{subfigure}
    \begin{subfigure}[t]{0.25\textwidth}
        \centering
        \includegraphics[width=\textwidth]{  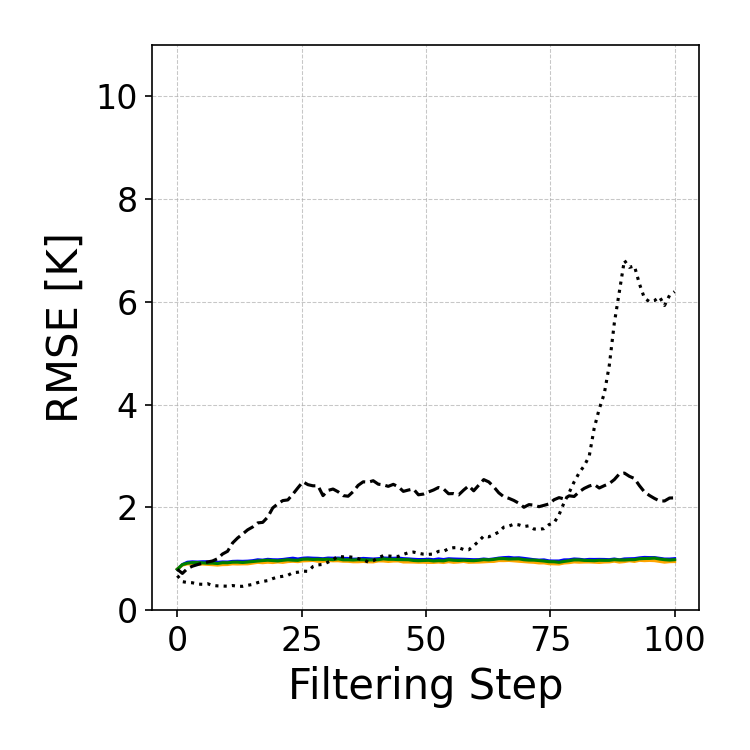}
        \caption{observed RMSE}
    \end{subfigure}
    \begin{subfigure}[t]{0.25\textwidth}
        \centering
        \includegraphics[width=\textwidth]{  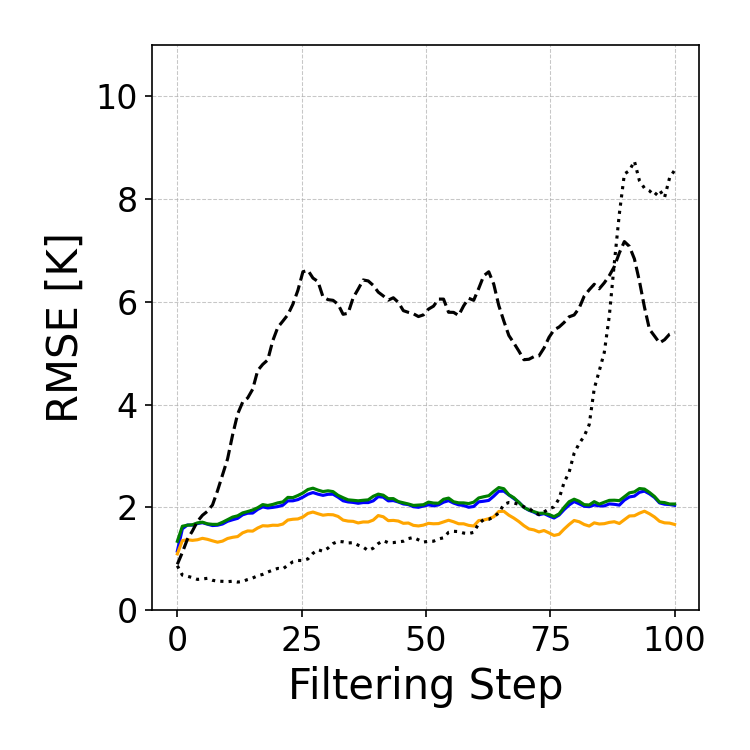}
        \caption{unobserved RMSE}
    \end{subfigure}
    \vspace{-0.5cm}
    \caption{RMSE of $(\mathbf{C}_{12})$}
    \label{fig:rmse linear 256 12hourly 25}
    \vspace{-0.3cm}
\end{figure}
\begin{figure}[h!]
    \centering
    \begin{subfigure}[t]{0.2\textwidth}
        \centering
        \includegraphics[width=\textwidth]{  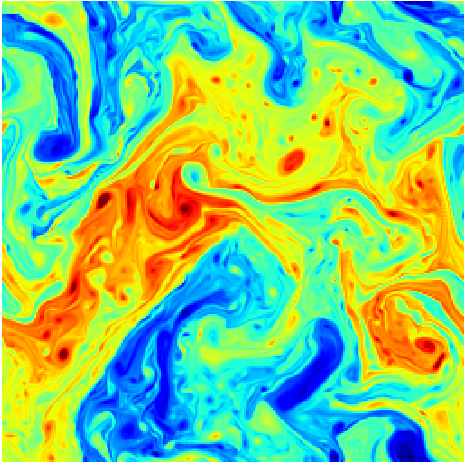}
        \caption{Truth}
    \end{subfigure}
    \begin{subfigure}[t]{0.2\textwidth}
        \centering
        \includegraphics[width=\textwidth]{  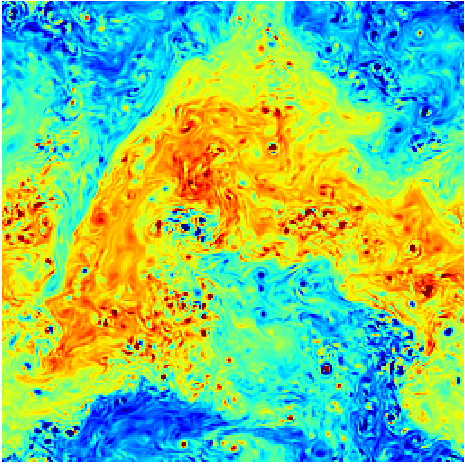}
        \caption{LETKF}
    \end{subfigure}
    \begin{subfigure}[t]{0.2\textwidth}
        \centering
        \includegraphics[width=\textwidth]{  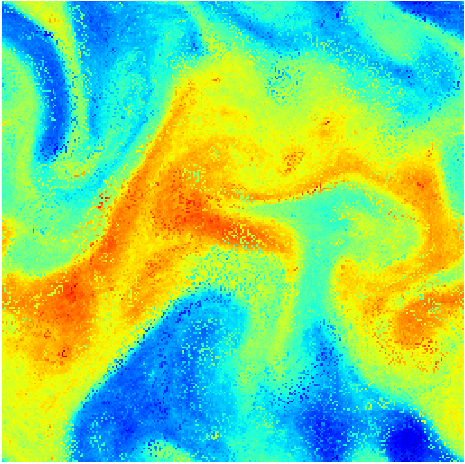}
        \caption{EnSF Only}
    \end{subfigure}\\
    \begin{subfigure}[t]{0.2\textwidth}
        \centering
        \includegraphics[width=\textwidth]{  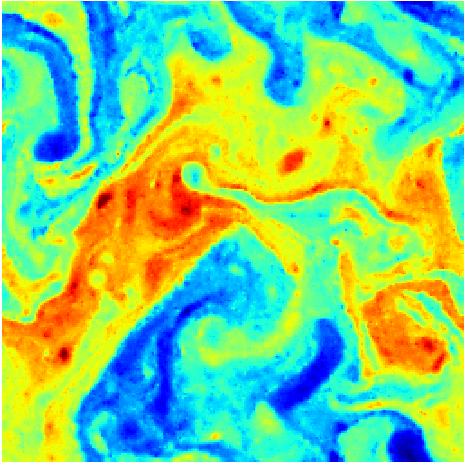}
        \caption{EnSF+Bi}
    \end{subfigure}
    \begin{subfigure}[t]{0.2\textwidth}
        \centering
        \includegraphics[width=\textwidth]{  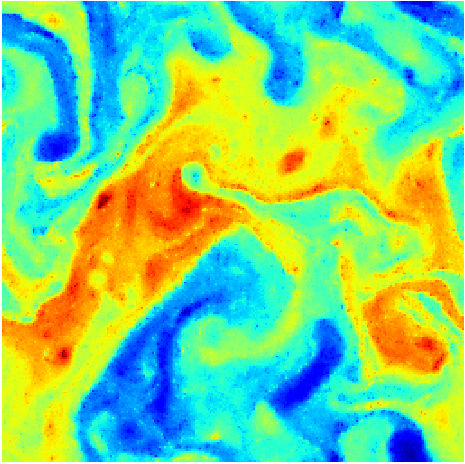}
        \caption{EnSF+DL}
    \end{subfigure}
    \begin{subfigure}[t]{0.2\textwidth}
        \centering
        \includegraphics[width=\textwidth]{  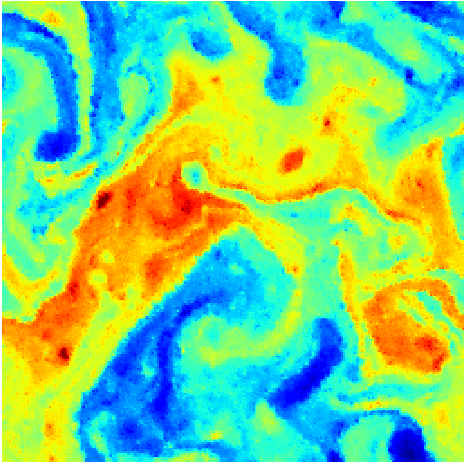}
        \caption{EnSF+NS}
    \end{subfigure}
    \vspace{-0.4cm}
    \caption{Snapshot at filtering step 100 of $(\mathbf{C}_{12})$}
    \label{snap:linear_N256_12hrly_25per}
    \vspace{-0.3cm}
\end{figure}
\begin{figure}[h!]
    \centering
    \begin{subfigure}[t]{0.8\textwidth}
        \centering
        \includegraphics[width=\textwidth]{ separate_legend.png}
    \end{subfigure} 
    \begin{subfigure}[t]{0.25\textwidth}
        \centering
        \includegraphics[width=\textwidth]{  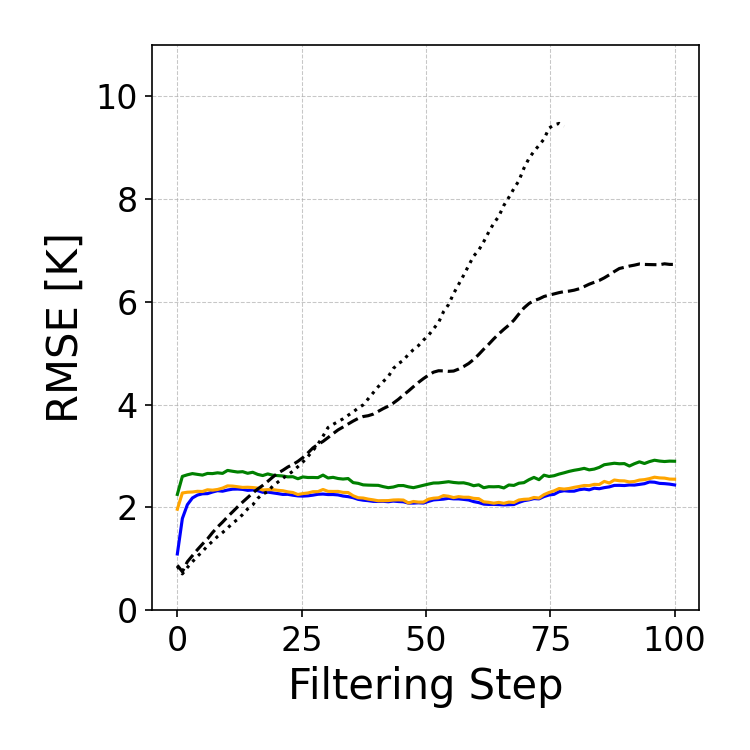}
        \caption{Total RMSE}
    \end{subfigure}
    \begin{subfigure}[t]{0.25\textwidth}
        \centering
        \includegraphics[width=\textwidth]{  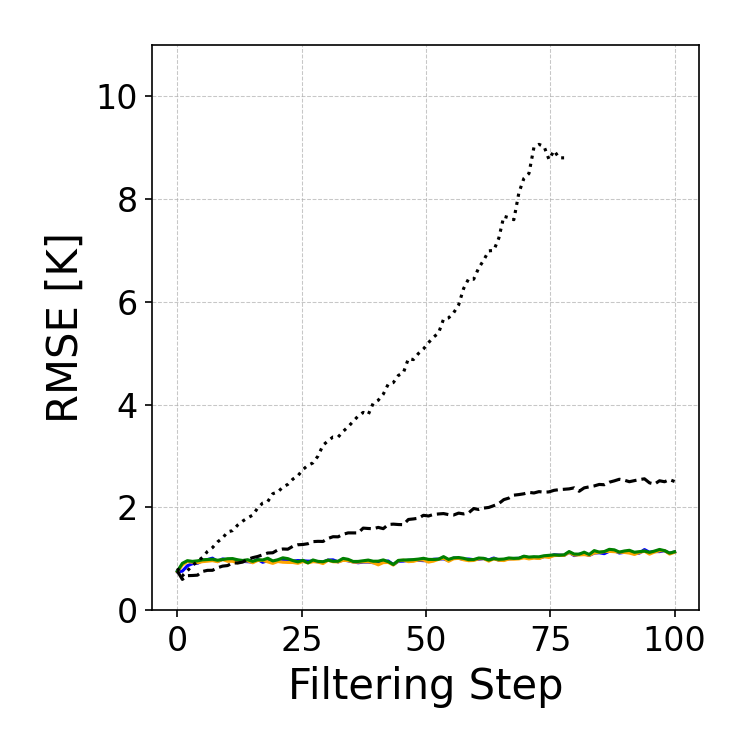}
        \caption{observed RMSE}
    \end{subfigure}
    \begin{subfigure}[t]{0.25\textwidth}
        \centering
        \includegraphics[width=\textwidth]{  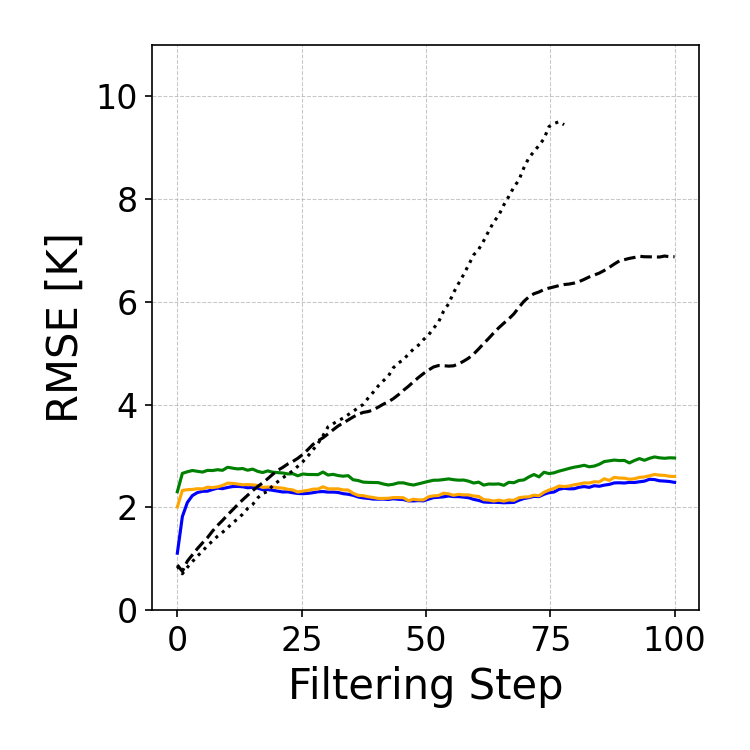}
        \caption{unobserved RMSE}
    \end{subfigure}
    \vspace{-0.5cm}
    \caption{RMSE of $(\mathbf{C}_{13})$}
    \label{fig:rmse arctan 256 3hourly 5}
    \vspace{-0.3cm}
\end{figure}
\begin{figure}[h!]
    \centering
    \begin{subfigure}[t]{0.2\textwidth}
        \centering
        \includegraphics[width=\textwidth]{  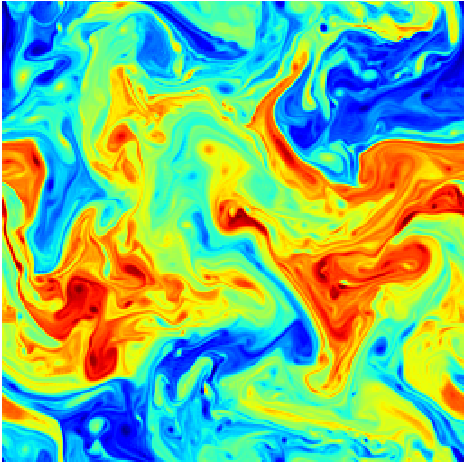}
        \caption{Truth}
    \end{subfigure}
    \begin{subfigure}[t]{0.2\textwidth}
        \centering
        \includegraphics[width=\textwidth]{  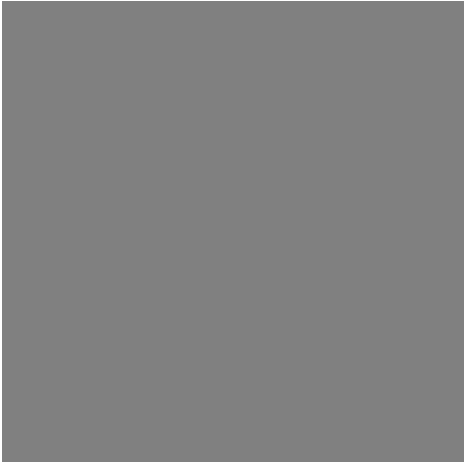}
        \caption{LETKF}
    \end{subfigure}
    \begin{subfigure}[t]{0.2\textwidth}
        \centering
        \includegraphics[width=\textwidth]{  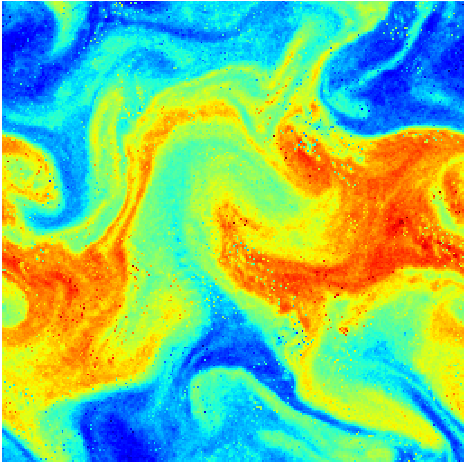}
        \caption{EnSF Only}
    \end{subfigure}\\
    \begin{subfigure}[t]{0.2\textwidth}
        \centering
        \includegraphics[width=\textwidth]{  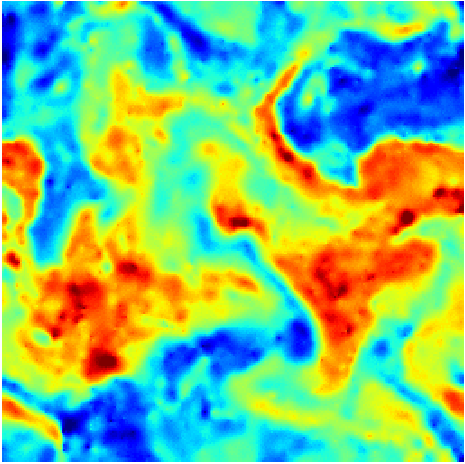}
        \caption{EnSF+Bi}
    \end{subfigure}
    \begin{subfigure}[t]{0.2\textwidth}
        \centering
        \includegraphics[width=\textwidth]{  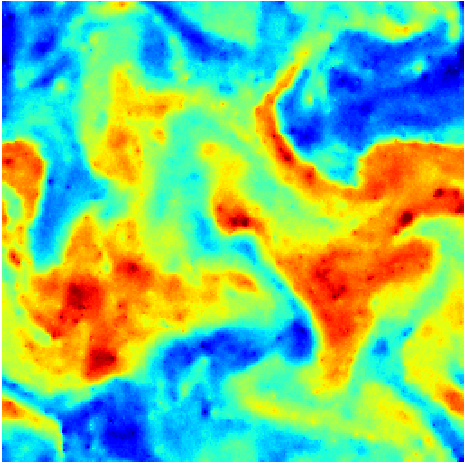}
        \caption{EnSF+DL}
    \end{subfigure}
    \begin{subfigure}[t]{0.2\textwidth}
        \centering
        \includegraphics[width=\textwidth]{  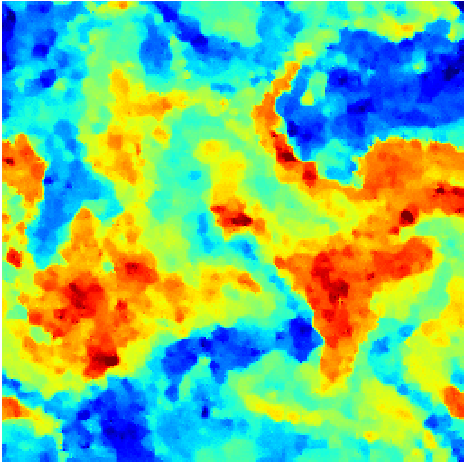}
        \caption{EnSF+NS}
    \end{subfigure}
    \vspace{-0.4cm}
    \caption{Snapshot at filtering step 100 of $(\mathbf{C}_{13})$}
    \label{snap:Arctan_N256_3hrly_5per}
    \vspace{-0.3cm}
\end{figure}
\begin{figure}[h!]
    \centering
    \begin{subfigure}[t]{0.8\textwidth}
        \centering
        \includegraphics[width=\textwidth]{ separate_legend.png}
    \end{subfigure} 
    \begin{subfigure}[t]{0.25\textwidth}
        \centering
        \includegraphics[width=\textwidth]{  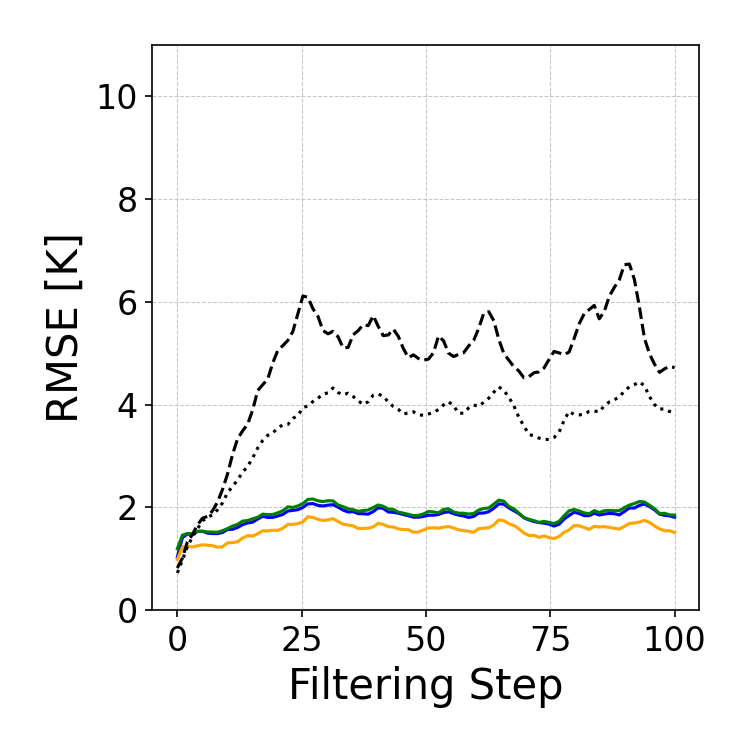}
        \caption{Total RMSE}
    \end{subfigure}
    \begin{subfigure}[t]{0.25\textwidth}
        \centering
        \includegraphics[width=\textwidth]{  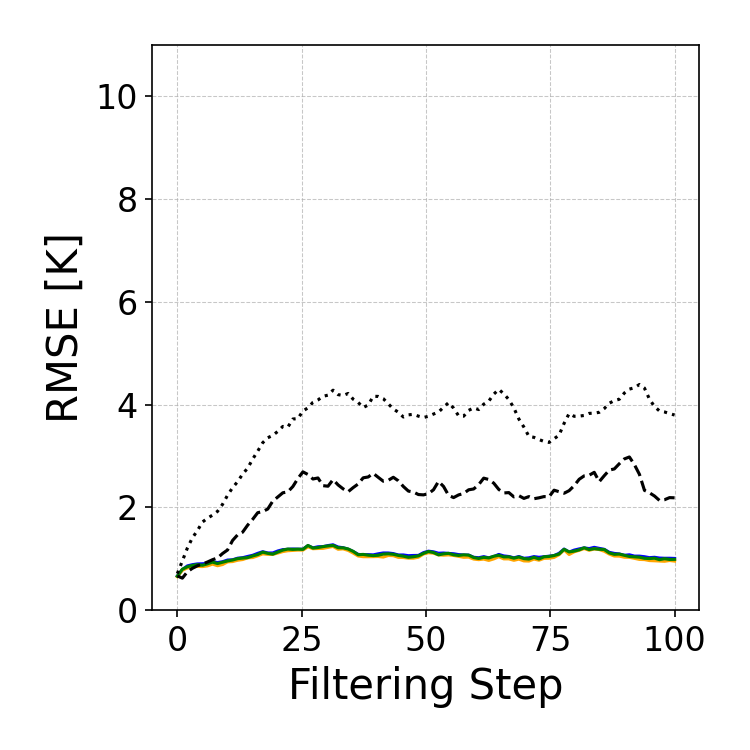}
        \caption{observed RMSE}
    \end{subfigure}
    \begin{subfigure}[t]{0.25\textwidth}
        \centering
        \includegraphics[width=\textwidth]{  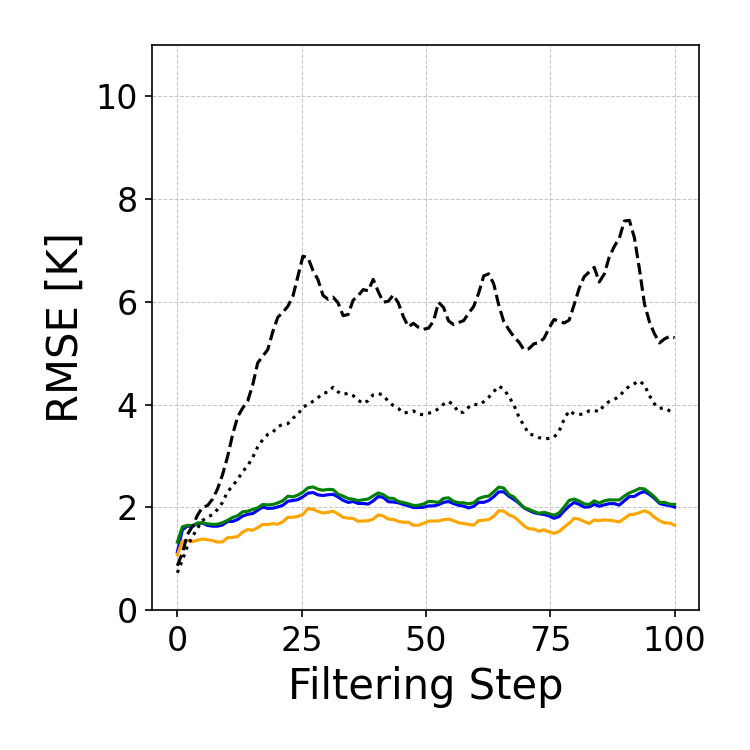}
        \caption{unobserved RMSE}
    \end{subfigure}
    \vspace{-0.5cm}
    \caption{RMSE of $(\mathbf{C}_{16})$}
    \label{fig:rmse arctan 256 12hourly 25}
    \vspace{-0.3cm}
\end{figure}
\begin{figure}[h!]
    \centering
    \begin{subfigure}[t]{0.2\textwidth}
        \centering
        \includegraphics[width=\textwidth]{  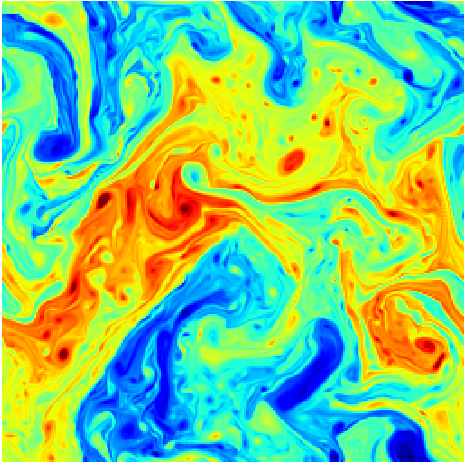}
        \caption{Truth}
    \end{subfigure}
    \begin{subfigure}[t]{0.2\textwidth}
        \centering
        \includegraphics[width=\textwidth]{  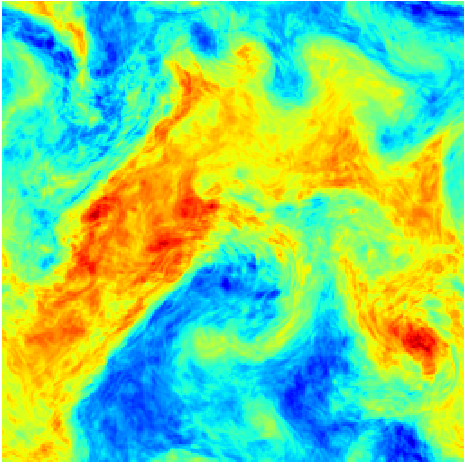}
        \caption{LETKF}
    \end{subfigure}
    \begin{subfigure}[t]{0.2\textwidth}
        \centering
        \includegraphics[width=\textwidth]{  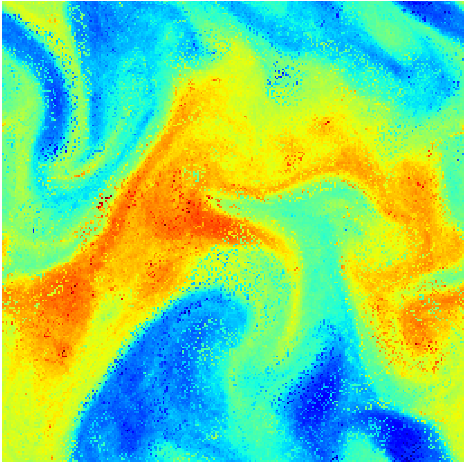}
        \caption{EnSF Only}
    \end{subfigure}\\
    \begin{subfigure}[t]{0.2\textwidth}
        \centering
        \includegraphics[width=\textwidth]{  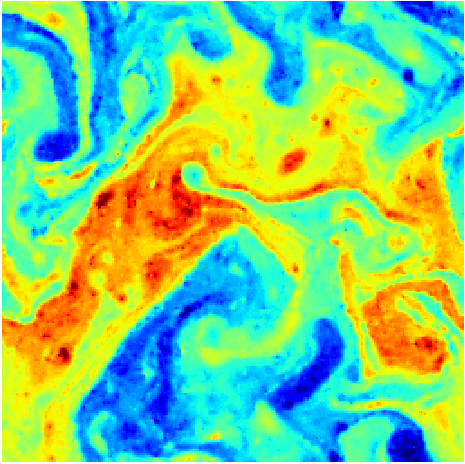}
        \caption{EnSF+Bi}
    \end{subfigure}
    \begin{subfigure}[t]{0.2\textwidth}
        \centering
        \includegraphics[width=\textwidth]{  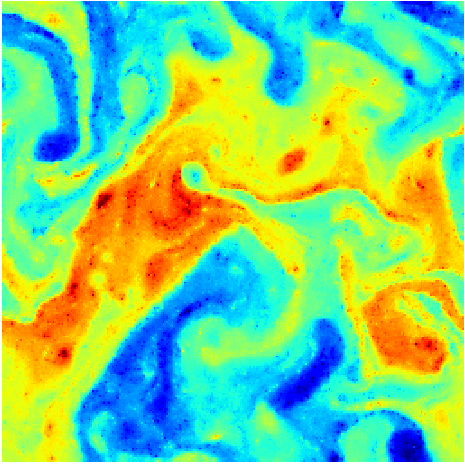}
        \caption{EnSF+DL}
    \end{subfigure}
    \begin{subfigure}[t]{0.2\textwidth}
        \centering
        \includegraphics[width=\textwidth]{  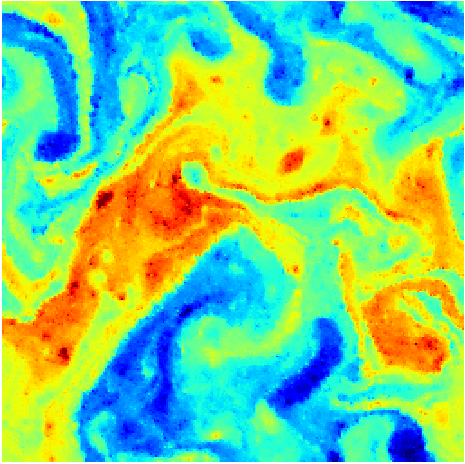}
        \caption{EnSF+NS}
    \end{subfigure}
    \vspace{-0.4cm}
    \caption{Snapshot at filtering step 100 of $(\mathbf{C}_{16})$}
    \label{snap:Arctan_N256_12hrly_25per}
    \vspace{-0.3cm}
\end{figure}
\end{document}